\documentclass[superscriptaddress,showpacs,nofootinbib]{revtex4-1}

\usepackage{amsmath,amsthm,amssymb,bm,hyperref,graphicx,color}
\usepackage[toc,page]{appendix}

\newcommand{\be}{\begin{equation}}
\newcommand{\ee}{\end{equation}}

\newtheorem{thm}{Theorem}

\newcommand{\lam}{\lambda}
\newcommand{\inti}{\int_{-\infty}^{+\infty}}

\newcommand{\6}{\partial}
\newcommand{\R}{\textsf{R}}
\newcommand{\Lo}{\textsf{L}}
\newcommand{\T}{\textsf{T}}
\newcommand{\Pe}{\textsf{P}}
\newcommand{\tr}{\textsf{t}}
\newcommand{\ov}{\overline}
\newcommand{\af}{\mathfrak{a}}
\newcommand{\fb}{\mathfrak{a}_{ph}}
\newcommand{\fc}{\mathfrak{a}_s}
\newcommand{\eq}{\varepsilon_0'(q)}
\newcommand{\la}{\lambda}
\newcommand{\m}{\mu}
\newcommand{\uu}{u'}

\begin{document}

\title{Correlation lengths of the repulsive one-dimensional Bose gas}

\author{Ovidiu I.~P\^{a}\c{t}u}
\affiliation{Institute for Space Sciences, Bucharest-M\u{a}gurele, R
077125, Romania}
\author{Andreas Kl\"umper}
\affiliation{Fachbereich C – Physik, Bergische Universit\"at  Wuppertal,
42097 Wuppertal, Germany}


\begin{abstract}
We investigate the large-distance asymptotic behavior of the static density-density
and field-field correlation functions in the one-dimensional Bose gas at finite
temperature. The asymptotic expansions of the Bose gas correlators
are obtained performing a specific continuum limit in the similar low-temperature
expansions  of the longitudinal and transversal correlation functions of the XXZ spin chain.
In the lattice system the correlation lengths are computed as ratios of the
largest and next-largest eigenvalues of the XXZ spin chain quantum transfer matrix.
In both cases, lattice and continuum, the correlation lengths are expressed in terms of
solutions of Yang-Yang type \cite{YY4} non-linear integral equations which are
easily implementable numerically.
\end{abstract}

\maketitle


\section{Introduction}

In the last decade we have witnessed significant advances in the
field of trapped ultracold gases \cite{BDZ} opening  new
avenues for the investigation of low-dimensional physical systems
which can be well approximated by integrable models. The paradigmatic
example is the Bose gas with contact interaction \cite{LL},
also known as the Lieb-Liniger model, whose
experimental realization  \cite{tMSK,tKWW1,tP,tLOH,tPRD,tAetal}
has spurred renewed interest in computing  physical properties
which are experimentally accessible. In particular, the correlation
functions, which can be measured using interference
\cite{tPAE,tIGD,tHetal,tDetal}, analysis of particle losses \cite{tLOH,tHetal1},
photoassociation \cite{tKWW2}, Bragg and photoemission spectroscopy
\cite{DADSC,tPetal,tCFFFI,tFetal1,tEetal}, density fluctuation statistics
\cite{AJKB,JABKB,Armijo1,Armijo2}, time-of-flight correlation statistics \cite{HDMBT} and
scanning electron microscopy \cite{GWEVBO} are extremely important.

Despite the integrability of the model the calculation of the
correlation functions is an extremely challenging problem which
remains unsolved to this day. Significant simplifications occur
in the case of infinite repulsion when the system is equivalent
to free fermions. In this case the correlators can be expressed
as Fredholm or Toeplitz determinants and the asymptotic behavior can be
extracted from the solution of an associated Riemann-Hilbert
problem \cite{Le1,Le2,VT1,VT2,JMMS,IIK,IIK0,IIK1,IIKV}. Similar
results, albeit in a non-rigorous fashion,  can be derived using
the replica method \cite{Ga1}.

The introduction of the algebraic Bethe ansatz (ABA) provided the
necessary tools to tackle the harder problem of calculating
the correlation functions of integrable models away from the
free fermion point \cite{BK1,IK1,Ka,KBI}. At zero temperature, the members
of the  Lyon group  (Kitanine, Kozlowski, Maillet, Slavnov and Terras),
making use of the results obtained in \cite{KMT1,KMT2,KMST1,KMST2,KKMST2}
derived in \cite{KKMST} the asymptotic behavior of the static density
correlators in the repulsive Lieb-Liniger
model and similar results for the longitudinal correlation of the XXZ
spin chain. The large-distance and long-time asymptotic analysis of the
density-density and  field-field correlators was performed in \cite{KT,KKK1,KKMSTaa}.
In all cases these exact results reproduce and generalize the predictions
of the Tomonaga-Luttinger liquid(TLL)/Conformal Field Theory (CFT) approach
\cite{BIK, BM1,BM2}. A method of determining the ``non-universal"
prefactors appearing in the TLL/CFT expansion was introduced
in \cite{SGCI1,SGCI2} and, also, very recently, in \cite{KKMSTa}.

The temperature dependent correlation functions of a system
characterized by a Hamiltonian $H$ are defined as
\be\label{corrT}
\langle\mathcal{O}\rangle_T=\frac{\sum\langle\Omega|
\mathcal{O}|\Omega\rangle e^{-E/T}}{\sum e^{-E/T}}\, ,
\ee
where $\mathcal{O}$ is a local operator, the sum is over all the
eigenstates $|\Omega\rangle$  of the Hamiltonian, and $E$ their
respective energies. The summation appearing in  (\ref{corrT})
makes the calculation of  temperature dependent correlation
functions extremely difficult. However, in the case of the interacting
Bose gas we can circumvent this problem in two ways. First, it
can be shown, see  Chap.~I of \cite{KBI} and references therein,
that in the thermodynamic limit (\ref{corrT}) can be replaced by
\be
\langle\mathcal{O}\rangle_T=\frac{\langle\Omega_T|\mathcal{O}|
\Omega_T\rangle}{\langle\Omega_T|\Omega_T\rangle}
\ee
where $|\Omega_T\rangle$ is any of the eigenstates corresponding to
thermal  equilibrium. This
allowed the  authors of \cite{KMS1,KMS2} to employ a method similar
with the zero  temperature analysis performed in \cite{KKMST} to
obtain the asymptotic expansion of the generating functional of
density  correlators. The second method
utilizes the quantum transfer matrix (QTM) and the connection between
the XXZ  spin chain and the one-dimensional Bose gas. Introduced
and developed in \cite{MS,K1}, the QTM, in
particular its spectrum, is an extremely important  tool in the
investigation of temperature dependent properties of lattice systems.
The free energy of the system is related to the largest eigenvalue of
the QTM \cite{K1,K2} and the correlation lengths of the Green's
functions can be obtained as ratios of the largest and  next-largest
eigenvalues \cite{K2,KSS}.
At the same time the QTM is a fundamental
ingredient in obtaining multiple integral representations for
temperature dependent correlation functions \cite{GKS1,GKS2,KK}.
Even though there is no QTM equivalent for continuous systems we can
use the fact that the one-dimensional Bose gas can be  obtained in a
specific continuum limit of the XXZ spin-chain \cite{SBGK,KS1}. Performing
this continuum limit in the non-linear integral equations (NLIEs)
characterizing the eigenvalues of the  XXZ spin-chain QTM we will obtain
the spectrum  of what we will call the ``continuum" QTM,  from which we
can calculate the thermodynamics and the correlation  lengths of the
continuum  system. It is precisely this method that we will use in
this paper to  obtain the asymptotic expansion of the temperature
dependent density-density  and field-field correlation functions in the
one-dimensional Bose gas. We should mention that the same  scaling limit
was used in \cite{P1} to obtain the k-body local correlators,
{\it i.e.,} correlation functions of the type $\langle(\Psi^\dagger(0))^k
(\Psi(0))^k\rangle_T$, for  $k\leq 4$ using,  however, a totally different
method than ours. For $k\leq 3$ local correlators were first calculated in
\cite{GS1,KGDS,CSZ1,CSZ2,KCI}. Other important results concerning the correlation
functions of the 1D Bose gas can be found in
\cite{Kb,KSa,KKS,EKL,IS,S1,KBI,CC1,CC2,CCS,GS2,IG1,KPKG,KKG1,SGDVK,CB1,DSGDDK,CCGOR,KMTa,KMTb}.

The plan of the paper is as follows. In the next section we
introduce the one-dimensional Bose gas and present the  asymptotic
expansions for the correlation functions which constitute the main
results of this paper. In Sec.~\ref{XXZ} we review the XXZ spin chain
and introduce the continuum limit which
allows for the derivation of the Bose gas results. In Sec.~\ref{sQTM}
we introduce the XXZ spin chain QTM and obtain
NLIEs for the largest and next-largest eigenvalues
from which the correlation lengths can be extracted. The validity of
the asymptotic expansions is checked in Sec.~\ref{CFT} by comparison with the
TLL/CFT predictions. Finally, the asymptotic behavior of the
correlators in the Bose gas is  obtained  by taking the continuum limit
in Sec.~\ref{sCL}. Some technical calculations are presented in several
appendices.


\section{The one-dimensional Bose gas and main result}

We consider a  one-dimensional system of bosons interacting via a
$\delta$-function potential with periodic boundary conditions.
The relevant Hamiltonian is
\be\label{hamLL}
H_{NLS}=\int_0^{l}\,  dx\,  \left[\6_x\Psi^\dagger(x)\6_x\Psi(x)+
c\Psi^\dagger(x)\Psi^\dagger(x)\Psi(x)\Psi(x)-\mu\Psi^\dagger(x)
\Psi(x)\right]\, ,
\ee
where $c>0$ is the coupling constant, $\mu$ the chemical potential,
$l$ the length of the system  and we have considered $\hbar=2m=1$,
with $m$ the mass of the particles. In (\ref{hamLL}) $\Psi^\dagger(x)$
and $\Psi(x)$ are  Bose fields satisfying the canonical commutation
relations
\[
[\Psi(x),\Psi^\dagger(x')]=\delta(x-x')\, ,\ \ \ \
[\Psi(x),\Psi(x')]=[\Psi^\dagger(x),\Psi^\dagger(x')]=0\, .
\]
The interacting one-dimensional Bose gas, also known as the Lieb-Liniger
or the quantum Non-Linear Schr\"odinger (NLS) model, is solvable by
Bethe ansatz \cite{LL,G,KBI}. In the case of $n$ particles  the energy
spectrum is given by
\be\label{enls}
\overline E(\{k\})=\sum_{j=1}^n \overline e_0(k_j)\, ,
\ \ \ \ \ \overline e_0(k)=k^2-\mu\, ,
\ee
with the quasimomenta $k_j$ satisfying the following set of Bethe ansatz equations (BAEs)
\be\label{BEnls}
e^{ik_j l}=\prod_{s\ne j}^n\frac{k_j-k_s+ic}{k_j-k_s-ic}\, ,
\ \ \ \ j=1,\cdots,n\, .
\ee
It is useful to present the logarithmic form of the BAEs
(\ref{BEnls})
\[
lk_j+\sum_{s=1}^n\overline\theta(k_j-k_s)=2\pi m_j\, ,
\]
where $m_j$ are integers or half-integers and the scattering phase
$\overline\theta(k)$
is defined by
\[
\overline\theta(k)=i\log\left(\frac{ic+k}{ic-k}\right)\, ,
\ \ \ \ \lim_{k\rightarrow\pm\infty}\overline\theta(k)=\pm \pi\, .
\]
At zero temperature and fixed number of particles $n$ the ground state
is obtained when the (half) integers take the values $m_j=j-(n+1)/2\,
, j=1,\cdots, n$ \cite{LL}. In the thermodynamic limit $l,n\rightarrow
\infty\, ,$ with their ratio finite $\overline D=n/l,$ the values of the momenta
$k_j$ condense in the interval $[-\overline q,\overline q]$ called the
Fermi zone or Dirac sea and the following integral equation for
the density of particles in momentum space can be derived:
\be\label{drnls}
\overline\rho(k)-\frac{1}{2\pi}\int_{-\overline q}^{\overline q}
\overline K(k-k')\overline \rho(k')\,
 dk'=\frac{1}{2\pi}\, ,\ \ \ \ \ \ \ \
\overline K(k-k')=\frac{d}{dk}\overline\theta(k-k')=
\frac{2c}{(k-k')^2+c^2}\, .
\ee
The Fermi momentum $\overline q$ can be obtained as a unique function
of $\overline D$, the
density of particles, via
$
\overline D=n/l=\int_{-\overline q}^{\overline q}\overline \rho(k)\,
dk\,.
$

At finite temperature the thermodynamics of the model was calculated
in \cite{YY4} (for a rigorous derivation, see \cite{DLP,KW}). The grand-canonical
potential per length is given by
\be\label{Pnls}
\phi(\mu,T)=-\frac{T}{2\pi}\inti \log\left(1+e^{-\overline\varepsilon(k)/T}
\right)\, dk\, ,
\ee
with $\overline\varepsilon(k)$, the dressed energy, satisfying the
Yang-Yang equation
\be\label{YYnls}
\overline\varepsilon(k)=k^2-\mu-\frac{T}{2\pi}\int_{\mathbb{R}}
\overline{K}(k-k')\log
\left(1+e^{-\overline{\varepsilon}(k')/T}\right)dk'\, .
\ee

\subsection{Main result}\label{MR}

The main result of this paper is the computation of the large-distance
asymptotic behavior  of the correlation functions in the 1D
Bose gas at finite temperature. Due to the fact that the derivation of
the asymptotic expansions is quite involved we prefer to present these
results in the beginning of the paper. The interested reader can find the
details in the following sections.

We will start with the static density-density correlation function,
$\langle j(x)j(0)\rangle_T$, with $j(x)=\Psi^\dagger(x)
\Psi(x)$. Consider the following set of functions $\overline u_i(k)$ satisfying the
nonlinear integral equations
\be\label{unls}
\overline{u}_i(k)=k^2-\mu+iT\sum_{j=1}^r\overline\theta(k-k^+_j)
-iT\sum_{j=1}^r\overline\theta(k-k^-_j)
-\frac{T}{2\pi}\int_{\mathbb{R}}\overline{K}(k-k')
\log\left(1+e^{-\overline{u}_i(k')/T}\right)dk'
\, .
\ee
The $2r$ parameters, $\{k^+_j\}_{j=1}^r\,  (\{k^-_j\}_{j=1}^r) $ appearing in
Eq.~(\ref{unls}) are
located in the upper (lower) half of the complex plane
and satisfy the constraint
\[
1+e^{-\overline u_i(k^\pm_j)/T}=0.
\]
For a given $r$, the previous equation has more than $2r$ solutions, the
subscript $i$ labels all the possible choices of solutions for all
$r=1,2,\cdots$. Note that the NLIEs (\ref{unls})
are almost identical with the Yang-Yang equation for the dressed energy
(\ref{YYnls}) with the exception of  the additional driving terms.
The large distance asymptotic expansion for the density-density correlation
function has the form
\be\label{AEdd}
\langle j(x)j(0)\rangle_T=const+\sum_i \tilde A_i\,
  e^{-\frac{x}{\xi^{(d)}[\overline{u}_i]}}\, ,
\ \ \ \ \ x\rightarrow\infty\, ,
\ee
where $\tilde A_i$ are distance independent amplitudes which cannot be
obtained using our method and the correlation lengths are given by
\be\label{cldd}
\frac{1}{\xi^{(d)}[\overline{u}_i]}=-\frac{1}{2\pi}\int_{\mathbb{R}}\log
\left(\frac{1+e^{-\overline{u}_i(k)/T}}{1+\, e^{-\overline{\varepsilon}(k)/T}}
\right)\, dk-i\sum_{j=1}^r k^+_j+i\sum_{j=1}^r k^-_j\, ,
\ee
with $\overline\varepsilon(k)$ the dressed energy satisfying (\ref{YYnls}).
Comparison with the TLL/CFT expansion (\ref{inte1}) and
other exact results (Chap.~XVII of \cite{KBI}) allows the identification of the
constant term with $\langle j(0)\rangle_T^2$. The leading  terms in the expansion
(\ref{AEdd}) are obtained considering $r=1$ in  Eq.~(\ref{unls}) with the
parameters $k^\pm_1$, satisfying $1+e^{\overline u_i(k^\pm_1)/T}=0$, closest
to the real axis.

A few remarks are in order.
Using a different method almost identical equations were obtained by
Kozlowski, Maillet and Slavnov \cite{KMS1,KMS2} for the generating functional
of density correlators, $\langle e^{\varphi\int_0^x j(x')\, dx'}\rangle_T$,
from which the density correlator can be obtained via $\langle
j(x)j(0)\rangle_T=\frac{1}{2}\frac{\6^2}{\6x^2}\frac{\6^2}{\6\varphi^2}\left.
\langle e^{\varphi\int_0^xj(x')\, dx'}\rangle_T\right|_{\varphi=0}$ \footnote{
It should be remarked that the authors of \cite{KMS1,KMS2}  noticed that their
results which were derived using the asymptotic analysis of
a generalized sine-kernel  Fredholm determinant can be interpreted in
the framework of the QTM which is the primary object of this paper
.}.
The only difference between our equations and the ones derived in \cite{KMS1,KMS2}
is the presence of a renormalized chemical potential $\mu\rightarrow
\mu+\varphi T$ in the r.h.s.~of Eq.~(\ref{unls}).  As we will show in Appendix
\ref{agenfunc} a slight modification of our method allows for the derivation
of the asymptotic expansion for the generating functional.  However, in order
to not confuse the reader, we prefer here and in the following sections to
focus on the density and field correlators (see below) because it allows for
an almost similar treatment.

In the case of the field-field correlation function $\langle \Psi^\dagger(x)
\Psi(0)\rangle_T$ we introduce the  set of functions
$\overline v_i(k)$ satisfying the NLIEs
\be\label{vnls}
\overline{v}_i(k)=k^2-\mu\pm i\pi T+iT\overline{\theta}(k-k_0)+iT\sum_{j=1}^r\overline{\theta}
(k-k^+_j)-iT\sum_{j=1}^r\overline{\theta}(k-k^-_j)
-\frac{T}{2\pi}\int_{\mathbb{R}}\overline{K}(k-k')
\log\left(1+e^{-\overline{v}_i(k')/T}\right)dk'\, .
\,
\ee
The functions $\overline v_i(k)$  depend on $2r+1$ parameters: $k_0$ and
$\{k_j^+\}_{j=1}^r\, $ located in the upper
half of the complex plane and $\{k_j^-\}_{j=1}^r\, $
located in the lower half of the complex plane, satisfying the constraints
\[
 1+e^{-\overline v_i(k_0)/T}=0\, ,\ \  1+e^{-\overline v_i(k^\pm_j)/T}=0.
\]
In Eq.~(\ref{vnls}) we will consider the plus sign in front of the $i\pi T$
term when $k_0$ is in the first quadrant of the complex plane
$\Re k_0\geq 0\, ,\Im k_0\geq 0,$ and the minus sign when $k_0$
is in the second quadrant of the complex plane $\Re k_0< 0\, ,\Im
k_0\geq 0.$ As in the case of the functions $\overline u_i(k)$ the
subscript $i$ labels all the possible choices of roots for all $r=0,1,2,\cdots$.
The large distance asymptotic expansion of the field-field correlation
function has the form
\be\label{AEff}
\langle \Psi^\dagger(x)\Psi(0)\rangle_T=\sum_i \tilde B_i\,
 e^{-\frac{x}{\xi^{(s)}[\overline{v}_i]}}\, ,
\ \ \ \ \ x\rightarrow\infty\, ,
\ee
where $\tilde B_i$ are distance independent amplitudes which cannot be
obtained using our method and the correlation lengths are given by
\be\label{inte3}
\frac{1}{\xi^{(s)}[\overline{v}_i]}=-\frac{1}{2\pi}\int_{\mathbb{R}}\log
\left(\frac{1+e^{-\overline{v}_i(k)/T}}{1+\,
e^{-\overline{\varepsilon}(k)/T}}\right)\, dk
-i k_0-i\sum_{j=1}^rk^+_j+i\sum_{j=1}^r k^-_j\, .
\ee
Eqs.~(\ref{vnls}) and (\ref{inte3}) are valid at intermediate and high temperature.
At low-temperature it is possible that $k_0$  dives below the real axis. In this case
the following modifications should be made: in both equations
the integral should be  taken along a contour which is  the real axis with an indentation
such that $k_0$ is above the contour (also the indentation does not contain a solution
of $1+e^{-\overline{\varepsilon}(k)/T}=0$) and in Eq.~(\ref{vnls}) the plus sign in front of the $i\pi T$ term
is considered when $k_0$ is in the fourth quadrant of the complex plane $\Re k_0\geq 0\, ,\Im k_0\leq 0,$
and the minus sign when $k_0$ is in the third quadrant of the complex plane
$\Re k_0< 0\, ,\Im k_0\leq 0.$ Also, $k_0$, which satisfies $1+e^{-\overline
v_i(k_0)/T}=0$, is the closest solution to the real axis in the lower half-plane.
To our knowledge, the asymptotic
expansion  (\ref{AEff}) is new in the literature (the authors of \cite{KMS1,KMS2}
did not consider the case of the field-field correlation functions).
Extensive numerical studies and the low-temperature analysis (see Sec.~\ref{Checknls})
show that  $\Re\left( 1/ \xi^{(d)}[\overline{u}_i]\right)>0$ and
$\Re\left( 1/ \xi^{(s)}[\overline{v}_i]\right)>0$ for all
$\ov u_i(k)$ and $\ov v_i(k)$. In Sec.~\ref{Checknls} we will also show
that  (\ref{AEdd}) and (\ref{AEff}) agree with
the TLL/CFT predictions and other exact results.

\section{The XXZ spin chain}\label{XXZ}

The asymptotic expansions presented in the previous section were derived
by taking a specific continuum limit in the equivalent expansions of the
low-temperature transversal and longitudinal correlation functions of the XXZ
spin chain. In order to obtain the asymptotic behavior of the correlators in
the lattice model we will investigate the spectrum of the QTM.
Therefore, it is useful to review the  Bethe ansatz solution of the XXZ
spin chain and the associated QTM.

The integrable spin-1/2 XXZ chain in external longitudinal
magnetic field $h$ is characterized by the following Hamiltonian
\be\label{ham}
H(J,\Delta,h)=H^{(0)}(J,\Delta)-hS_z\, ,
\ee
where
\be\label{com}
H^{(0)}(J,\Delta)=J\sum_{j=1}^L\left[\sigma_x^{(j)}\sigma_x^{(j+1)}+\sigma_y^{(j)}\sigma_y^{(j+1)}
                  +\Delta(\sigma_z^{(j)}\sigma_z^{(j+1)}-1)\right]\, ,
                  \ \ \ \ S_z=\frac{1}{2}\sum_{j=1}^L\sigma_z^{(j)}\, .
\ee
We assume periodic boundary conditions and the number of lattice sites $L$ to be
even. The Hamiltonian (\ref{ham}) acts on the Hilbert space
$\mathcal{H}=(\mathbb{C}^2)^{\otimes L},$ $J>0$ fixes the energy scale
and $\Delta$ is the anisotropy. In Eq.~(\ref{com}), $\sigma_{x,y,z}^{(j)}$ are
local spin operators which act nontrivially only on the $j$-th lattice site
$\sigma_{x,y,x}^{(j)}=\mathbb{I}_2^{\otimes (j-1)}\otimes\sigma_{x,y,z}\otimes
\mathbb{I}_2^{\otimes (L-j)}$ with $\sigma_{x,y,z}$ the Pauli matrices
\[
\sigma_x=\left(\begin{array}{lr}0&1\\
                                 1&0
               \end{array}\right)\ \ \ \ \ \ \
\sigma_x=\left(\begin{array}{lr}0&-i\\
                                i& 0
               \end{array}\right)\ \ \ \ \ \ \
\sigma_x=\left(\begin{array}{lr} 1&0\\
                                 0&-1
               \end{array}\right)\, ,
\]
and $\mathbb{I}_2$ the $2$-by-$2$ unit matrix. $S_z$ commutes with $H^{(0)}(J,\Delta)$
and, therefore, does not affect the integrability of the model. Also, due  to the similarity
transformation $H(J,\Delta,h)\rightarrow V H(J,\Delta,-h) V^{-1}$ with
$V=\prod_{j=1}^L\sigma_x^{(j)},$ it is sufficient to consider only the case of positive magnetic field.
Another consequence is  that the  thermodynamics of the model does not depend on the sign of $h$.
In this paper we are  going to consider the massless regime of the XXZ spin chain
$|\Delta|<1$, parametrized by $\Delta=\cos \eta$ with $0<\eta<\pi$ and the magnetic field
$h$ smaller than the critical value $h_c=8J\cos^2(\eta/2)$.

The  Hamiltonian (\ref{ham}) is integrable and was solved by Yang and Yang in \cite{YY1,YY2,YY3}
with the help of the coordinate Bethe ansatz (for an ABA solution see \cite{EPAPS}).
The energy spectrum of the XXZ spin chain in magnetic
field is given by
\be\label{magnon}
E(\{\la\})=\sum_{j=1}^ne_0(\la_j)-h\frac{L}{2}\, ,\ \ \ \
e_0(\la)=\frac{2J\sinh^2(i\eta)}{\sinh(\la+i\eta/2)\sinh(\la-i\eta/2)}+h\, .
\ee
with the $\{\la_j\}_{j=1}^n$ parameters satisfying the Bethe equations
\be\label{BE}
\left(\frac{\sinh(\la_j-i\eta/2)}{\sinh(\la_j+i\eta/2)}\right)^L=
\prod_{s\ne j}^n\frac{\sinh(\la_j-\la_s-i\eta)}{\sinh(\la_j-\la_s+i\eta)}\, ,
\ \ \ \ \ \ \ j=1,\cdots,n\, .
\ee

\subsection{Ground-state properties}

The ground state of the XXZ spin chain at finite magnetization  is constructed
essentially in the same way as in the case of the Lieb-Liniger model. This means
that the (half) integers $m_j$ which appear in the logarithmic form of the
Bethe equations (\ref{BE})
\be\label{BEl}
Lp_0(\la_j)-\sum_{k=1}^n\theta(\la_j-\la_k)=2\pi m_j\, ,\ \ \ \  j=1,\cdots,n\,
\ee
fill all the possible values in the symmetric interval $-(n-1)/2\leq m_j\leq (n-1)/2.$
In Eq.~(\ref{BEl}) we have introduced the bare momentum $p_0(\la)$ and the scattering phase
$\theta(\la)$
\be\label{momentum}
p_0(\la)=i\log\left(\frac{\sinh(i\eta/2+\la)}{\sinh(i\eta/2-\la)}\right)\, ,\ \ \ \
 \theta(\la)=i\log\left(\frac{\sinh(i\eta+\la)}{\sinh(i\eta-\la)}\right)\, ,
\ee
where the branches of the logarithm are specified by the conditions
$\lim_{\la\rightarrow \infty}p_0(\la)=\pi-\eta\, $
and $\lim_{\la\rightarrow \infty}\theta(\la)=\pi-2\eta\, .$ The ground state
is characterized by real Bethe roots $\la_j$ which are contained in the
interval $[-q,q]$ called the Fermi zone. If we call every down spin a particle,
then the thermodynamic limit is characterized  by $L\rightarrow \infty\, ,$
$n\rightarrow \infty$ with constant density of particles $D=\lim_{L,n\rightarrow \infty}n/L.$
In the thermodynamic limit the Bethe roots fill densely the interval $[-q,q]$ and we can
introduce the spectral density of particles $\rho(\la)$ which satisfies the following
integral equation
\be\label{density}
\rho(\la)+\frac{1}{2\pi}\int_{-q}^qK(\la-\m)\rho(\m)\, d\m=\frac{1}{2\pi}p_0'(\la)\, , \ \ \
K(\la)=\theta'(\la)=\frac{\sin(2\eta)}{\sinh(\la+i\eta)\sinh(\la-i\eta)}\, .
\ee
The average density of particles is then $D=\int_{-q}^q\rho(\la)d\la$ from which the
Fermi boundary $q$ can be obtained.

In the presence of a magnetic field the magnetization of the ground state
is no longer fixed, it depends on the magnitude of $h$. In this case
the boundary of the Fermi zone $q$ can be defined by the requirement that
the energy of a hole at the Fermi boundary should be zero, $\varepsilon_0(\pm q)=0,$
where the dressed energy $\varepsilon_0(\la)$ satisfies the integral equation
\be\label{dressede}
\varepsilon_0(\la)+\frac{1}{2\pi}\int_{-q}^qK(\la-\m)\varepsilon_0(\m)\, d\m
=h-2Jp_0'(\la)\sin\eta\equiv e_0(\la)\, .
\ee
It can be shown that in the massless phase $(|\Delta|<1)$ considered in this paper
and $h$ smaller than the critical magnetic field $h_c,$ Eq.~(\ref{dressede}) has
a unique solution. When the magnetic field is vanishing the Fermi boundary
goes to infinity.

An important role in the analysis performed in Sec.~\ref{CFT} is played
by the dressed charge $Z(\la)$ defined by the following integral equation
\be\label{defz}
Z(\la)+\frac{1}{2\pi}\int_{-q}^qK(\la-\m)Z(\m)\, d\m=1\, ,\ \ \ \ \ \ Z(\pm q)=\mathcal{Z},
\ee
and the resolvent of the operator $I+\frac{1}{2\pi}K$ which satisfies
\be\label{defres}
R(\la,\m)+\frac{1}{2\pi}\int_{-q}^qK(\la-\nu)R(\nu,\m)\, d\nu=\frac{1}{2\pi}K(\la-\m)\, .
\ee
We will also make use of the dressed phase $F(\la|\m)$ defined by
\be\label{defdp}
F(\la|\m)+\frac{1}{2\pi}\int_{-q}^qK(\la-\nu)F(\nu|\m)\, d\nu=\frac{1}{2\pi}\theta(\la-\m)\, ,
\ee
and is connected with the dressed charge via
\be\label{dpident}
Z(\la)=1+F(\la|q)-F(\la|-q)\, ,\ \ \ \ \frac{1}{\mathcal{Z}}=1+F(q|q)+F(q|-q)\, .
\ee
A proof of the identities (\ref{dpident}) can be found in  \cite{KS,S},

\subsection{Continuum limit of the XXZ spin chain}\label{Continuum}

The fact that the Hamiltonian of the one-dimensional Bose gas can be obtained performing a
certain continuum limit in the Hamiltonian of the  XXZ spin chain was discovered long time
ago \cite{KS1}. In \cite{SBGK} it was  shown  that the Yang-Yang thermodynamics,
(\ref{Pnls}), (\ref{YYnls}), of the 1D  Bose gas can be obtained by
performing the same limit in the thermodynamics of the lattice model derived using the
QTM formalism. This is to be expected if we take into account that both models are
integrable and  that the BAEs and the energy spectrum of the Bose gas
can be obtained from the BAEs and energy spectrum of the XXZ spin-chain in
the continuum limit.
Moreover, the authors of \cite{SBGK}, see also \cite{SGK3}, derived
multiple integral representations for the correlation functions of the Bose gas from
equivalent expressions for the XXZ spin chain. In this paper we will employ a similar technique
to derive the large-distance asymptotic behavior of temperature dependent Green's functions
in the Bose gas from equivalent results for the XXZ spin chain.

The XXZ spin chain is characterized by five parameters: lattice constant $\delta$, number of
lattice sites $L$, strength of the interaction $J$, anisotropy $\Delta=\cos\eta$ and
magnetic field $h$. The Bose gas is  characterized by four  parameters: mass of the
particles $m=1/2$, physical length $l$, coupling strength $c$ and chemical potential $\mu$.
First, we will show how we can obtain the BAEs of the Bose gas (\ref{BEnls}) from (\ref{BE}).

Let $\epsilon\rightarrow 0$
be a small parameter.
The desired continuum limit is obtained considering
$\eta=\pi-\epsilon$, $\delta\rightarrow 0$ like $\mathcal{O}(\epsilon^2)$,
$L$ even, $L\rightarrow \infty$  like $\mathcal{O}(1/\epsilon^2)$ with $L\delta=l$ and
$c=\epsilon^2/\delta.$
Performing this limit together with the  reparametrization
of the Bethe roots $\la_j=\delta k_j/\epsilon$ in (\ref{BE}) we find
\begin{align*}
\left(\frac{\cosh(\frac{\delta}{\epsilon}k_j+i\frac{\epsilon}{2})}
{\cosh(\frac{\delta}{\epsilon}k_j-i\frac{\epsilon}{2})}\right)^L
&=\prod_{s\ne j}^n\frac{\sinh(\frac{\delta}{\epsilon}k_j-\frac{\delta}{\epsilon}k_s+i\epsilon)}
{\sinh(\frac{\delta}{\epsilon}k_j-\frac{\delta}{\epsilon}k_s-i\epsilon)}\, ,\\
\left(\frac{1+i\frac{\delta}{2}k_j}
{1-i\frac{\delta}{2}k_j}\right)^{\frac{l}{\delta}}
&=\prod_{s\ne j}^n\frac{\sinh(\frac{\delta}{\epsilon}k_j-\frac{\delta}{\epsilon}k_s+i\epsilon)}
{\sinh(\frac{\delta}{\epsilon}k_j-\frac{\delta}{\epsilon}k_s-i\epsilon)}\, ,\\
e^{ik_j l}&=\prod_{s\ne j}^n\frac{k_j-k_s+ic}{k_j-k_s-ic}\, ,  \ \ \ \ j=1,\cdots, n\, ,
\end{align*}
which are exactly the BAEs for the Bose gas  (\ref{BEnls}).
Performing the same limits in $e_0(\la)$, see (\ref{magnon}), we find
\be\label{i1}
e_0(\la)\rightarrow 2J\delta^2 k^2-\left(2J\epsilon^2+\frac{J}{2}\epsilon^4-h\right)+\mathcal{O}(\epsilon^6)\, .
\ee
In order to obtain the energy spectrum (\ref{enls})  of the Bose gas from (\ref{magnon})
(we neglect the zero point energy $hL/2$), we need to consider  $J\rightarrow \infty$
like $\mathcal{O}(1/\epsilon^4)\, ,$  $h\rightarrow \infty$ like  $\mathcal{O}(1/\epsilon^2)$
with $2J\delta^2=1\ $ and $\mu=(2J\epsilon^2+\frac{J}{2}\epsilon^4-h)$ finite. This means that
by performing the thermodynamic limit followed by the continuum limit in the canonical
partition function of the XXZ spin chain (modulo the zero point energy) we obtain  the
grand-canonical partition function of the Lieb-Liniger model
\be\label{i2}
Z_{XXZ}( h,\beta)\equiv\lim_{L\rightarrow \infty}\sum_{\{\la\}}e^{-\beta E(\{\la\})}\, \rightarrow
Z_{NLS}(\mu,\beta)\equiv\lim_{l\rightarrow \infty}\sum_{\{k\}}e^{-\beta \overline E(\{k\})}\, .
\ee
In the following sections we will use a slightly modified scaling limit
compared with the one presented before and utilized in \cite{SBGK}.
Eq.~(\ref{i2}) can also be obtained if we consider $J=1/2$, the continuum model at
inverse temperature $\overline\beta$ related to the inverse  temperature of
the lattice model via $\beta=\overline\beta/\delta^2$, and $h\rightarrow 0$
like $\mathcal{O}(\epsilon^2)$ such that $\mu=(\epsilon^2/\delta^2+\epsilon^4/(4\delta^2)-h/\delta^2)$
is finite. Then
\begin{align*}
\beta e_0(\la)&\rightarrow \beta\left[ 2J\delta^2 k^2-\left(2J\epsilon^2+\frac{J}{2}\epsilon^4-h\right)\right]\, ,\\
                &=\overline\beta (k^2-\mu)=\overline\beta\overline e_0(k)\, .
\end{align*}
This shows that the thermodynamic properties and the correlation functions of the Bose gas
at any temperature can be obtained from the thermodynamic properties and correlation
functions of the XXZ spin chain at low-temperature and vanishing magnetic field.
In the next sections we will use this continuum limit, summarized in Table \ref{table1},
to derive the correlation lengths of the Bose gas from the
low-temperature spectrum of the XXZ-QTM.
\begin{table}[h]
\caption{\label{table1} Parameters for the XXZ spin chain and the one-dimensional Bose gas.}
\begin{center}
\begin{tabular}{|l|l|}
  \hline
  XXZ spin chain & One-dimensional  Bose gas \\
  \hline \hline
 lattice constant $\delta=\mathcal{O}(\epsilon^2)$            & particle mass $m=1/2$\\
 number of lattice sites $L=\mathcal{O}(1/\epsilon^2)$        & physical length $l=L\delta$\\
 interaction strength $J=1/2$                                            & repulsion strength $c=\epsilon^2/\delta$\\
 magnetic field $h= \mathcal{O}(\epsilon^2)$                  & chemical potential $\mu=(\epsilon^2/\delta^2+\epsilon^4/(4\delta^2)-h/\delta^2)$\\
 inverse temperature $\beta$                                            & inverse temperature $\overline\beta=\beta \delta^2$\\
 anisotropy $\Delta=\cos\eta=\epsilon^2/2-1$                            &                                                       \\
  \hline
\end{tabular}
\end{center}
\end{table}

\


\section{The low-temperature spectrum of the XXZ spin chain quantum transfer matrix }\label{sQTM}

In this section we are going to investigate the low-temperature spectrum of the
XXZ spin chain QTM \cite{MS,K1}. A short review of the relevant facts about the
QTM can be found in \cite{GKS1,EPAPS}. The QTM is important for two reasons: first,
the largest eigenvalue, which we will denote by $\Lambda_0(\la)$, completely
characterizes the thermodynamics of the system via
\[
f(h,T)=-\frac{1}{\beta}\log\Lambda_0(0)\, ,
\]
with $f(h,T)$ the free energy per lattice site. In general, the largest
eigenvalue of the QTM can be expressed in terms of some finite number of
auxiliary functions  satisfying non-linear integral equations \cite{K3}.
This is a very efficient thermodynamic  description for the model contrasting
with the Thermodynamic Bethe Ansatz (TBA)\cite{T} approach which relies on the
string hypothesis and provides an infinite number of NLIEs.
The second reason is given by the fact that the correlation lengths of various
Green's functions can be obtained as ratios of the largest and next-largest
eigenvalues of the QTM \cite{KSS,EPAPS}. This is a consequence of the finite gap
between the largest eigenvalue and the rest of the spectrum of the QTM. In the
next sections we are going to study the low-temperature spectrum
of the QTM in order to obtain the asymptotic expansion of the longitudinal
and transversal correlation functions in the XXZ spin chain. Performing the
continuum limit presented in Sec.~\ref{Continuum} we are going to arrive
at the results presented in Sec.~\ref{MR}.

The QTM  is constructed with the help of the XXZ  trigonometric $\R$-matrix
\be\label{rmat}
\R(\la,\m)\equiv\left(\begin{array}{cccc} \R_{11}^{11}(\la,\m) & \R_{12}^{11}(\la,\m) & \R_{21}^{11}(\la,\m) & \R_{22}^{11}(\la,\m)\\
                                     \R_{11}^{12}(\la,\m) & \R_{12}^{12}(\la,\m) & \R_{21}^{12}(\la,\m) & \R_{22}^{12}(\la,\m)\\
                                     \R_{11}^{21}(\la,\m) & \R_{12}^{21}(\la,\m) & \R_{21}^{21}(\la,\m) & \R_{22}^{21}(\la,\m)\\
                                     \R_{11}^{22}(\la,\m) & \R_{12}^{22}(\la,\m) & \R_{21}^{22}(\la,\m) & \R_{22}^{22}(\la,\m)
                  \end{array}\right)
          =   \left(\begin{array}{cccc} 1 & 0         & 0         & 0\\
                                       0 & b(\la,\m) & c(\la,\m) & 0\\
                                       0 & c(\la,\m) & b(\la,\m) & 0\\
                                       0 & 0         & 0         & 1
                  \end{array}\right)\, ,
\ee
where
\be\label{defbc}
b(\la,\m)=\frac{\sinh(\la-\m)}{\sinh(\la-\m+i\eta)}\, , \ \ \ \ \  c(\la,\m)=\frac{\sinh(i\eta)}{\sinh(\la-\m+i\eta)}\, .
\ee
We introduce two types of $\Lo$-operators $\Lo_j(\la,-u')\, ,\tilde\Lo_j(u',\la)
\in\mbox{End}\left((\mathbb{C}^2)^{\otimes(N+1)}\right)$ defined as
\be\label{LQTM}
\Lo_j(\la,-u')=\sum_{a,b,a_1,b_1=1}^2\R_{b\,b_1}^{aa_1}
(\la,-u')e_{ab}^{(0)} e_{a_1b_1}^{(j)}\, ,\ \ \ \ \ \
\tilde\Lo_j(u',\la)=\sum_{a,b,a_1,b_1=1}^2\R_{a_1\, b}^{b_1\, a }
(u',\la)e_{ab}^{(0)} e_{a_1b_1}^{(j)}\, ,
\ee
where $u'=iu\, , $ $u=-2J\sin\eta\frac{\beta}{N},$  $N$ is the Trotter number and
$e_{ab}^{(j)}$  the canonical basis in $\mbox{End}\left((\mathbb{C}^2)^{\otimes(N+1)}\right),$
{\it i.e.}, $e_{ab}^{(0)}=e_{ab}\otimes\mathbb{I}_2^{\otimes L}\, $ and $e_{ab}^{(i)}
=\mathbb{I}_2\otimes \mathbb{I}_2^{\otimes (i-1)}\otimes e_{ab}\otimes\mathbb{I}_2^{\otimes(N-i)}$ with $e_{ab}$
the $2$-by-$2$ matrices with all the elements zero except the one at the intersection of the
$a$-th row and $b$-th column which is equal to one. The monodromy matrix of the QTM is defined as
\[
\T^{QTM}(\la)=\Lo_{N}(\la,-u')\tilde\Lo_{N-1}(u',\la)\cdots
\Lo_{2}(\la,-u')\tilde\Lo_{1}(u',\la)\, .
\]
and provides a representation
of the Yang-Baxter algebra
\be\label{YBA}
\check\R(\la,\m)[\T^{QTM}(\la)\otimes\T^{QTM}(\m)]=[\T^{QTM}(\m)\otimes\T^{QTM}(\la)]\check\R(\la,\m)\, ,
\ee
with $\check\R_{b_1\, b_2}^{a_1a_2}(\la,\m)=\R_{b_1b_2}^{a_2a_1}(\la,\m)$. Using the explicit
expression of the  $\Lo$-operators in the auxiliary space
\[
\Lo_j(\la,-u')=\left(\begin{array}{lr} e_{11}^{(j)}+b(\la,-u')e_{22}^{(j)} & c(\la,-u')e_{21}^{(j)}\\
                                     c(\la,-u') e_{12}^{(j)} & b(\la,-u')e_{11}^{(j)}+e_{22}^{(j)}
                   \end{array}\right)\, ,\ \ \ \
\tilde\Lo_j(u',\la)=\left(\begin{array}{lr} e_{11}^{(j)}+b(u',\la)e_{22}^{(j)} & c(u',\la)e_{12}^{(j)}\\
                                     c(u',\la) e_{21}^{(j)} & b(u',\la)e_{11}^{(j)}+e_{22}^{(j)}
                   \end{array}\right)\, ,
\]
where now $e_{ab}^{(j)}$ is the canonical basis in $\mbox{End}\left((\mathbb{C}^2)^{\otimes N}\right),$
it is easy to see that
\be\label{pseudoQTM}
|\Omega\rangle=\underbrace{\left(\begin{array}{c} 1\\
0 \end{array}\right)\otimes\left(\begin{array}{c} 0\\1
\end{array}\right)\otimes\cdots
\otimes\left(\begin{array}{c} 1\\0 \end{array}\right)\otimes
\left(\begin{array}{c} 0\\1 \end{array}\right)}_{\mbox{ N factors}}\, ,
\ee
satisfies the conditions of a pseudovacuum (is  an eigenvector of
$A^{QTM}(\la)$ and $D^{QTM}(\la)$ and the action of $\T^{QTM}(\la)$ on it is triangular)
for the monodromy matrix of
the QTM and
\be\label{QTMp}
\T^{QTM}(\la)|\Omega\rangle=\left(\begin{array}{cc} A^{QTM}(\la)|\Omega\rangle & B^{QTM}(\la)|\Omega\rangle \\
                                             C^{QTM}(\la)|\Omega\rangle & D^{QTM}(\la)|\Omega\rangle
                       \end{array}\right)
                    =\left(\begin{array}{cr} \left(b(u',\la)\right)^{N/2}|\Omega\rangle & B^{QTM}(\la)|\Omega\rangle \\
                                              0 & \left(b(\la,-u')\right)^{N/2}|\Omega\rangle
                    \end{array}\right)\, .
\ee
The presence of the magnetic field in the Hamiltonian (\ref{ham}) is easily
taken into account by the following transformation of the monodromy matrix
\be\label{QTMm}
\T^{QTM}(\la)\rightarrow\T^{QTM}(\la)
\left(\begin{array}{cc} e^{\frac{\beta h}{2}} & 0\\
 0 & e^{-\frac{\beta h}{2}}
 \end{array}\right)\, .
\ee
The quantum transfer matrix $\tr^{QTM}(\la)$ is defined as the trace in
the auxiliary space of the monodromy matrix $\tr^{QTM}(\la)=\mbox{tr}_0
\T^{QTM}(\la).$ The existence of the pseudovacuum (\ref{pseudoQTM}) and
the fact that $\T^{QTM}(\la)$ provides a representation of the Yang-Baxter
algebra ensures that the eigenvalues of the QTM can be obtained using
the ABA. As shown in \cite{GKS1,EPAPS}  the solutions of the eigenvalue equation
\[
\tr^{QTM}(\la)|\{\la\}\rangle\equiv\tr^{QTM}(\la)B^{QTM}(\la_1)\cdots
B^{QTM}(\la_p)|\Omega\rangle=\Lambda(\la)|\{\la\}\rangle\, ,
\]
are given by
\be\label{eQTM}
\Lambda(\la)=b(u',\la)^{N/2}e^{\beta h/2}
\prod_{j=1}^p\frac{\sinh(\la-\la_j-i\eta)}{\sinh(\la-\la_j)}
+b(\la,-u')^{N/2}e^{-\beta h/2}\prod_{j=1}^p\frac{\sinh(\la-\la_j+i\eta)}
{\sinh(\la-\la_j)}\, ,
\ee
provided that the parameters $\{\la_j\}_{j=1}^p$ satisfy the Bethe equations
\be\label{BEQTM}
\left(\frac{b(u',\la_j)}{b(\la_j,-u')}\right)^{N/2}=e^{-\beta h}
\prod_{j\ne k}^p
\frac{\sinh(\la_j-\la_k+i\eta)}{\sinh(\la_j-\la_k-i\eta)}\, ,
\ \ \ \ j=1,\cdots,p\, .
\ee

The asymptotic expansion of the longitudinal correlation function is given by \cite{GKS1,EPAPS,KSS,KSc}
\be
\langle\sigma_z^{(1)}\sigma_z^{(m+1)}\rangle_T=const+\sum_i\,  A_i\,
e^{-\frac{m}{\xi^{(d)}_i}}\,  \, ,\ \ \ \ m\rightarrow \infty\, ,
\ee
where $A_i$ are unknown amplitudes, $1/\xi^{(d)}_i=\log(\Lambda_0(0)/\Lambda_i^{(ph)}(0))$
and the sum is over all the next-largest eigenvalues $\Lambda_i^{(ph)}(0)$ in the
$N/2$ sector. We remind the reader that an eigenvalue of the QTM is
said to be in the $M$
sector if $p=M$ in Eqs.~(\ref{eQTM}) and (\ref{BEQTM}). The asymptotic expansion of
the transversal correlation
function is given by
\be
\langle\sigma_+^{(1)}\sigma_-^{(m+1)}\rangle_T=\sum_i\,  B_i\,
e^{-\frac{m}{\xi^{(s)}_i}}\,  \, ,\ \ \ \ m\rightarrow \infty\, ,
\ee
where $B_i$ are unknown amplitudes, $1/\xi^{(s)}_i=\log(\Lambda_0(0)/\Lambda_i^{(s)}(0))\, ,$
$\sigma^{(j)}_{\pm}=(\sigma_x^{(j)}\pm i\sigma_y^{(j)})/2$
and the sum is over all the next-largest eigenvalues $\Lambda_i^{(s)}(0)$ in the
$N/2-1$ sector.

The QTM method can also be utilized to investigate the generating
functional for the $\sigma_z$ correlators, {\it i.e.,} $\langle e^{\{\varphi\sum_{n=1}^me^{(n)}_{22}\}}\rangle_T$,
from which the longitudinal correlation function can be obtained via
$\langle\sigma_z^{(1)}\sigma_z^{(m+1)}\rangle_T=
(2D_m^2\6_\varphi^2-4D_m\6_\varphi+1)\langle e^{\{\varphi\sum_{n=1}^me^{(n)}_{22}\}}\rangle_T|_{\varphi=0}$
with $D_m$ the lattice derivative defined as $D_ma_m=a_m-a_{m-1}$ for any sequence $(a_n)_{n\in\mathbb{N}}$.
In this case the asymptotic expansion is
\begin{align}\label{int50}
\langle e^{\{\varphi\sum_{n=1}^me^{(n)}_{22}\}}\rangle_T&=
\sum_{i}C_i e^{-\frac{m}{\xi^{(\varphi)}_i}}\, ,\ \ \ \ \ m\rightarrow \infty\, ,
\end{align}
In (\ref{int50}) (see \cite{EPAPS}) the sum is over all the eigenvalues of
$\tr^{QTM}_\varphi(0)=A^{QTM}(0)+e^\varphi D^{QTM}(0)$
in the $N/2$ sector denoted by $\Lambda_i^{(\varphi)}(0)$ and
$1/\xi^{(\varphi)}_i=\log(\Lambda_0(0)/\Lambda_i^{(\varphi)}(0))$.

\subsection{Non-linear integral equations for the largest eigenvalue}\label{SLE}

Deriving NLIEs for the largest and next-largest eigenvalues
of the QTM requires a different method from the one utilized  in the
computation of the  ground state and low-lying excitations of the transfer
matrix in the massless regime. This is due to the fact that in the
Trotter limit, $N\rightarrow \infty$, the distribution of  Bethe roots (\ref{BEQTM}) in the
complex plane presents an accumulation point at the origin and  isolated
solutions which makes it impossible to  introduce the ``density of roots"
like in the case of the ground state of the transfer matrix.
Fortunately, the Bethe roots appear only in some strips of the complex plane
which allows for the introduction of some auxiliary functions which
satisfy functional equations. Thanks to fundamental properties of the
gross distribution of $\{\la_j\}_{j=1}^p$ the auxiliary functions enjoy
certain analyticity properties which allow to transform the functional
equations in terms of non-linear integral equations. Eventually
the eigenvalues of the QTM can be expressed in terms of these auxiliary
functions. This method, which was developed in \cite{KBP,K1,KPe,K2}, will be
our main tool in investigating the spectrum of the QTM.

The largest eigenvalue of the QTM lies in the  $N/2$ sector.
We will employ the  following notations
\[
\phi_{\pm}(\la)=\left(\frac{\sinh(\la\pm iu)}{\sin \eta}\right)^{N/2}\, ,\ \
q(\la)=\prod_{j=1}^{N/2}\sinh(\la-\la_j)\, ,
\]
which allows to express the eigenvalues of the QTM (\ref{eQTM}) as
\be\label{Lamle}
\Lambda_0(\la)=\frac{\phi_-(\la)}{\phi_-(\la-i\eta)}\frac{q(\la-i\eta)}{q(\la)}e^{\frac{\beta h}{2}}+
\frac{\phi_+(\la)}{\phi_+(\la+i\eta)}\frac{q(\la+i\eta)}{q(\la)}e^{-\frac{\beta h}{2}}\, .
\ee
Below, the NLIE and integral expression for the largest eigenvalue
will be determined following \cite{K3}.

\subsubsection{Integral equation for the auxiliary function}\label{Sleie}

\begin{figure}
\includegraphics[width=1\linewidth]{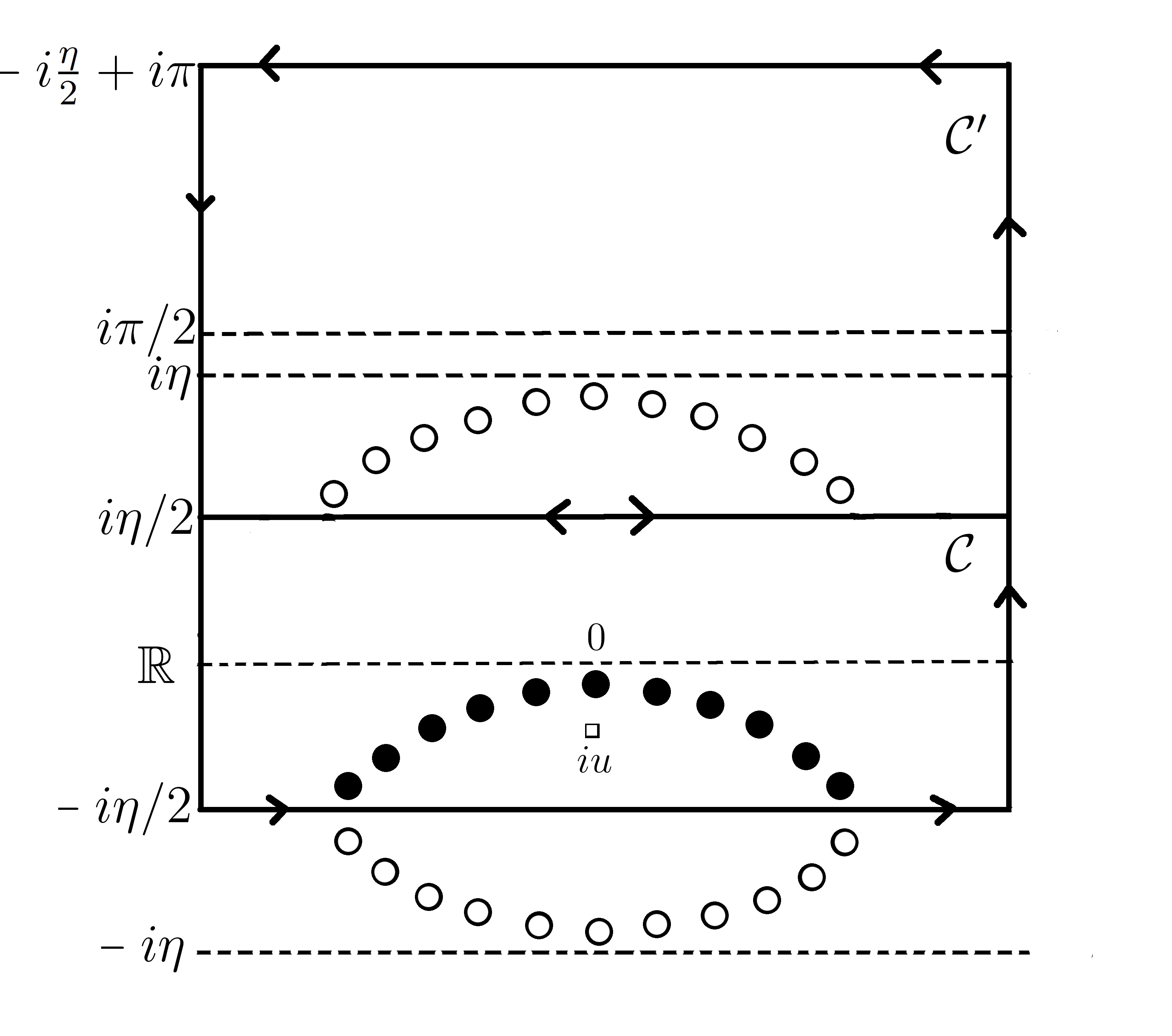}
\caption{
Typical distribution of Bethe roots ($\bullet$) and holes ($\circ$) in the strip $|\Im \lam|<\eta\, ,\ \eta\in[0,\pi/2)$
characterizing the largest eigenvalue of the QTM at low-temperatures. All the other roots
and holes can be obtained using the $i\pi$ periodicity. The contour $\mathcal{C}$ surrounds all the Bethe
roots and the pole of the auxiliary function $\mathfrak{a}(\la)$  at $iu$.}

\label{LEfig}
\end{figure}

An extremely important role in the following will be played by the  auxiliary
function $\mathfrak{a}(\la)$, which is periodic of period $i\pi$ and defined by
\be\label{defaa}
\mathfrak{a}(\la)=\frac{\phi_+(\la)}{\phi_-(\la)}\frac{\phi_-(\la-i\eta)}{\phi_+(\la+i\eta)}
\frac{q(\la+i\eta)}{q(\la-i\eta)}e^{-\beta h}\, .
\ee
We note that the  Bethe equations (\ref{BEQTM}) can be rewritten as
$\mathfrak{a}(\la_j)=-1\, ,j=1,\cdots,N/2.$ However, the equation $\mathfrak{a}(\la)=-1$
has $3N/2$ solutions in a period strip, of which, only $N/2$ are given by the Bethe roots
$\{\la_j\}_{j=1}^{N/2}.$ The additional  $N$ solutions are called holes and we will denote
them by $\{\la_j^{(h)}\}_{j=1}^N$.  A typical distribution of Bethe roots and holes for
$\eta\in(0,\pi/2)$ and low temperatures is presented in Fig.~\ref{LEfig}.
Let $\mathcal{C}$ be a rectangular contour with positive orientation,
centered at the origin, extending
to infinity,  with the upper (lower) edges
parallel to the real axis through
$\pm i(\eta-\epsilon)/2,$
with $\epsilon\rightarrow 0$.
It is important to note that this contour is independent of the Trotter number and
the following considerations are valid for all $N$. Inside the contour,
the function $1+\mathfrak{a}(\la)$ has $N/2$ zeros at the Bethe roots and a pole of
order $N/2$ at $iu$, which means that $\log(1+\mathfrak{a}(\la))$ has no winding number
around the contour allowing us to define ($\la$ is
located outside of $\mathcal{C}$)
\be\label{f}
f(\la)\equiv\frac{1}{2\pi i}\int_{\mathcal{C}}\frac{d}{d\la}\left(\log\sinh(\la-\m)\right)
\log(1+\mathfrak{a}(\m))d\m=\frac{1}{2\pi i}\int_{\mathcal{C}}\log\sinh(\la-\m)
\frac{\mathfrak{a}'(\m)}{1+\mathfrak{a}(\m)}d\m\, .
\ee
For the evaluation of the r.h.s.~of Eq.~(\ref{f}) we will use the following theorem:
\begin{thm}\label{T}\cite{WW}
Let $\mathcal{C}$ be a contour in the complex plane, and let $g(\la)$ be a function
analytic
and non-zero
inside and on $\mathcal{C}$. Let $\phi(\la)$ be another function which is
analytic inside and on $\mathcal{C}$ except at a finite number of
points; let the zeros
of $\phi(\la)$ in the interior of $\mathcal{C}$ be $a_1,a_2,\cdots$ and let their degrees
of multiplicity be $r_1,r_2,\cdots$; and let its poles in the interior of $\mathcal{C}$
be $b_1,b_2,\cdots$ and let their degrees of multiplicity be $s_1,s_2,\cdots.$ Then
\[
\frac{1}{2\pi i}\int_{\mathcal{C}}g(\la)\frac{\phi'(\la)}{\phi(\la)}d\la=
\sum_{i\in\mbox{zeros}}r_ig(a_i)
-\sum_{i\in\mbox{poles}}s_ig(b_i)\, ,
\]
\end{thm}
obtaining
\be\label{int}
f(\la)=\sum_{j=1}^{N/2}\log\sinh(\la-\la_j)-\frac{N}{2}\log\sinh(\la-iu)=
\log q(\la)-\log\phi_-(\la)-\frac{N}{2}\log\sin\eta\, .
\ee
Eq.~(\ref{int}) provides an integral representation for $q(\la)$ in terms of
$\log(1+\af(\la))$. Taking the logarithm in Eq.~(\ref{defaa}) and using (\ref{int})
we can derive
\[
\log\af(\la)=-\beta h+\log\left(\frac{\phi_+(\la)}{\phi_-(\la)}\frac{\phi_-(\la+i\eta)}
{\phi_+(\la+i\eta)}\right)+f(\la+i\eta)-f(\la-i\eta)\, ,
\]
which is a nonlinear integral equation  of convolution type for the auxiliary
function $\af(\la),$  valid for all $N$. Using
\be\label{Trotterl}
\lim_{N\rightarrow\infty}\log\left(\frac{\phi_+(\la)}{\phi_-(\la)}\right)=
-2J\beta\sinh(i\eta)\coth\la\, ,
\ee
the Trotter limit, $N\rightarrow\infty,$ can be performed with the final result
\be\label{NLIEale}
\log\af(\la)=-\beta h-\beta\frac{2J\sinh^2(i\eta)}{\sinh(\la+i\eta)\sinh\la}-
\frac{1}{2\pi}\int_{\mathcal{C}}\frac{\sin(2\eta)}{\sinh(\la-\m+i\eta)
\sinh(\la-\m-i\eta)}\log(1+\af(\m))d\m\, .
\ee
Eq.~(\ref{NLIEale}) was obtained under the assumption that $\eta\in(0,\pi/2).$ It is also
valid for  $\eta\in(\pi/2,\pi)$ but in this case  $\mathcal{C}$ is a rectangular contour
centered at zero, extending
to infinity, with the upper (lower) edges parallel to the real
axis through $\pm i(\pi-\eta-\epsilon)/2$ with $\epsilon\rightarrow 0$.

\subsubsection{Integral expression for the largest eigenvalue}\label{Sleir}

It remains to obtain an integral expression for the largest eigenvalue $\Lambda_0(\la)$
in terms of the auxiliary function $\af(\la).$ Consider $\eta\in(0,\pi/2)$
for which the distribution of roots and holes is presented in Fig.~\ref{LEfig}. First, we note that
Eq.~(\ref{Lamle}) can be rewritten as
\be\label{int2}
\Lambda_0(\la)=\frac{p(\la)}{\phi_-(\la-i\eta)\phi_+(\la+i\eta)q(\la)}\, ,
\ee
with $p(\la)=\phi_-(\la)\phi_+(\la+i\eta)q(\la-i\eta)e^{\beta h/2}+
\phi_+(\la)\phi_-(\la-i\eta)q(\la+i\eta)e^{-\beta h/2}.$ The function $p(\la)$
is quasi-periodic  $p(\la+i\pi)=(-1)^{3N/2}p(\la)$ and   $\lim_{\la\rightarrow\infty}
p(\la)/(\sinh \la)^{3N/2}=(e^{\beta h/2}+e^{-\beta h/2})/(\sin \eta)^N.$ The zeros
of the function $p(\la)$ in a strip of width $i\pi$  are the solutions of the
equations  $\af(\la)=-1$
yielding
\be\label{int3}
p(\la)=c\, \prod_{j=1}^{N/2}\sinh(\la-\la_j)
\prod_{j=1}^N\sinh(\la-\la_j^{(h)})\, ,
\ee
where $\{\la_j\}_{j=1}^{N/2}$ and $\{\la_j^{(h)}\}_{j=1}^N$ are the Bethe
roots and holes, respectively, and $c$ is a constant.
Defining  $q^{(h)}(\la)=\prod_{j=1}^N\sinh(\la-\la_j^{(h)})$ and using (\ref{int3})
to replace $p(\la)$ in Eq.~(\ref{int2}) we obtain
\be\label{Lh}
\Lambda_0(\la)=c\, \frac{q^{(h)}(\la)}{\phi_-(\la-i\eta)\phi_+(\la+i\eta)}
\, ,
\ee
which provides an alternative expression for the largest eigenvalue in terms of
the holes and not Bethe roots. Below, we will show how an  integral representation
of $\log q^{(h)}(\la)$ in terms of the auxiliary function $\af(\la)$ can be calculated.

We consider a new rectangular contour with positive orientation $\mathcal{C}'$
(see Fig.~\ref{LEfig}) extending
to infinity, with the upper (lower) edges parallel
to the real axis
through $i(\eta-\epsilon)/2$ and $-i(\eta-\epsilon)/2+i\pi$
with $\epsilon\rightarrow 0$. The lower edge of the contour $\mathcal{C}'$  at
$i(\eta-\epsilon)/2$ coincides with the upper edge of $\mathcal{C}$ but has
opposite orientation. Now we can prove the following identity
\be\label{integralr}
\int_{\mathcal{C}+\mathcal{C}'}d(\la-\m)\frac{\af'(\m)}{1+\af(\m)}d\m=0\, ,\ \ \ \
d(\la-\m)=\frac{d}{d\la}\log\sinh(\la-\m)\, .
\ee
First, we notice that the contributions of the two contours parallel to the real axis
through $i(\eta-\epsilon)/2$ cancel each other due to the opposite orientation.
Then it can be easily verified using the definition of $\af(\la)$
(\ref{defaa})
that the functions appearing in (\ref{integralr}) are all periodic of period
$i\pi,$ which means that the upper and lower edges of
$\mathcal{C}+\mathcal{C}'$ do not contribute to the integral. Finally, the
contributions of the sides parallel to the imaginary axis are also zero as it
can be seen from
\[
\lim_{\Re\m\rightarrow\pm\infty}d(\la-\m)=\mp 1\, ,\ \ \
\frac{\af'(\m)}{1+\af(\m)}=\frac{\af'(\m)}{\af\,
  (\m)}\frac{1}{1+\af^{-1}(\m)}\, ,\ \ \ \lim_{\Re\m\rightarrow\pm\infty}\frac{1}{1+\af^{-1}(\m)}\rightarrow\frac{1}{1+e^{\beta h}}\, ,\ \ \
\lim_{\Re\m\rightarrow\pm\infty}\frac{\af'(\m)}{\af\, (\m)}=0\, .
\]

Consider $\la$ close to the real axis. Then making use of (\ref{integralr})
we find
\be\label{int4}
\frac{1}{2\pi i}\int_{\mathcal{C}}d(\la-\m)\frac{\af'(\m)}{1+\af(\m)}d\m=
-\frac{1}{2\pi i}\int_{\mathcal{C'}}d(\la-\m)\frac{\af'(\m)}{1+\af(\m)}d\m\, .
\ee
The r.h.s.~of (\ref{int4}) can be calculated using  Theorem \ref{T} by taking into
account that the function $1+\af(\la)$ has inside the contour $\mathcal{C}',$ $N$
zeros at the holes $\{\la_j^{(h)}\}_{j=1}^N$ or $\{\la_j^{(h)}\}_{j=1}^N+i\pi$,
$N/2$ simple poles at
$\{\la_j\}_{j=1}^{N/2}+i\eta$ and a pole of order $N/2$ at $-iu-i\eta+i\pi$
with the result
\be\label{int5}
\frac{1}{2\pi i}\int_{\mathcal{C}}d(\la-\m)\frac{\af'(\m)}{1+\af(\m)}d\m=
-\left(\sum_{j=1}^Nd(\la-\la_j^{(h)})-\sum_{j=1}^{N/2}d(\la-\la_j-i\eta)
-\frac{N}{2}d(\la+iu+i\eta)\right)\, .
\ee
Using again Theorem {\ref{T}} and the fact that the function $1+\af(\la)$
has inside the contour $\mathcal{C},$ $N/2$ zeros at the Bethe roots $\{\la_j\}_{j=1}^{N/2}$
and a pole of order $N/2$ at $iu$ we find
\be\label{int6}
\frac{1}{2\pi i}\int_{\mathcal{C}}d(\la-\m-i\eta)\frac{\af'(\m)}{1+\af(\m)}d\m=
\sum_{j=1}^{N/2}d(\la-\la_j-i\eta)-\frac{N}{2}d(\la-iu-i\eta)\, .
\ee
The integral representation for $\log q^{(h)}(\la)$ is obtained by taking the
difference of Eqs.~(\ref{int5}) and (\ref{int6}), integrating by parts, and then
integrating
w.r.t.~$\la$ with the result
\be\label{int7}
\frac{1}{2\pi i}\int_{\mathcal{C}}\left[d(\la-\m)-d(\la-\m-i\eta)\right]\log(1+\af(\m))d\m
=-\log q^{(h)}(\la)+\log\left(\phi_+(\la+i\eta)\phi_-(\la-i\eta)\right)+c\, ,
\ee
where $c$ is a constant of integration. Making use of this integral representation
in Eq.~(\ref{Lh}) the largest eigenvalue of the QTM can be written as
\[
\log\Lambda_0(\la)=c+\frac{1}{2\pi i}\int_{\mathcal{C}}\frac{\sinh(i\eta)}
{\sinh(\la-\m-i\eta)\sinh(\la-\mu)}\log(1+\af(\m))d\m\, .
\]
The constant of integration is computed  using the behavior of the involved functions at infinity.
Performing the change  of variables $z=\la-\m$ in the integral, and using
$\lim_{\la\rightarrow\infty}\log\Lambda_0(\la)=\log(e^{\beta h/2}+e^{-\beta h/2})$
and $\lim_{\la\rightarrow\infty}\log(1+\af(\la))=\log(1+e^{-\beta h})$, we find that
the constant of integration is $\beta h/2.$ The final result for the largest eigenvalue
of the QTM evaluated at $0$ is
\be\label{LEf}
\log\Lambda_0(0)=\frac{\beta h}{2}+\frac{1}{2\pi }\int_{\mathcal{C}}\frac{\sin\eta}
{\sinh(\m+i\eta)\sinh(\mu)}\log(1+\af(\m))d\m\, .
\ee
The integral expression (\ref{LEf}), which was obtained for $\eta\in(0,\pi/2)$, is also
valid for $\eta\in(\pi/2,\pi)$ but, as in the case of the NLIE for the auxiliary function
(\ref{NLIEale}), the contour $\mathcal{C}$ should be replaced by a similar rectangular contour
with the upper (lower) edges parallel to the real axis
through $\pm i(\pi-\eta-\epsilon)/2$
with $\epsilon\rightarrow 0$.

\subsubsection{Final form of the integral equations}\label{SLElt}

The NLIE (\ref{NLIEale}) and integral expression (\ref{LEf}) are in fact correct for
all temperatures \cite{K3}. This is due to the fact that, even at high temperatures, the
Bethe roots are contained in the strip $|\Im \la|<(\eta-\epsilon)/2$ for $\eta\in(0,\pi/2)$
(or the strip $|\Im \la|<(\pi-\eta-\epsilon)/2$  for $\eta\in(\pi/2,\pi)$), which means that
the reasoning of the previous sections is still valid,  producing the same results. However,
in order to obtain the thermodynamic properties and correlation lengths of the Bose gas we
will be interested only in the low-temperature limit in which some simplifications of
(\ref{NLIEale}) and (\ref{LEf}) appear.

Let us consider $\eta\in(\pi/2,\pi)$. In this case, the upper edge  of the contour
$\mathcal{C}$, which we will call $\mathcal{C}_+$, is a parallel line to the real axis
through $i(\pi-\eta-\epsilon)/2$ (for the following discussion the presence of the
$\epsilon$ term is irrelevant). Then for $\la\in\mathcal{C}_+\, ,$ $\la=x+i(\pi-\eta)/2$ with
$x$ real,  the driving term
on the r.h.s.~of (\ref{NLIEale}) is negative and equal to
\[
-\beta h- \beta\frac{2J\sin^2\eta}{\cosh(x+i\eta/2)\cosh(x-i\eta/2)}\, ,
\]
which means that in the low-temperature limit $\beta\rightarrow\infty\, , $
$(h,J>0)$, the contribution of the  upper part of the contour is negligible
and we can restrict the free argument $\la$ and the integration variable to
the lower part of the  contour. We can shift this line to the line parallel
to the real axis through $-i\eta/2$ without crossing any poles of the driving term
obtaining
\be\label{int8}
\log\af(\la-i\eta/2)=-\beta h-\beta\frac{2J\sinh^2(i\eta)}{\sinh(\la+i\eta/2)\sinh(\la-i\eta/2)}-
\frac{1}{2\pi}\int_{\mathbb{R}}\frac{\sin(2\eta)}{\sinh(\la-\m+i\eta)
\sinh(\la-\m-i\eta)}\log(1+\af(\m-i\eta/2))d\m\, .
\ee
Applying a similar reasoning to the integral expression of the largest
eigenvalue (\ref{LEf}), after the shift at $-i\eta/2$, we find
\be\label{int9}
\log\Lambda_0(0)=\frac{\beta h}{2}+\frac{1}{2\pi }\int_{\mathbb{R}}\frac{\sin\eta}
{\sinh(\m+i\eta/2)\sinh(\m-i\eta/2)}\log(1+\af(\m-i\eta/2))d\m\, .
\ee
Let us introduce the function $\varepsilon (\la)$ satisfying
$e^{-\varepsilon(\la)/T}=\af(\la-i\eta/2)$ where $T=1/\beta$ is the
temperature. Then noticing that the driving term in Eq.~(\ref{int8})
is the magnon energy (\ref{magnon}), and using (\ref{momentum}) and
(\ref{density}),
the NLIE for the auxiliary function and the integral
expression for the largest eigenvalue at low-temperatures can be written
as
\begin{subequations}\label{LElt}
\begin{align}
\varepsilon(\la)&=e_0(\la)+\frac{T}{2\pi}\int_{\mathbb{R}}K(\la-\m)\log
\left(1+e^{-\varepsilon(\m)/T}\right)d\m\, ,\label{LElta} \\
\log\Lambda_0(0)&=\frac{h}{2T}+\frac{1}{2\pi}\int_{\mathbb{R}}p_0'(\la)
\log\left(1+e^{-\varepsilon(\la)/T}\right)d\la\, .\label{LElte}
\end{align}
\end{subequations}
These equations are very similar to the Yang-Yang equation for the
excitation energy
and the grand-canonical potential of the Bose gas \cite{YY4}. Following \cite{SBGK}, in Sec.~\ref{sCL}
we will show that the Yang-Yang thermodynamics can be obtained from Eqs.~(\ref{LElt})
if we perform the scaling limit presented in Sec.\ref{Continuum}.
Even though Eqs.~(\ref{LElt}) were
obtained for $\eta\in(\pi/2,\pi),$ they are also valid for
$\eta\in(0,\pi/2),$ which can be proved by using appropriate contour
manipulations. As these are beyond the scope of this paper, we confine ourselves
to assuming the validity.
This assumption will be verified in
Sec.~\ref{CFT} where we will show that they  reproduce the TLL/CFT predictions for the free
energy and asymptotic behavior of the  correlation functions.

\subsection{Integral equations for the next-largest eigenvalues in the $N/2$ sector}

Computing  the correlation lengths for the Green's function
$\langle\sigma_z^{(1)}\sigma_z^{(m+1)}\rangle_T $ requires the derivation of
integral equations for the next-largest eigenvalues  of the QTM in the $N/2$
sector.  This means that, as in the case of the
largest eigenvalue, we will have $N/2$ Bethe roots and $N$ holes.
In the previous section we have derived  the NLIE for the auxiliary function
and the integral expression for the largest eigenvalue making use of the fact
that the Bethe roots were
located in a relevant strip (modulo the periodicity)
of the complex plane which was independent of the Trotter number and
temperature. In the case of the next-largest eigenvalues in the $N/2$ sector at
low-temperatures,
some of the Bethe roots are found outside of this strip and an equal number
of holes are inside the strips. We will employ the same method used
in Sec.~\ref{SLE} but modified in such a way that  these Bethe roots and holes
are  properly taken into account. The calculations are presented in Appendix
\ref{Anleph}. At low-temperatures, the next-largest eigenvalues of the QTM in the
$N/2$ sector are given by
\be\label{NLEphgf}
\log\Lambda^{(ph)}_i(0)=\frac{h}{2T}+i\sum_{j=1}^r p_0(\la^+_j)-i\sum_{j=1}^r p_0(\la^-_j)
             +\frac{1}{2\pi}\int_{\mathbb{R}}p_0'(\la)\log\left(1+e^{-u_i(\la)/T}\right)d\la\, ,
\ee
with the auxiliary functions $u_i(\la)$ satisfying the NLIEs
\be\label{NLIEphgf}
u_i(\la)=e_0(\la)-iT\sum_{j=1}^r\theta(\la-\la^+_j)+iT\sum_{j=1}^r\theta(\la-\la^-_j)
+\frac{T}{2\pi}\int_{\mathbb{R}}K(\la-\m)\log\left(1+e^{-u_i(\m)/T}\right)d\m
\, .
\ee
The parameters $\{\la^+_j\}_{j=1}^r$ and $\{\la^-_j\}_{j=1}^r,$ which belong to the upper,
resp., lower half-plane are fixed by the constraint $1+e^{-u_i(\la^\pm_j)/T}=0.$
In  Eqs.~(\ref{NLEphgf}) and (\ref{NLIEphgf}), $r$ can take the values $1,2\cdots$.
The subscript $i$ enumerates the sets of parameters $\{\la^{\pm}_j\}_{j=1}^r$ satisfying the constraint
$1+e^{-u_i(\la^{\pm}_j)/T}=0$.


\subsection{Integral equations for the next-largest eigenvalues in the $N/2-1$ sector}\label{Ss}

The next-largest eigenvalues in the $N/2-1$ sector are relevant for the computation
of the correlation lengths  of the Green's function $\langle\sigma_+^{(1)}\sigma_-^{(m+1)}\rangle_T.$
The eigenvalues in this sector are characterized by $p=N/2-1$ in Eqs.~(\ref{eQTM}) and
(\ref{BEQTM}). Some of the features encountered in the study of the largest and $N/2$
sector eigenvalues are also present in this case.

At low-temperatures, the next-largest eigenvalue in this sector  has $N/2-1$ Bethe
roots  and, possibly, a  hole in a certain strip of the complex plane. Eigenvalues
with decreasing magnitude are obtained by moving pairs of Bethe roots/holes outside/inside
the strip. This means that it is sufficient to obtain integral equations  for the cases
with  one hole or no hole inside the strip, the equations for the other eigenvalues  are
obtained by adding extra driving terms of the type encountered in Eqs. (\ref{NLEphgf})
and (\ref{NLIEphgf}). The necessary calculations are presented in Appendix \ref{Anles}.
We distinguish two cases.
For $\la_0$  in the upper half plane, the next-largest
eigenvalues in the $N/2-1$ sector at low-temperatures have the integral representation
\be\label{NLEsgf}
\log\Lambda^{(s)}_i(0)=\frac{h}{2T}-i\pi+ip_0(\la_0)+i\sum_{j=1}^r p_0(\la^+_j)-i\sum_{j=1}^rp_0(\la^-_j)
             +\frac{1}{2\pi}\int_{\mathbb{R}}p_0'(\la)\log\left(1+e^{-v_i(\la)/T}\right)d\la\, ,
\ee
with the auxiliary functions $v_i(\la)$ satisfying the NLIEs
\be\label{NLIEsgf}
v_i(\la)=e_0(\la)\pm i\pi T-iT\theta(\la-\la_0)-iT\sum_{j=1}^r\theta(\la-\la^+_j)+iT\sum_{j=1}^r\theta(\la-\la^-_j)
+\frac{T}{2\pi}\int_{\mathbb{R}}K(\la-\m)\log\left(1+e^{-v_i(\m)/T}\right)d\m
\, .
\ee
The $2r+1$ parameters $\la_0\, ,\{\la^+_j\}_{j=1}^r$ and $\{\la_j^-\}_{j=1}^r$ which belong to the upper,
resp., lower half-plane are fixed by the constraints $1+e^{-v_i(\la_0)/T}=0\, ,1+e^{-v_i(\la^\pm_j)/T}=0.$
On the r.h.s.~of Eq.~(\ref{NLIEsgf}) the plus (minus) sign in front of the $i\pi T$ term is considered
when $\la_0$ is in the first (second) quadrant of the complex plane $\Re \la_0\geq 0\, ,
\Im \la_0\geq 0$ ($\Re \la_0< 0\, , \Im \la_0\geq 0$).
For $\la_0$ in the lower half-plane Eqs.~(\ref{NLEsgf}) and (\ref{NLIEsgf}) remain valid but the integration
contour now is the real axis with an indentation such that $\la_0$ is above the contour. Also, the
plus (minus) sign in front of the $i\pi T$ term of (\ref{NLIEsgf}) is considered
when $\la_0$ is in the fourth (third) quadrant of the complex plane $\Re \la_0\geq 0\, ,
\Im \la_0\leq 0$ ($\Re \la_0< 0\, , \Im \la_0\leq 0$). In this case $\la_0$, which satisfies $1+e^{-
v_i(\la_0)/T}=0$, is the closest solution to the real axis in the lower half-plane.



\section{Comparison with the TLL/CFT predictions}\label{CFT}

In Sec.~\ref{SLElt} we have derived  an integral expression (\ref{LElt}),
for the  largest eigenvalue of the QTM in terms of an auxiliary function which
obeys a NLIE very similar
to the Yang-Yang equation (\ref{YYnls}).
Eq.~ (\ref{LElt}) and the similar integral
representations for the next-largest eigenvalues (\ref{NLEphgf}) and (\ref{NLEsgf}),
are valid only at low-temperatures and, in the course of the derivation,
we have made some assumptions which were not fully justified for some
values of the anisotropy. Here, we will show that our results are in perfect
agreement with the  predictions of the Tomonaga-Luttinger liquid \cite{H1,H2} and Conformal Field Theory
\cite{BPZ,C1,VW,C2,BCN,A,BIK}, confirming the validity of our assumptions.

\subsection{Low-temperature behavior of the free energy}

At low-temperatures CFT predicts \cite{A} that the free energy per lattice site
scales like
\be\label{Affleck}
f(h,T)=\epsilon_0(h)-\frac{\pi T^2c}{6v_F}+O(T^3)\, ,\ \ \
\epsilon_0(h)=-\frac{h}{2}+\int_{-q}^qe_0(\la)\rho(\la)d\la\, ,
\ee
where $\epsilon_0(h)$ is the density of the ground state energy, $c$ is the
conformal charge, (not to be confused with the coupling constant of the Bose gas),
which is equal to one in the case  of the XXZ spin chain, and $v_F$ is the Fermi
velocity defined as
\be\label{vf}
v_F=\frac{\eq}{2\pi\rho(q)}\, ,
\ee
with $\eq$ the derivative of the dressed energy (\ref{dressede})
evaluated at the Fermi boundary $q$.

Let us show that the free energy  per lattice site obtained from Eq.~(\ref{LElte}),
via $f(h,T)=-T\log\Lambda_0(0)$, satisfies (\ref{Affleck}).
Performing an analysis similar
to the one in Appendix A  of \cite{YY4}
or Chap.~I of \cite{KBI} it can be shown  that for $\eta\in(0,\pi)$ and
$h<h_c=8J\cos^2(\eta/2)$, the NLIE  (\ref{LElta}) for the auxiliary function
$\varepsilon(\la)$ has two zeros on the real axis  which we will denote by $\pm q(T).$
Also $\varepsilon(\la)$ is negative on $\left(-q(T),q(T)\right)$ and positive outside of
this interval. Let us denote $\lim_{T\rightarrow 0}  q(T)= \tilde q$ and
$\lim_{T\rightarrow 0}\varepsilon(\la)=\tilde\varepsilon_0(\la)$. Then using
\be\label{int25}
\lim_{T\rightarrow 0}T\log\left(1+e^{-\varepsilon(\la)/T}\right)=
\left\{\begin{array}{ccl} -\tilde\varepsilon_0(\la) &\ \ \ \ \ \
 &\la\in(-\tilde q,\tilde q)\, ,\\
                               0                       &\ \ \ \ \ \
                               &\la\in(-\infty,-\tilde q)\cup(\tilde q,+\infty)\, ,
       \end{array}
\right.
\ee
we find that in the low-temperature limit the NLIE (\ref{LElta}) transforms in
the linear equation for the dressed energy (\ref{dressede}). The equation
for the dressed energy has a unique solution for $\eta\in(0,\pi)$ which means that
$\lim_{T\rightarrow 0}\varepsilon(\la)=\varepsilon_0(\la)$ and
$\lim_{T\rightarrow 0} q(T)=  q$. In order to show that Eq.~(\ref{LElte}) gives
the correct free energy satisfying (\ref{Affleck}) we need a more accurate
estimation of integrals containing the factor $\log\left(1+e^{-\varepsilon(\la)/T}\right)$.
In the following we are going to assume that for low temperatures the auxiliary
function has the expansion \cite{JM,MN,S1}
\be\label{eexpansion}
\varepsilon(\la)=\varepsilon_0(\la)+\varepsilon_1(\la)T+\varepsilon_2(\la)T^2+O(T^3)\, .
\ee
Then it can be shown (see Appendix A of \cite{KMS2} or \cite{KKK2}) that, for any function $f(\la)$,
bounded on the real axis and differentiable in the vicinity of $\pm q$  we have :
\begin{align}\label{eexp}
&\lim_{T\rightarrow 0}T\int_{\mathbb{R}}f(\la)\log\left(1+e^{-\varepsilon(\la)/T}\right)\, d\la=
-\int_{-q}^qf(\la)\varepsilon(\la)d\la\nonumber\\
& \ \ \ \ \ \ \ \ \ \ \ \ \ \ \ \ \ \ \ \ \ \ \ \ \ \ \ \ \ \ \
+\frac{T^2\pi^2}{6\eq}(f(q)+f(-q))+\frac{T^2\varepsilon_1^2(q)f(q)}{2\eq}+
\frac{T^2\varepsilon_1^2(-q)f(-q)}{2\eq}+O(T^3)\, .
\end{align}
We should mention that a more compact, but maybe not as transparent, method of
investigating the low-temperature limit of the QTM spectrum was employed in
\cite{KSc}. All the results derived here and in the following sections can
also be obtained utilizing the results of the aforementioned paper.
Using (\ref{eexp}) and substituting (\ref{eexpansion}) in the equation for
the auxiliary function
(\ref{LElta}) we obtain
\begin{align*}
&\sum_{k=0}^2T^k\varepsilon_k(\la)+\frac{1}{2\pi}\sum_{k=0}^2T^k
\int_{-q}^qK(\la-\m)\varepsilon_k(\m)\, d\m=e_0(\la)\,\\
&\ \ \ \ \ \ \ \ \ \ \ \ \ \ \ \
-\frac{\pi T^2}{12\eq}(K(\la-q)+K(\la+q))-\frac{T^2\varepsilon_1^2(q)}
{4\pi\eq}K(\la-q)-\frac{T^2\varepsilon_1^2(-q)}{4\pi\eq}K(\la+q)+O(T^3)\, .
\end{align*}
Equating terms of the same order in temperature we find
\be\label{e2}
\varepsilon_1(\la)=0\, ,\ \ \ \ \varepsilon_2(\la)=\frac{\pi^2}{6\eq}(R(\la,q)+R(\la,-q))\, ,
\ee
with the resolvent $R(\la,\m)$ defined in (\ref{defres}). The low-temperature expansion
of the free energy per lattice site is calculated using the asymptotic formula (\ref{eexp})
in Eq.~(\ref{LElte}) with the result
\be\label{int26}
f(h,T)=-\frac{h}{2}+\frac{1}{2\pi}\int_{-q}^q p_0'(\la)(\varepsilon_0(\la)+
\varepsilon_2(\la)T^2)\, d\la-\frac{\pi T^2}{12\eq}p_0'(q)+O(T^3)\, ,
\ee
where we have used the fact that $p_0'(\la)$ is even, $p_0'(-q)=p_0'(q)$. Using
Eq.~(\ref{e2}) and the identity
\be
\int_{-q}^qp_0'(\la)R(\la,\pm q)\, d\la=p_0'(q)-2\pi\rho(q)\, ,
\ee
(for a proof, see Appendix  \ref{Aidentities}), (\ref{int26}) takes the form
\be
f(h,T)=-\frac{h}{2}+\frac{1}{2\pi}\int_{-q}^qp_0'(\la)\varepsilon_0(\la)\,
d\la-\frac{\pi T^2}{6}\frac{2\pi\rho(q)}{\eq}+O(T^3)\, .
\ee
Using the identity (\ref{ident2}) and the definition of the Fermi velocity
(\ref{vf}), it is easy to see that this expression is  identical with (\ref{Affleck}).

\subsection{Low-temperature asymptotic behavior of the longitudinal correlation}\label{CFTdd}

The Tomonaga-Luttinger liquid theory and CFT predict the following asymptotic behavior
of the longitudinal correlation function at low-temperatures (we consider only
the leading-order of the oscillatory terms) \cite{BIK}
\be\label{asymptd}
\langle\sigma^{(1)}_z\sigma^{(m+1)}_z\rangle_T=\langle\sigma^{(1)}_z\rangle^2_T-
\frac{(T\mathcal{Z}/v_F)^2}{2\sinh^2(\pi T m/v_F)}+
\sum_{l\in\mathbb{Z}^*}A_l\ e^{2i ml k_F }\left(\frac{\pi T/v_F}{\sinh(\pi T m/v_F)}
\right)^{2l^2\mathcal{Z}^2}\, , \ \ \ \ \ m\rightarrow\infty\, ,
\ee
with the Fermi momentum defined by $k_F=\pi D$ and $A_l$  are coefficients that
do not depend on $T$. The analysis of the QTM spectrum \cite{EPAPS}
shows that the asymptotic  behavior of the correlation function can be expressed as
\be\label{asymptdQTM}
\langle\sigma^{(1)}_z\sigma^{(m+1)}_z\rangle_T
=const+\sum_i \tilde A_i\,  e^{-\frac{m}{\xi^{(d)}[u_i]}}\, ,
\ \ \ \ \ m\rightarrow\infty\, ,
\ee
where the sum is over all the correlation lengths, $ 1/\xi^{(d)}[u_i]=\log(\Lambda_0(0)/
\Lambda^{(i)}_{ph}(0))$, determined as the ratio of the largest
and next-largest eigenvalues in the $N/2$ sector. Using (\ref{LElte}) and (\ref{NLEphgf})
we obtain the following explicit expression  for the correlation lengths
\be\label{cld}
\frac{1}{\xi^{(d)}[u_i]}=-\frac{1}{2\pi}\int_{\mathbb{R}}p_0'(\la)\log
\left(\frac{1+e^{-u_i(\la)/T}}{1+\, e^{-\varepsilon(\la)/T}}\right)\, d\la
-i\sum_{j=1}^rp_0(\la^+_j)+i\sum_{j=1}^rp_0(\la^-_j)\, ,
\ee
where the functions $\varepsilon(\la)$  and $u_i(\la)$ satisfy Eqs.~(\ref{LElta})
and (\ref{NLIEphgf}). In the rest of this section we will show that  (\ref{asymptdQTM}) is
equivalent to (\ref{asymptd}) in the conformal limit.

The analysis of the correlation lengths (\ref{cld}) in the limit $T\rightarrow 0$
is very similar
to the one performed by Kozlowski, Maillet and Slavnov for the
correlation  lengths of the Bose gas, and, for this reason, we are going to  use
some of the notations and terminology employed in \cite{KMS2}. In the following
we are going to discard the subscript $i$ for the auxiliary function $u_i(\la)$.
We are considering an arbitrary auxiliary function $u(\la)$ satisfying
Eq.~(\ref{NLIEphgf}), with $2r$ parameters $\{\la^+_j\}_{j=1}^r$,
$\{\la^-_j\}_{j=1}^r$,
located in the
upper (lower) half-plane, which  also satisfy the constraint $1+e^{-u(\{\la^\pm_j\})/T}=0$.
The first observation that we are going to make is that  $\lim_{T\rightarrow 0}u(\la)
=\varepsilon_0(\la)$. In analogy with the case of the auxiliary function $\varepsilon(\la)$
we expect that all the solutions of the equation $1+e^{-u(\la)/T}=0$ will collapse
at $\pm q$ in the limit $T\rightarrow 0$. We will say that the solutions that collapse
at $q$ $(-q)$ belong to the right (left) series. If we assume that $u(\la)$ has the expansion
\be\label{uexpansion}
u(\la)=\varepsilon_0(\la)+u_1(\la)T+u_2(\la)T^2+O(T^3)\, ,
\ee
then a formula similar
to (\ref{eexp}) can be derived as in \cite{KMS2}
\begin{align}\label{uexp}
&\lim_{T\rightarrow 0}T\int_{\mathbb{R}}f(\la)\log\left(1+e^{-u(\la)/T}\right)\, d\la=
-\int_{-q}^qf(\la)u(\la)d\la\nonumber\\
& \ \ \ \ \ \ \ \ \ \ \ \ \ \ \ \ \ \ \ \ \ \ \ \ \ \ \ \ \ \ \
+\frac{T^2\pi^2}{6\eq}(f(q)+f(-q))+\frac{T^2u_1^2(q)f(q)}{2\eq}+
\frac{T^2 u_1^2(-q)f(-q)}{2\eq}+O(T^3)\, .
\end{align}
We are going to consider that the roots $\{\la^\pm\}$  are distributed in the
following manner: $r_p^+$ roots $\la^+$ and $r_h^+$ roots $\la^-$ belonging to the
right series (collapse at $q$); $r_p^-$ roots $\la^+$ and $r_h^-$ roots $\la^-$
belonging to the left series (collapsing at $-q$), where $r_p^\pm$ and $r_h^\pm$  satisfy
the constraints
\[
r_p^++r_p^-=r_h^++r_h^-=r\, ,\ \ \ \ \ \ r_p^+-r_h^+=r_h^--r_p^-=l\, ,
\]
with $l$ integer, satisfying $-r\leq l\leq r$. More explicitly, at sufficiently
low temperatures we have
\begin{subequations}\label{int26b}
\begin{align}
\{\la^+_j\}_{j=1}^r&=\{q+iT\alpha_k^+\}_{k=1}^{r_p^+}
\cup\{-q+iT\alpha_k^-\}_{k=1}^{r_p^-}\, ,\ \ \ \ \Re(\alpha_k^\pm)>0\, ,\\
\{\la^-_j\}_{j=1}^r&=\{q-iT\, \beta_k^+\}_{k=1}^{r_h^+}
\cup\{-q-iT\, \beta_k^-\}_{k=1}^{r_h^-}\, ,  \ \ \ \ \Re(\beta_k^\pm)>0\, ,
\end{align}
\end{subequations}
where  $\alpha_k^\pm$ and $\beta_k^\pm$ satisfy
$
1+e^{-u(\pm q+iT\alpha_k^\pm)/T}=1+e^{-u(\pm q-iT\beta_k^\pm)/T}=0\, .
$
The leading Taylor coefficients can be parameterized by a set of integers,
$p_k^\pm$ and $s_k^\pm$ via
\[
u(\pm q+iT\alpha_k^\pm)=\pm 2\pi iT(p_k^\pm-\frac{1}{2})\, ,\ \ \ \ \ \ \
 u(\pm q-iT\beta_k^\pm)=\mp 2\pi iT(s_k^\pm-\frac{1}{2})\, .
\]
Using the expansion (\ref{uexpansion}), $\varepsilon(\pm q)=0\, $
and $\varepsilon_0'(-q)=-\varepsilon_0'(q)$ we find
\be\label{par}
\alpha_k^\pm=\frac{2\pi}{\eq}\left(p_k^\pm-\frac{1}{2}\right)\pm i\frac{u_1(\pm q)}{\eq}\, ,\ \ \ \ \ \
\beta_k^\pm=\frac{2\pi}{\eq}\left(s_k^\pm-\frac{1}{2}\right)\mp i\frac{u_1(\pm q)}{\eq}\, .
\ee
Substituting the parametrization (\ref{int26b}) in Eq.~(\ref{NLIEphgf}) and expanding
the driving terms up to the second order in $T$ we find
\be\label{int27}
u(\la)=e_0(\la)+\frac{T}{2\pi}\int_{\mathbb{R}}K(\la-\m)\log\left(1+e^{-u(\la)/T}\right)
\, d\m+g_1(\la)T+g_2(\la)T^2+O(T^3)\, ,
\ee
with
\[
g_1(\la)=-il(\theta(\la-q)-\theta(\la+q))\, ,
\]
and
\[
g_2(\la)=-K(\la-q)\left(\sum_{k=1}^{r_p^+}\alpha_k^++\sum_{k=1}^{r_h^+}\beta_k^+\right)
-K(\la+q)\left(\sum_{k=1}^{r_p^-}\alpha_k^-+\sum_{k=1}^{r_h^-}\beta_k^-\right)\, .
\]
We can now use (\ref{uexp}) in (\ref{int27}) obtaining a system of equations for the
unknown functions $u_1(\la)$ and $u_2(\la)$. For $u_1(\la)$ it reads
\[
u_1(\la)+\frac{1}{2\pi}\int_{-q}^qK(\la-\m)\, u_1(\m)\, d\m=-il(\theta(\la-q)-\theta(\la+q))\, .
\]
Comparison with the integral equation for the dressed phase (\ref{defdp}) shows that
$u_1(\la)=-2\pi i l(F(\la|q)-F(\la|-q))$. Using the first identity in (\ref{dpident})
we can obtain an expression in terms of the dressed charge $u_1(\la)=2\pi i l(1-Z(\la)).$
The equation for $u_2(\la)$ is
\[
u_2(\la)+\frac{1}{2\pi}\int_{-q}^qK(\la-\m)\, u_2(\m)\, d\m=g_2(\la)+
K(\la-q)\left(\frac{\pi}{12\eq}+\frac{u_1^2(q)}{4\pi\eq}\right)
+K(\la+q)\left(\frac{\pi}{12\eq}+\frac{u_1^2(-q)}{4\pi\eq}\right)\, ,
\]
with the solution
\[
u_2(\la)=-R(\la,q)\left[\sum_{k=1}^{r_p^+}\alpha_k^++\sum_{k=1}^{r_h^+}\beta_k^+
-\frac{1}{2\eq}\left(\frac{\pi^2}{3}+u_1^2(q)\right)\right]
-R(\la,-q)\left[\sum_{k=1}^{r_p^-}\alpha_k^-+\sum_{k=1}^{r_h^-}\beta_k^-
-\frac{1}{2\eq}\left(\frac{\pi^2}{3}+u_1^2(-q)\right)\right]\, .
\]
Using (\ref{eexp}) and (\ref{uexp}) in (\ref{cld}) and expanding the $p_0(\la^\pm)$
terms up to the first order in $T$ we obtain
\be\label{int27a}
\frac{1}{\xi^{(d)}[u]}=-2ilk_F-T\frac{u_1^2(q)\rho(q)}{\eq}+2\pi T\rho(q)\left(
\sum_{k=1}^{r_p^+}\alpha_k^+
+\sum_{k=1}^{r_p^-}\alpha_k^-
+\sum_{k=1}^{r_h^+}\beta_k^+
+\sum_{k=1}^{r_h^-}\beta_k^-
\right)+O(T^2)\, .
\ee
In deriving (\ref{int27a}) we have used also the identity
$\int_{-q}^q p_0'(\la)Z(\la)\, d\la=\int_{-q}^q\rho(\la)\, d\la$
and (\ref{ident1}). Finally, using (\ref{par}) we find
\be\label{int27b}
\frac{1}{\xi^{(d)}[u]}=-2ilk_F+\frac{2\pi T}{v_F}\left(l^2\mathcal{Z}^2-l^2-r+
\sum_{k=1}^{r_p^+}p_k^+
+\sum_{k=1}^{r_p^-}p_k^-
+\sum_{k=1}^{r_h^+}s_k^+
+\sum_{k=1}^{r_h^-}s_k^-
\right)+O(T^2)\, .
\ee
The second term in the expansion (\ref{asymptd}) is obtained for $r=1,l=0$ and
$p_1^+=p_1^-=1$ (or $s_1^+=s^-_1=1$). The next leading terms are obtained for
$r=l\, ,l=1,2,\cdots$
and the integers  $p^+_k$ and $s^-_k$ (or $p^-_k$ and $s^+_k$) taking values from $1$ to $l$.

There is, however, one caveat. If we assume $\Re(\alpha_k^\pm)>0\, ,\Re(\beta_k^\pm)>0$,
then,  (\ref{par}) together with $u_1(\pm q)=2\pi i l(1-\mathcal{Z})$ impose some
constraints on the allowed values of  $p^{\pm}_k$ and $s^{\pm}_k$.  A relatively
straightforward analysis shows that for $\eta\in(\pi/2,\pi)\, ,(\mathcal{Z}>1)$,
which is the region most interesting for us, the allowed values for $p^{\pm}_k$ and $s^{\pm}_k$
contain $\{1,2,\cdots\}$.
For $\eta\in(0,\pi/2)$ we have $1/\sqrt{2}<\mathcal{Z}<1$. In this case, for $\mathcal{Z}$
close to $1/\sqrt{2}$, the integers $p_k^\pm$ and $s_k^\pm$ can take the values $\{1, 2, \cdots\}$
only for $l=0,\pm 1$. For $\mathcal{Z}\rightarrow 1$ the value of $l$ for which the allowed values
of $p_k^\pm$ and $s_k^\pm$ contain $\{1, 2, \cdots\}$ increases. This means that under  the aforementioned
assumptions, in the worst case scenario, our equations can reproduce only the $l=0,\pm 1$
terms of the CFT expansion.

\subsection{Low-temperature asymptotic behavior of the transversal correlation}\label{CFTff}

In the case of the transversal correlation TLL and CFT predict
the following asymptotic behavior at low-temperatures \cite{BIK}
\be\label{asympts}
\langle\sigma^{(1)}_+\sigma^{(m+1)}_-\rangle_T=
\sum_{l\in\mathbb{Z}}B_l\ e^{2i ml k_F }\left(\frac{\pi T/v_F}{\sinh(\pi T m/v_F)}
\right)^{\frac{1}{2\mathcal{Z}^2}+2l^2\mathcal{Z}^2}\, , \ \ \ \ \ m\rightarrow\infty\, ,
\ee
with $B_l$  coefficients that do not depend on $T$. The analysis of the correlation
functions in the framework of the QTM  \cite{EPAPS} showed that
\be\label{asymptsQTM}
\langle\sigma^{(1)}_+\sigma^{(m+1)}_-\rangle_T=\sum_i \tilde B_i\,  e^{-\frac{m}{\xi^{(s)}[v_i]}}\, ,
\ \ \ \ \ m\rightarrow\infty\, ,
\ee
where the sum is over all the correlation lengths $ 1/\xi^{(s)}[v_i]=\log(\Lambda_0(0)/
\Lambda^{(i)}_{s}(0))$, determined as the ratio of the largest
and next-largest eigenvalues in the $N/2-1$ sector.
Using (\ref{LElte}) and (\ref{NLEsgf})
we obtain the following explicit expression  for the correlation lengths (we neglect
the $i\pi$ term which produces an $(-1)^m$ factor)
\be\label{cls}
\frac{1}{\xi^{(s)}[v_i]}=-\frac{1}{2\pi}\int_{\mathbb{R}}p_0'(\la)\log
\left(\frac{1+e^{-v_i(\la)/T}}{1+\, e^{-\varepsilon(\la)/T}}\right)\, d\la
-ip_0(\la_0)-i\sum_{j=1}^rp_0(\la^+_j)+i\sum_{j=1}^rp_0(\la^-_j)\, ,
\ee
where $\la_0$ is in the upper half-plane and the functions $\varepsilon(\la)$  and
$v_i(\la)$ satisfy Eqs.~(\ref{LElta}) and (\ref{NLIEsgf}). For $\la_0$ in the lower
half-plane the integration contour is the real axis with an indentation such that
$\la_0$ is above the contour.

First, we will consider the case with $\la_0$ in the
upper half-plane.
It is sufficient to consider the conformal limit of the following correlation
length
\be\label{clss}
\frac{1}{\xi^{(s)}[v]}=-\frac{1}{2\pi}\int_{\mathbb{R}}p_0'(\la)\log
\left(\frac{1+e^{-v(\la)/T}}{1+\, e^{-\varepsilon(\la)/T}}\right)\, d\la
-ip_0(\la_0)\, ,
\ee
with $v(\la)$ satisfying the equation (\ref{NLIEsgf}) with $r=0$. The behavior
of the correlation length (\ref{cls}) will be obtained by summing the contributions
of (\ref{clss}) and (\ref{cld}) derived in the previous section.
We notice that $\lim_{T\rightarrow 0}v(\la)=\varepsilon_0(\la),$ therefore,
we are going to consider the following expansion
\be\label{vexpansion}
v(\la)=\varepsilon_0(\la)+v_1(\la)T+v_2(\la)T^2+O(T^3)\, ,
\ee
with $v_1(\la)$ and $v_2(\la)$ unknown functions. Also (\ref{uexp}) is still valid
if we replace $u(\la)$ with $v(\la)$. We are going to consider that $\la_0$ is
located
in the first quadrant (this means that we are going to have a plus sign
in front of the $i\pi T$ term in Eq.~(\ref{NLIEsgf})). The same result is obtained
if we consider $\la_0$ in the second quadrant. Then at sufficiently low-temperatures
we have
\be\label{int00}
\la_0=q+i\alpha_0^+T\, ,\ \ \ \alpha_0^+=
\frac{2\pi}{\eq}\left(p_0^+-\frac{1}{2}\right)+i\frac{v_1(q)}{\eq}\, ,
\ee
with $p_0^+$ an integer parameterizing  the leading Taylor coefficient $\alpha_0^+.$
Using this parametrization in Eq.~(\ref{NLIEsgf}), and expanding to the second order in
$T$ we find
\be\label{int28}
v(\la)=e_0(\la)+\frac{T}{2\pi}\int_{\mathbb{R}}K(\la-\m)\log\left(1+e^{-v(\m)/T}\right)
\, d\m+[i\pi -i\theta(\la-q)]T-K(\la-q)\alpha_0^+T^2+O(T^3)\, ,
\ee
The integral equations satisfied by $v_1(\la)$  and $v_2(\la)$ are obtained by substituting
(\ref{uexp}), modified for the $v(\la)$ function, in (\ref{int28}). For $v_1(\la)$ it reads
\[
v_1(\la)+\frac{1}{2\pi}\int_{-q}^qK(\la-\m)\, v_1(\m)\, d\m=i\pi -i\theta(\la-q)\, .
\]
Comparison with the integral equation for the dressed phase (\ref{defdp}) and dressed
charge (\ref{defz}) shows that $v_1(\la)=i\pi Z(\la)-2i\pi  F(\la|q))$. Using the first
identity in (\ref{dpident}) the solution to this equation can be rewritten as
$v_1(\la)=i\pi(1-F(\la|q)-F(\la|-q)).$
The equation for $v_2(\la)$ is
\[
v_2(\la)+\frac{1}{2\pi}\int_{-q}^qK(\la-\m)\, v_2(\m)\, d\m=
K(\la-q)\left(-\alpha_0^++\frac{\pi}{12\eq}+\frac{v_1^2(q)}{4\pi\eq}\right)
+K(\la+q)\left(\frac{\pi}{12\eq}+\frac{v_1^2(-q)}{4\pi\eq}\right)\, ,
\]
with the solution
\[
v_2(\la)=-R(\la,q)\left[-2\pi\alpha_0^+
-\frac{1}{2\eq}\left(\frac{\pi^2}{3}+v_1^2(q)\right)\right]
-R(\la,-q)\left[
-\frac{1}{2\eq}\left(\frac{\pi^2}{3}+v_1^2(-q)\right)\right]\, .
\]
Using (\ref{eexp}) and (\ref{uexp}) in (\ref{clss}) and expanding the $p_0(\la^+_0)$
term up to the first order in $T$ we obtain
\be\label{int29}
\frac{1}{\xi^{(s)}[v]}=2\pi T \rho(q)\left(\alpha_0^+-\frac{v_1^2(q)}
{4\pi\eq}-\frac{v_1^2(-q)}{4\pi\eq}\right)
+O(T^2)\, .
\ee
In deriving (\ref{int29}) we have also used the fact that $F(\la|q)+F(\la|-q)$
is an odd function of $\la,$ ($F(-\la|-\m)=-F(\la|\m)$) and (\ref{ident1}).
The final result follows from (\ref{int00}) and the use of the  second identity in (\ref{dpident})
\be\label{ad1}
\frac{1}{\xi^{(s)}[v]}=\frac{2\pi T}{v_F}\left(\frac{1}{4\mathcal{Z}^2}+p_0^+-1\right)
+O(T^2)\, .
\ee

The case with $\la_0$ in the lower half-plane can be treated along similar lines
if we notice that (\ref{uexp}) remains valid even if  on the l.h.s we have an integral
over a modified contour. Considering $\la_0$ in the fourth quadrant then (\ref{int00})
is still valid but, in this case, $\Re \alpha_0^+<0$. We find
$v_1(\la)=i\pi(1-F(\la|q)-F(\la|-q))$ and the same expression (\ref{ad1}) for the
correlation length.
The  difference between the two cases is given by the range of allowed values for the
integer $p_0^+$.  The condition $\Re\alpha_0^+>0$ together with $v_1(q)=1-F(q|q)-F(q|-q)$
and $F(q|q)+F(q|-q)=-1+1/\mathcal{Z}$ imply that the allowed values for $p_0^+$ are
$\{2,3,\cdots\}$ for $\eta\in(\pi/2,\pi)\,  (\mathcal{Z}>1)$ and $\{1,2,\cdots \}$ for
$\eta\in(0,\pi/2)\, (1/\sqrt{2}<\mathcal{Z}<1)$. This shows that while for $\eta\in(0,\pi/2)$
the leading term of the expansion can be obtained with $\la_0$ in the upper half-plane this
is no longer true for $\eta\in(\pi/2,\pi)$. Imposing
$\Re\alpha_0^+<0$ we have $p_0^+=\{1\}$ for  $\eta\in(\pi/2,\pi)$ ( this also means
that   $\la_0$ is the solution lying in the lower half-plane of $1+e^{-v(\la_0)/T}=0$ closest to the real axis)
for which (\ref{ad1}) reproduces the leading term of the CFT expansion. Summarizing:
the leading term of the expansion is obtained for $\la_0$ in the upper (lower) half-plane
for $\eta\in(0,\pi/2)(\eta\in(\pi/2,\pi))$.

In a similar fashion we can treat the general case (\ref{cls}). As an example, for $\la_0$ in the first quadrant
and $l$ ``particle-hole"  pairs $\{\la_1^+,\cdots,\la_l^+\}$ in the second quadrant and $\{\la_1^-,\cdots,\la_l^-\}$
in the fourth quadrant we obtain
\be
\frac{1}{\xi^{(s)}[v]}=2ilk_F+\frac{2\pi T}{v_F}\left(\frac{1}{4\mathcal{Z}^2}+l^2\mathcal{Z}^2-l^2-1+p_0^++\sum_{k=1}^l(p_k^-+s_k^+) \right)+O(T^2)\, .
\ee
This distribution reproduces all the terms $l=1,2,\cdots$ appearing in the CFT expansion (\ref{asympts}) for $\eta\in(\pi/2,\pi)$.


\section{Continuum limit}\label{sCL}

In the previous sections we have obtained NLIEs and integral representations
for the auxiliary functions and eigenvalues of the QTM in the $N/2$ and
$N/2-1$ sector, valid at low-temperatures. In \cite{SBGK} it was shown that
by performing the continuum limit presented in Sec.~\ref{Continuum},
the Yang-Yang thermodynamics of the one-dimensional
Bose gas can be obtained from the largest eigenvalue of the
QTM. The next natural step is to perform the same limit in the equations
for the next largest eigenvalues obtaining the spectrum of what we can
call the ``continuum" quantum transfer matrix. The ratio of the largest
and next-largest eigenvalues of this ``continuum" QTM will give the correlation
lengths of the density-density and field-field correlation functions of the
Bose gas. The correspondence between the correlation functions in the two models
is presented in Table~\ref{table2}.
It should be noted that the results obtained for the Bose gas are
valid at all temperatures and are not restricted to low-temperatures as in the
case of
similar results obtained for the XXZ spin chain.

\begin{table}[h]
\caption{\label{table2} Correspondence in the continuum limit between the correlation functions of the XXZ spin chain and the one-dimensional Bose gas.}
\begin{center}
\begin{tabular}{|l|l|}
  \hline
  XXZ spin chain & One-dimensional  Bose gas \\
  \hline \hline
 $\langle\sigma^{(1)}_z\sigma^{(m+1)}_z\rangle_T$ & $\langle j(x)j(0)\rangle_T$\\
 $\langle\sigma^{(1)}_+\sigma^{(m+1)}_-\rangle_T$ & $\langle \Psi^\dagger(x)\Psi(0)\rangle_T$\\
 $\langle e^{\{\varphi\sum_{n=1}^me^{(n)}_{22}\}}\rangle_T$ & $\langle e^{\varphi\int_0^x j(x')\, dx'}\rangle_T$\\
  \hline
\end{tabular}
\end{center}
\end{table}

We start by showing how we can obtain (\ref{Pnls}), (\ref{YYnls}) from (\ref{LElt}).
Performing the continuum limit presented in Sec.~\ref{Continuum}, (this also includes
the reparametrization $\la=\delta k/\epsilon$ ) we obtain
\[
p_0(\la)\rightarrow \delta k\, ,\ \ \ \
p_0'(\la)\rightarrow \epsilon\, ,\ \ \ \
\theta(\la)\rightarrow -\overline\theta(k)\, ,\ \ \ \
\theta'(\la)=K(\la)\rightarrow
 -\frac{\epsilon}{\delta}\frac{2c}{k^2+c^2}=-\frac{\epsilon}{\delta}\overline K(k)\, ,\ \ \ \
\]
and $ e_0(\la)/T\rightarrow  \overline e_0(k)/\overline T$, where $\overline T$ is the
temperature in the Bose gas. Using these results and denoting by $\overline \varepsilon(k)/\overline T$
the continuum limit of $\varepsilon(\la)/T$ we see that  Eq.~(\ref{LElta})
transforms
into the Yang-Yang equation (\ref{YYnls}). The grand-canonical potential of the Bose
gas per unit length $\phi(\mu,\overline T)$ is related to the free energy of the
Heisenberg spin chain per lattice constant through the relation $\phi(\mu,\overline
T)=(f(h,T)+h/2)/\delta^3$.
Using $f(h,T)=-T\log\Lambda_0(0)$  it is easy to see that
(\ref{Pnls}) can be obtained from (\ref{LElte}) in the continuum limit.

We define the eigenvalues of the ``continuum" QTM by
\[
\log\overline\Lambda(0)=\frac{1}{\delta}\left(\log\Lambda(0)-\frac{h}{2T}\right)\, ,
\]
where on the r.h.s.~of this relation the continuum limit of $\log\Lambda(0)$ is understood. For the
eigenvalues in the $N/2$ sector we obtain
\be
\log\overline \Lambda^{(ph)}_i(0)=i\sum_{j=1}^r k^+_j-i\sum_{j=1}^r k^-_j
             +\frac{1}{2\pi}\int_{\mathbb{R}}\log\left(1+e^{-\overline u_i(k)/T}\right)dk\, ,
\ee
with $\overline u_i(k)$ satisfying (\ref{unls})
and for the eigenvalues in the $N/2-1$ sector
we find (the $i\pi$ term on the r.h.s.~of Eq.~(\ref{NLEsgf})  is irrelevant in the continuum limit)
\be
\log\overline\Lambda^{(s)}_i(0)=ik_0+i\sum_{j=1}^r k^+_j-\sum_{j=1}^r k^-_j
             +\frac{1}{2\pi}\int_{\mathbb{R}}\log\left(1+e^{-\overline v_i(k)/T}\right)dk\, ,
\ee
with $\overline v_i(k)$ satisfying (\ref{vnls}) and $k_0$ in the upper half-plane. When $k_0$ is
in the lower half-plane the integration contour has to be changed accordingly. The correlation lengths of the density-density
correlation function, $\langle j(x)j(0)\rangle_T,$ are obtained as
ratios of the largest ``continuum" eigenvalue,
$\log\overline \Lambda_0(0)=\frac{1}{2\pi}\int_{\mathbb{R}}\log\left(1+e^{-\overline\varepsilon(k)/T}\right)dk$,
and the next-largest ``continuum" eigenvalues in the $N/2$ sector justifying (\ref{cldd}).
In the case of the field-field correlation function, $\langle \Psi^\dagger(x)\Psi(0)\rangle_T,$
the correlation lengths are obtained using the next-largest eigenvalues in the $N/2-1$ sector
with the result (\ref{inte3}). The case of the generating functional is treated in Appendix \ref{agenfunc}.

\subsection{Checking the results}\label{Checknls}

The asymptotic expansions (\ref{AEdd}) and (\ref{AEff}) which are valid for all
temperatures should reproduce the TLL/CFT results \cite{BM1,BM2}:
\be\label{inte1}
\langle j(x)j(0)\rangle_T-\langle j(0)\rangle^2_T=-
\frac{(T\overline{\mathcal{Z}}/v_F)^2}{2\sinh^2(\pi T x/v_F)}+
\sum_{l\in\mathbb{Z}^*}\tilde A_l\ e^{2i xl k_F }\left(\frac{\pi T/v_F}{\sinh(\pi T x/v_F)}
\right)^{2l^2\overline{\mathcal{Z}}^2}\, , \ \ \ \ \ x\rightarrow\infty\, ,
\ee
\be\label{inte2}
\langle\Psi^\dagger(x)\Psi(0)\rangle_T=
\sum_{l\in\mathbb{Z}}\tilde B_l\ e^{2i xl k_F }\left(\frac{\pi T/v_F}{\sinh(\pi T x/v_F)}
\right)^{\frac{1}{2\overline{\mathcal{Z}}^2}+2l^2\overline{\mathcal{Z}}^2}\, , \ \ \ \ \ x\rightarrow\infty\, .
\ee
in the $T\rightarrow 0$ limit. In Eqs.~(\ref{inte1}) and (\ref{inte2}), $v_F$ is the Fermi velocity
defined in (\ref{vfnls}), $k_F=\pi \overline D$ is the Fermi momentum and $\overline{\mathcal{Z}}$ is the
dressed charge evaluated at $\overline q$ (see \ref{dcnls}). The agreement with the conformal results is proved
in Appendix \ref{ACFTnls}.

In the impenetrable limit, $c\rightarrow \infty$, the leading term of the asymptotic
expansion for the field-field correlation function was computed by Its, Izergin and Korepin
by solving an associated Riemann-Hilbert problem \cite{IIK1}, Chap.~XVI. of \cite{KBI},
This gives us another opportunity to check the validity of our results by comparison
with another exact result. The leading term in the expansion (\ref{AEff}) is obtained
when $r=0$ in Eq.~(\ref{vnls}). We will consider that $k_0$ is in the first quadrant,
$\Re k_0\geq 0\, ,\Im k_0> 0$. Taking into account that
\[
\lim_{c\rightarrow \infty} \overline K(k)=\lim_{c\rightarrow \infty} \overline \theta (k)=0
\]
the equations (\ref{vnls}) for the auxiliary function $\overline v(k)$ (we drop the subscript $i$)
and dressed energy (\ref{YYnls}) become
\be\label{inte4}
\overline v(k)=k^2-\mu+i\pi T\, , \ \ \ \ \ \ \ \ \overline\varepsilon(k)=k^2-\mu\, .
\ee
The asymptotic behavior depends on the sign of the chemical potential. We will consider first
the case of negative chemical potential. In this case the solution of the equation $1+e^{-\overline v(k_0)/T}$
which is closest to the real axis and
located in the first quadrant is $k_0=i\sqrt{|\mu|}$. Using (\ref{inte4})
and this value for $k_0,$ the correlation length (\ref{inte3}) can be rewritten as
\[
\frac{1}{\xi^{(s)}[\overline v]}=\frac{1}{2\pi}\int_{\mathbb{R}}
\log\left(\frac{e^{(k^2-\mu)/T}+1}{e^{(k^2-\mu)/T}-1}\right)\, dk +\sqrt{|\mu|}\, ,\ \ \ \ \ \ \ \mu<0
\]
which is precisely the result obtained in \cite{IIK1} for negative chemical potential. In the
case of positive chemical potential we have $k_0=\sqrt{\mu}$ and the correlation length is
\begin{align*}
\frac{1}{\xi^{(s)}[\overline v]}&=\frac{1}{2\pi}\int_{\mathbb{R}}
\log\left(\frac{e^{(k^2-\mu)/T}+1}{e^{(k^2-\mu)/T}-1}\right)\, dk -i\sqrt{\mu}\, ,\\
&=\frac{1}{2\pi}\int_{\mathbb{R}}
\log\left|\frac{e^{(k^2-\mu)/T}+1}{e^{(k^2-\mu)/T}-1}\right|\, dk \, ,
\end{align*}
coinciding with the result derived in \cite{IIK1}.

\subsection{Numerical results}
In this section we present some numerical solutions to the non-linear integral
equations derived above. Quite generally, we truncate the real axis to a
finite symmetric interval and use a uniform discretization. The convolution
type integrals are carried out by Fourier transforms. In ``momentum space''
convolutions are done by simple products of the Fourier transforms of the
functions resulting in an efficient numerical algorithm. The integral equation
for the function and the subsidiary equations for the discrete excitation
parameters are solved by iterations which turn out to be quickly convergent.

The results obtained in this paper for the Hamiltonian (\ref{hamLL})
are given in dimensionless units. Restoring physical units is a simple
task which can be accomplished in the following way. For particles
of mass $m$  and contact interaction strength $g$ we introduce
a length scale $a$ via $c=m g a/\hbar^2$. Then, the units of temperature,
chemical potential, density of particles, reciprocal correlation length, wavenumber  and specific heat
are $T_0=\hbar^2/(2 m a^2 k_B)$, $\mu_0= \hbar^2/(2 m a^2),$ $n_0=\xi_0^{-1}=k_0=1/a$,
$c_0=k_B/a$. The physical data presented in the figures of this section is
given in these units for three values of the chemical potential $\mu=-1,
0, +1$ and fixed value of the dimensionless coupling $c=2$ which is realized for any
parameter values of $m$ and $g$ with a suitably chosen $a=2 \hbar^2/(m g)$.

\begin{figure}
\includegraphics*{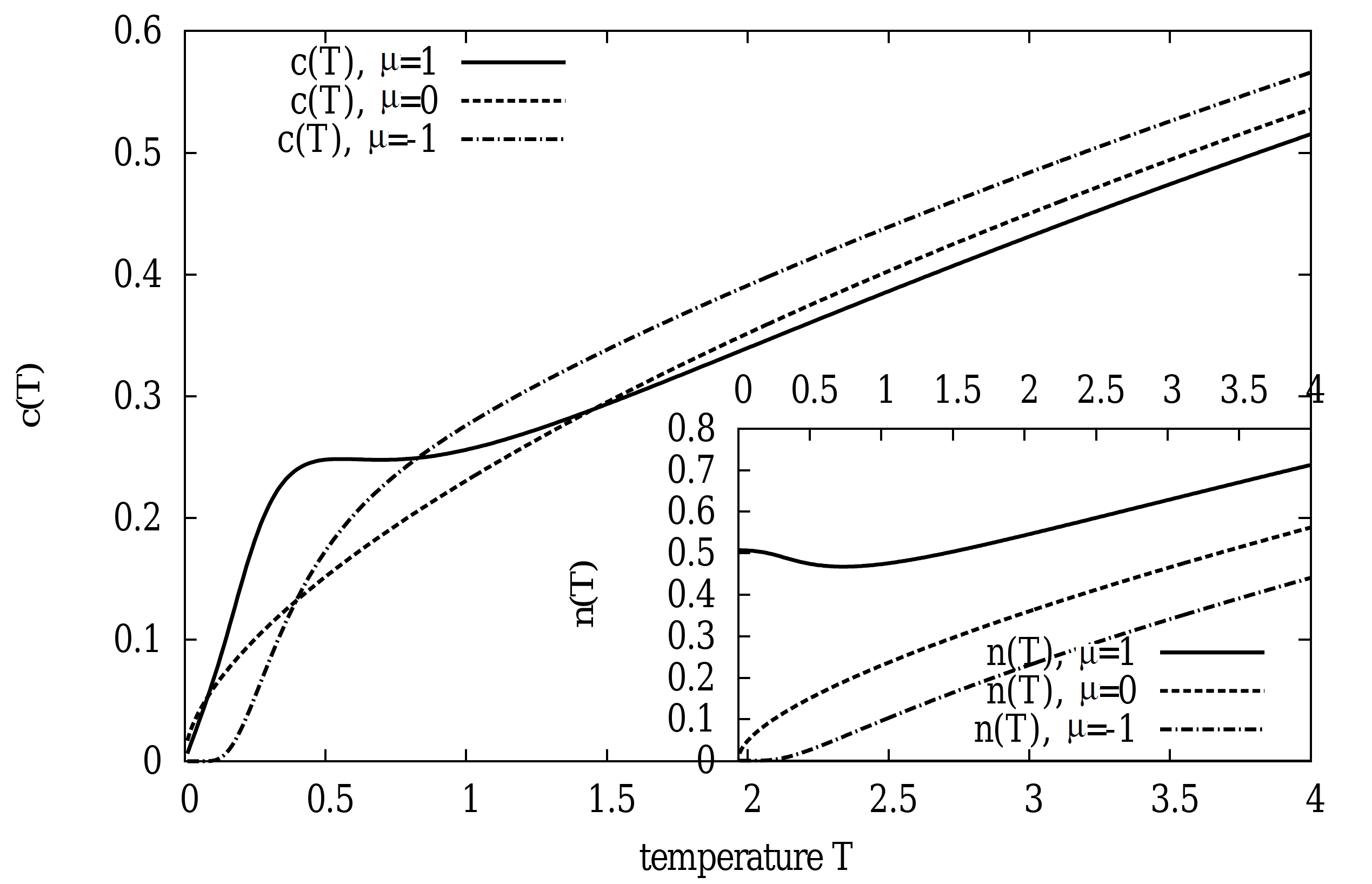}
\caption{
Thermodynamics: Specific heat as a function of temperature
$T$ for three characteristic cases of the chemical potential $\mu=-1, 0, +1$
and fixed interaction strength $c=2$. (Inset) Particle densities as
functions of temperature. (All quantities in units of $c_0,$ $T_0,$ $\mu_0$ and $n_0$.)
}
\label{FigCN}
\end{figure}

The specific heat and the particle density in grand-canonical
ensemble are shown in Fig.\ref{FigCN}. Note that negative chemical potentials
like $\mu=-1$ correspond to the {\em dilute phase} as the particle density vanishes
at low temperatures exponentially as does the specific heat, $c(T), n(T)
\simeq \exp(\mu/T)$. Positive chemical potentials like $\mu=+1$ correspond to
the {\em dense phase} with finite particle density at low temperatures and linear
dependence of the specific heat on temperature. The ``critical'' chemical
potential $\mu=0$ separates the dilute and dense phases and shows a square
root dependence of specific heat and particle density on temperature $c(T),
n(T) \simeq T^{1/2}$.

\begin{figure}
\includegraphics*{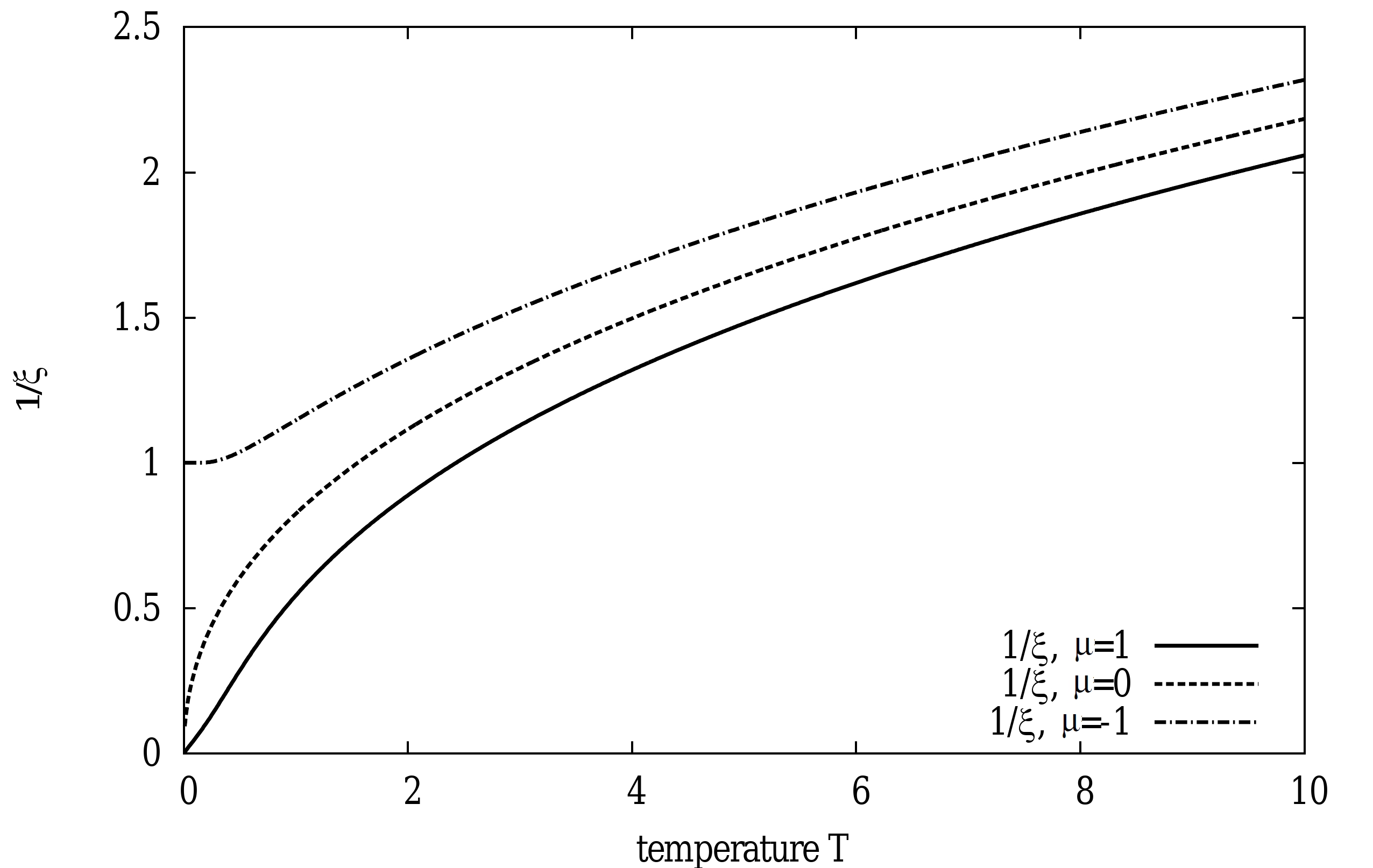}
\caption{
Green's function: Reciprocal correlation length
  $1/\xi$ as a function of  temperature $T$ for three characteristic cases
  of the chemical potential $\mu=-1, 0, +1$ and fixed interaction strength
  $c=2$. (All quantities in units of $\xi_0^{-1},$ $T_0$ and  $\mu_0$.)}
\label{Fighbw}
\end{figure}

Next, we like to present our results for the leading correlation length of the
Green's function. We calculate the distribution shown for the {\em dense
  phase} in Fig.\ref{NLEsfig} by means of the above non-linear integral
equation (\ref{vnls}). First of all, we realize that due to the coupling of
all roots and holes, a backflow effect sets in and the distribution shown in
Fig.\ref{NLEsfig} symmetrizes. And second, for lower temperatures all hole
parameters including $k_0$ are below the real axis and all roots are
above. For the numerical treatment of this distribution a straight integration
contour is more suitable than the indented contour that allowed for a uniform
treatment of the CFT properties. Choosing a straight contour for the case of
$k_0$ below the real axis makes the contribution of $k_0$ to the driving
term disappear, but imposes a severe change on the asymptotics of
$\log\left(1+e^{-\overline{v}_i(k)/T}\right)$. This function converges to $0$
for $k\to -\infty$, but to $-2\pi$i for $k\to +\infty$. This modified
asymptotics can be enforced numerically and yields the results shown in
Fig.\ref{Fighbw}.
Note  that  there is no oscillating $k_F$ factor for this correlation function.
For $\mu<0$ the low temperature limit of $1/\xi(T)$ is finite,
for the critical value $\mu=0$ we see a $1/\xi(T) \simeq T^{1/2}$ behaviour
and for $\mu>0$ the CFT behaviour ${1/\xi(T) \simeq 2\pi \frac {T}{v_F} \frac
  1{4{\cal Z}^2} }$ sets in at low temperatures.

\begin{figure}
\includegraphics*{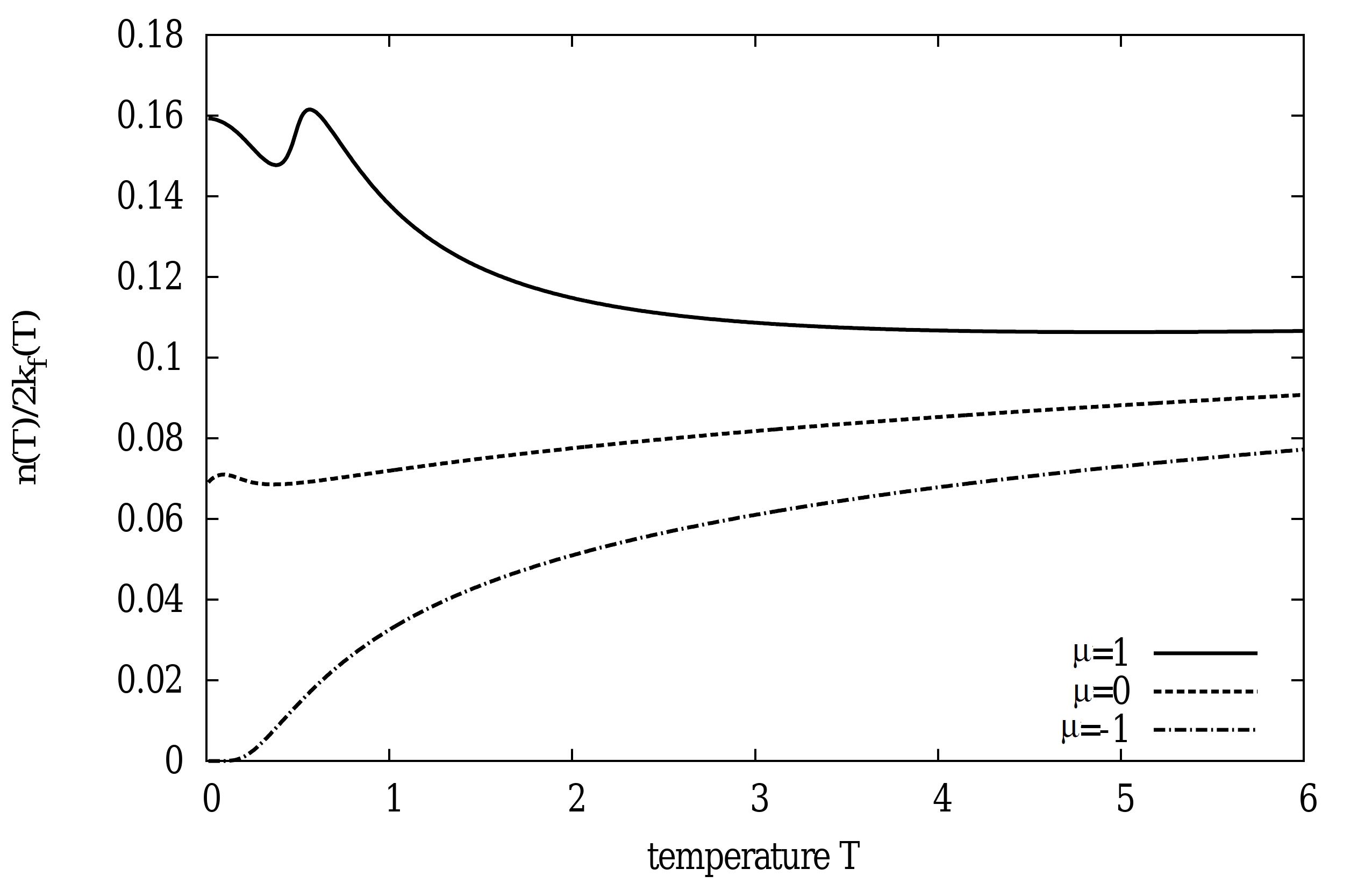}
\caption{
Density-density correlation function (2nd leading term): Particle number
$n(T)$ divided by the wave number $2k_F(T)$ as a function of temperature $T$. Note for the case
$\mu=+1$ the universal $T\to 0$ limit
$1/2\pi$. (Temperature in units of   $T_0$.)
}
\label{Fignkf}
\end{figure}

Finally, we present our results for the density-density correlator. The
leading term is given by a ``particle-hole excitation'' at one Fermi point
without $2k_F$ oscillations at low temperature, see
(\ref{inte1}). Interestingly, for this leading contribution there is a
cross-over scenario at elevated temperatures from non-oscillating to
oscillating behaviour. The detailed study of this phenomenon is beyond the
scope of this publication. Therefore, we restrict ourselves to the study of
the next-leading contribution with $2k_F$ oscillations at low temperatures
with roots and holes as illustrated in Fig.\ref{NLEphfig}.
\begin{figure}
\includegraphics*{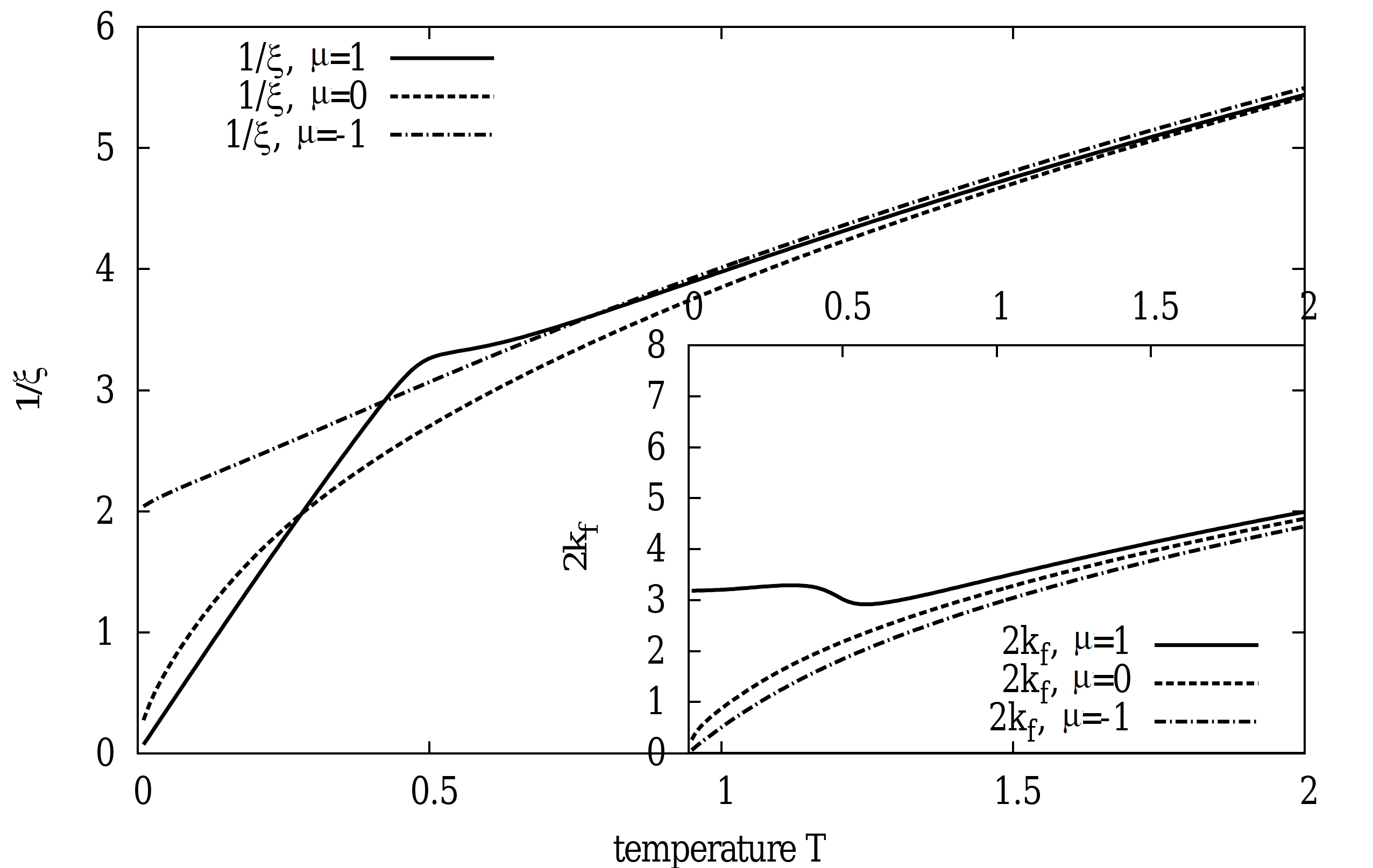}
\caption{
Density-density correlation function (2nd leading term):
 Reciprocal correlation length $1/\xi$ as a function of temperature $T$
  for three characteristic cases of the chemical potential $\mu=-1, 0, +1$ and
  fixed interaction strength $c=2$. (Inset) Wave number $2k_F$ of the
  oscillating factor. (All quantities in units of $\xi^{-1}_0,$ $T_0,$ $\mu_0$ and $k_0$.)
  }
\label{Figphbw}
\end{figure}
For $\mu<0$ the low temperature limit of $1/\xi(T)$ is finite,
for the critical value $\mu=0$ we see a $1/\xi(T) \simeq T^{1/2}$ behaviour
and for $\mu>0$ the CFT behaviour ${1/\xi(T) \simeq 2\pi \frac {T}{v_F} {{\cal Z}^2}
}$ sets in at low temperatures. The oscillations $2k_F$ vanish at low $T$ for
$\mu\le 0$. In the dense phase ($\mu>0$) at low $T$ we expect the universal
relation $2k_F(T)\simeq
2\pi n(T)$ which is nicely satisfied at very low $T$, see Fig.\ref{Fignkf},
but shows a non-trivial temperature dependence at elevated $T$.
We like to note that the dressed charge for $\mu=+1$ and $c=2$ takes
the value ${\cal Z}=1.38$ consistent with the low-temperature behaviour of the
correlators shown above.

\section{Conclusions}

Using the spectrum of the XXZ spin chain QTM and a specific continuum limit we
have derived  the asymptotic expansions of the temperature dependent
density-density and  field-field  correlation functions in the
interacting one-dimensional Bose gas. As a by-product we have also
obtained
similar expansions, valid at low-temperatures, for the
longitudinal and transversal correlation functions  in the XXZ
spin chain. One could naturally expect that similar results can be derived
in the case of the spinorial 1D Bose gas \cite{LGYE} which can be obtained as
the continuum limit of the $U_q(\hat {sl}(3))$ Perk-Schultz spin chain \cite{BVV,dV}.
This subject  will be deferred to a future publication.

\section{Acknowledgements}

The authors would like to thank Frank G\"ohmann for numerous comments
and   discussions.  Financial
support from the  VolkswagenStiftung and the PNII-RU-TE-2012-3-0196 grant  of the
Romanian National Authority for Scientific Research is gratefully acknowledged.

{\it Note added in proof-} Recently we  become aware of
\cite{DGK}  where some of the results for the XXZ spin
chain were rederived and  generalized.




\begin{appendices}

\section{Derivation of the integral equations for the next-largest eigenvalues in the $N/2$ sector}\label{Anleph}

\subsection{Integral equation for the auxiliary function}

\begin{figure}
\includegraphics[width=1\linewidth]{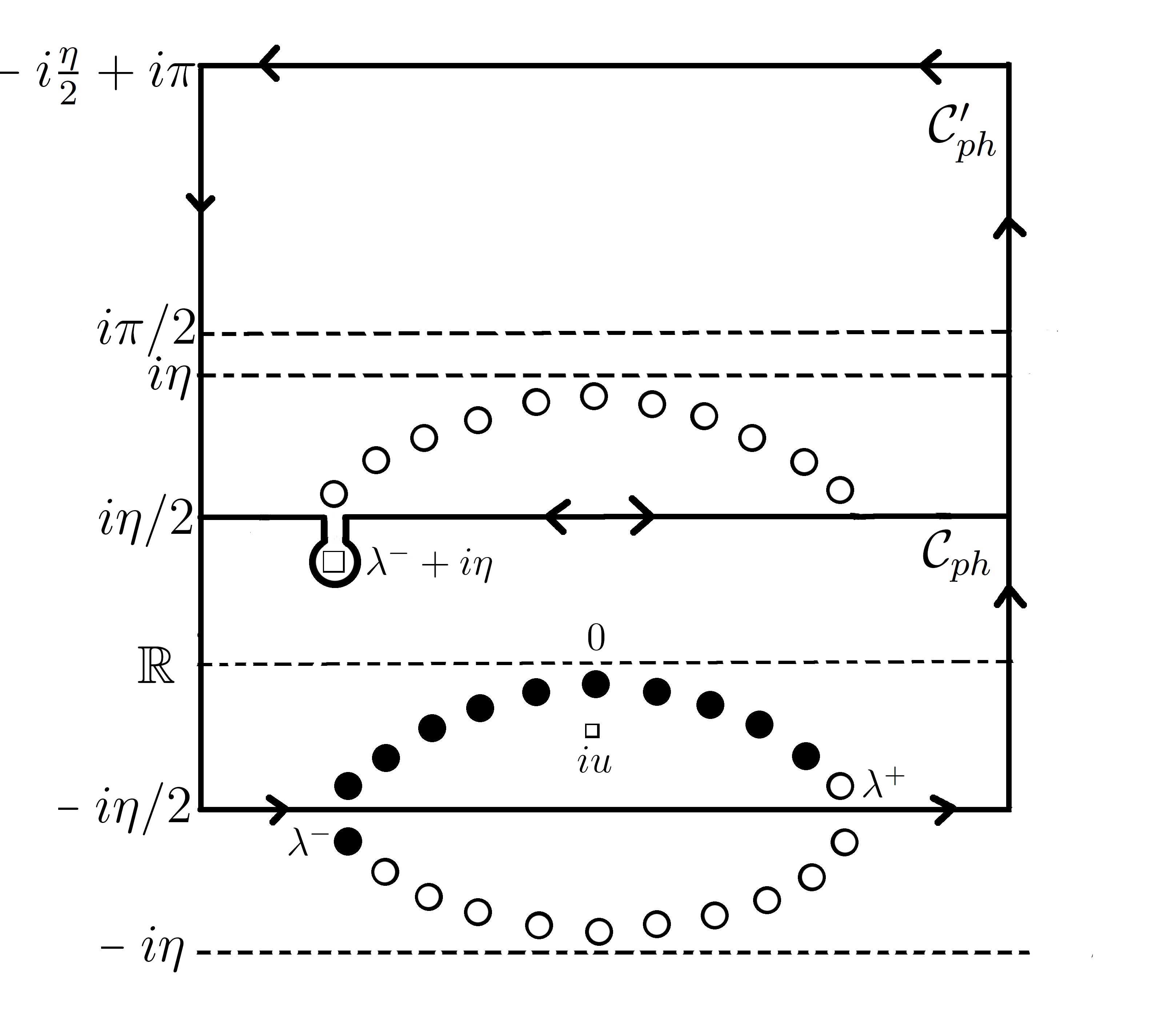}
\caption{
Typical distribution of Bethe roots ($\bullet$) and holes ($\circ$) in the strip $|\Im \lam|<\eta\, ,\eta\in[0,\pi/2)$ characterizing
one of the next-largest eigenvalues of the QTM in the $N/2$ sector at low-temperature. All the other roots
and holes can be obtained using the $i\pi$ periodicity. The indentation of the contour $\mathcal{C}_{ph}$
excludes $\lambda^-+i \eta$ and a hole located close to it. The lower edge of the contour  $\mathcal{C}_{ph}'$
is the same as the upper edge of $\mathcal{C}_{ph}$  but with different orientation.
The contour  $\mathcal{C}_{ph}$ surrounds all the Bethe roots except $\la^-$ the hole
located at $\la^+$ and the pole of the auxiliary function $\fb(\la)$ at $iu$.
}
\label{NLEphfig}
\end{figure}

For reasons of clarity we are going to consider first the simplest case in which only one
Bethe root/hole is outside/inside the relevant strip in the complex plane. The
generalization to the case of $r$ pairs is a natural extension of this particular example.
A typical distribution of roots and holes for  $\eta\in(0,\pi/2)$ and low-temperatures
is presented in  Fig.~\ref{NLEphfig},
where  we have denoted by $\la^-$ and $\la^+$ the Bethe root,
respectively, the  hole outside (inside) the strip $|\Im\la|<\eta/2.$ It should be
emphasized that this ``particle-hole'' distribution is valid only at low-temperatures,
at higher  temperatures the next-largest eigenvalues in the $N/2$ sector are characterized
by  the so-called 1-string type and 2-string type solutions \cite{KMSS}. The eigenvalue
and the  auxiliary function $\fb(\la)$ corresponding to the distribution presented in
Fig.~\ref{NLEphfig} are  described by the same formulas as in (\ref{Lamle}) and (\ref{defaa}).
It is useful to present  $q(\la)$ in the following form
\[
q(\la)=\sinh(\la-\la^-)\prod_{j=1}^{\frac{N}{2}-1}\sinh(\la-\la_j)\, ,
\]
where $\{\la_j\}_{j=1}^{N/2-1}$ are the $N/2-1$ Bethe roots inside the strip $|\Im\la|<\eta/2.$ The equation
$\fb(\la)+1=0$  has $3N/2$ solutions, of which $N/2$ are the Bethe roots and $N$
are holes denoted by $\{\la^{(h)}_j\}_{j=1}^{N-1}$ and $\la^+.$

We introduce the rectangular contour $\mathcal{C}_{ph}$ (see Fig.~\ref{NLEphfig})
centered at the origin, extending
to infinity
with the edges parallel to the real axis
through $\pm i(\eta-\epsilon)/2\, ,$
$\epsilon\rightarrow 0$ which presents an indentation of the upper edge such that
$\la^-+i\eta$ is not in the interior of the contour.  Inside the contour
$\mathcal{C}_{ph}$ the function $1+\fb(\la)$ has $N/2$ zeros at the Bethe roots $\{\la_j\}_{j=1}^{N/2-1}$
and hole  $\la^+$ and a pole of order $N/2$ at $iu$. Therefore, the function
$\log(1+\fb(\la))$ has no winding number around the contour (the presence of the
indentation ensures that the function $\log(1+\fb(\la))$ does not have an
extra pole at $\la^-+i\eta$)  allowing us to define ($\la$ is
located outside the contour $\mathcal{C}_{ph}$)
\be\label{ff}
f_{ph}(\la)\equiv\frac{1}{2\pi i}\int_{\mathcal{C}_{ph}}\frac{d}{d\la}\left(\log\sinh(\la-\m)\right)
\log(1+\fb(\m))d\m=\frac{1}{2\pi i}\int_{\mathcal{C}_{ph}}\log\sinh(\la-\m)
\frac{\fb'(\m)}{1+\fb(\m)}d\m\, ,
\ee
which can be evaluated using Theorem \ref{T} with the result
\be\label{int10}
f_{ph}(\la)=\sum_{j=1}^{N/2-1}\log\sinh(\la-\la_j)+\log\sinh(\la-\la^+)-\frac{N}{2}\log\sinh(\la-iu)\, .
\ee
Eq.~(\ref{int10}) which can be rewritten as $\log q(\la)=f_{ph}(\la)+\log\sinh(\la-\la^-)-\log\sinh(\la-\la^+)
+\log\phi_-(\la)+N/2\log\sin\eta$
providing an integral representation for $\log q(\la).$
Taking the logarithm of Eq.~(\ref{defaa}) and using this integral representation we find
\[
\log\fb(\la)=-\beta h+\log\left(\frac{\phi_+(\la)}{\phi_-(\la)}\frac{\phi_-(\la+i\eta)}
{\phi_+(\la+i\eta)}\right)+ \log\left(\frac{\sinh(\la-\la^-+i\eta)}{\sinh(\la-\la^--i\eta)}
\right)-\log\left(\frac{\sinh(\la-\la^++i\eta)}{\sinh(\la-\la^+-i\eta)}
\right)+f_{ph}(\la+i\eta)-f_{ph}(\la-i\eta)\, .
\]
Performing the Trotter limit, $N\rightarrow\infty$, with the help of Eq.~(\ref{Trotterl}) we
obtain the NLIE for the auxiliary function
\begin{align}\label{NLIEaph}
\log\fb(\la)&=-\beta e_0(\la+i\eta/2)
+i\theta(\la-\la^+)
-i\theta(\la-\la^-)
-\frac{1}{2\pi}\int_{\mathcal{C}_{ph}}K(\la-\m)\log(1+\fb(\m))d\m\, .
\end{align}
Eq.~(\ref{NLIEaph}) was obtained assuming $\eta\in(0,\pi/2).$ It remains
valid also for $\eta\in(\pi/2,\pi)$ if the contour $\mathcal{C}_{ph}$ is replaced by
a similar rectangular contour with the upper (lower) edges parallel to the real axis
through $\pm i(\pi-\eta-\epsilon)/2\, , \epsilon\rightarrow 0$  but, in this case,
without the indentation.

\subsection{Integral expression for the next-largest eigenvalue in the $N/2$ sector}

The integral expression for the next-largest eigenvalue in the $N/2$ sector
is obtained in a similar fashion as in the largest eigenvalue case. The starting
point is, again, the  representation (\ref{Lh}) of the eigenvalue in terms of the
holes where it is useful to denote $q^{(h)}(\la)$ as
\[
q^{(h)}(\la)=\sinh(\la-\la^+)\prod_{j=1}^{N-1}\sinh(\la-\la_j^{(h)})\, .
\]
In order to obtain an integral expression for $q^{(h)}(\la)$ we introduce a
rectangular contour $\mathcal{C}_{ph}'$ (see Fig.~\ref{NLEphfig}) extending
to infinity
with the edges parallel to the real axis
through $i(\eta-\epsilon)/2$ and
$-i(\eta-\epsilon)/2+i\pi.$ The edge at $i(\eta-\epsilon)/2$ presents an indentation
such that $\la_-+i\eta$ is contained in the interior of $\mathcal{C}_{ph}'$ and
is identical with the upper edge of the contour $\mathcal{C}_{ph}$ but with
opposite orientation. Then the following identity
\be\label{integralrph}
\int_{\mathcal{C}_{ph}+\mathcal{C}_{ph}'}d(\la-\m)\frac{\fb'(\m)}{1+\fb(\m)}d\m=0\, ,
\ee
can be proved in exactly the  same way as its largest eigenvalue counterpart
(\ref{integralr}).
For $\la$ close to the real axis using (\ref{integralrph}) and Theorem \ref{T} we
obtain
\begin{align}\label{int11}
\frac{1}{2\pi i}\int_{\mathcal{C}_{ph}}d(\la-\m)\frac{\fb'(\m)}{1+\fb(\m)}d\m&=-
\left(\sum_{j=1}^{N-1}d(\la-\la_j^{(h)})+d(\la-\la^-)-d(\la-\la^--i\eta)\right.\nonumber\\
                &\ \ \ \ \ \ \ \ \ \ \ \ \ \ \ \ \ \ \ \ \ \ \ \ \ \
                \left.-\sum_{j=1}^{N/2-1}d(\la-\la_j-i\eta)-\frac{N}{2}d(\la+iu+i\eta)\right)\, .
\end{align}
In deriving Eq.~(\ref{int11}), we have used the fact that, inside the contour $\mathcal{C}_{ph}',$
the function $1+\fb(\la)$ has: $N$ zeros at $\la^-+i\pi,$ $\{\la_j^{(h)}\}_{j=1}^{N-1}$ or
$\{\la_j^{(h)}\}_{j=1}^{N-1}+i\pi\, ,$ $N/2-1$ simple poles at $\{\la_j\}_{j=1}^{N/2-1}+i\eta$,
a simple pole at $\la^-+i\eta$ and a pole of order $N/2$ at  $-iu-i\eta+i\pi$. Using again
Theorem \ref{T} and the fact that inside the contour $\mathcal{C}_{ph}$ the function $1+\fb(\la)$
has: $N/2$ zeros at the Bethe roots $\{\la_j\}_{j=1}^{N/2-1}$ and hole $\la^+$, and a pole of order $N/2$ at $iu$
we find
\be\label{int12}
\frac{1}{2\pi i}\int_{\mathcal{C}_{ph}}d(\la-\m-i\eta)\frac{\fb'(\m)}{1+\fb(\m)}d\m=
\sum_{j=1}^{N/2-1}d(\la-\la_j-i\eta)+d(\la-\la^+-i\eta)-\frac{N}{2}d(\la-iu-i\eta)\, .
\ee
Taking the difference of Eqs.~(\ref{int11}) and (\ref{int12}), integrating by parts,
and then integrating w.r.t.~$\la$ we obtain the following representation
\begin{align}\label{int13}
\log q^{(h)}(\la)&=\log\left(\frac{\sinh(\la-\la^+)}{\sinh(\la-\la^+-i\eta)}\right)-
                   \log\left(\frac{\sinh(\la-\la^-)}{\sinh(\la-\la^--i\eta)}\right)
                   +\log\left(\phi_+(\la+i\eta)\phi_-(\la-i\eta)\right)\nonumber\\
&\ \ \ \ \ \ \ \ \ \ \ \ \ \ \ \
-\frac{1}{2\pi i}\int_{\mathcal{C}_{ph}}\left[d(\la-\m)-d(\la-\m-i\eta)\right]\log(1+\fb(\m))\, d\m
+c\, ,
\end{align}
with $c$ a constant of integration. Finally, the integral expression for the next-largest eigenvalue
of the QTM in the $N/2$ sector is obtained by replacing (\ref{int13}) in (\ref{Lh}) with the result
\be\label{NLEph}
\log\Lambda_{ph}(0)=\frac{\beta h}{2}+\log\left(\frac{\sinh\la^+}{\sinh(\la^++i\eta)}\right)-
                   \log\left(\frac{\sinh\la^-}{\sinh(\la^-+i\eta)}\right)+
\frac{1}{2\pi }\int_{\mathcal{C}_{ph}}p_0'(\m+i\eta/2)\log(1+\fb(\m))d\m\, .
\ee
The constant of integration, $\beta h/2,$ was calculated using the behavior of the involved
functions at infinity, like in the case of the largest eigenvalue. Eq.~(\ref{NLEph}) is also
valid in the domain $\eta\in(\pi/2,\pi)$ if the contour $\mathcal{C}_{ph}$ is replaced
by a rectangular contour, extending
to infinity, with the edges parallel to the real axis
through $\pm i(\pi-\eta-\epsilon)/2\, ,$ $\epsilon\rightarrow 0$.

\subsection{Final form of the integral equations}\label{Aphlt}

We consider $\eta\in(0,\pi/2)$. In the low-temperature limit we are going to neglect
the contribution from the upper edge of the contour as we did in Sec.~\ref{SLElt}.
If in Eq.~(\ref{NLIEaph}) we  restrict the free parameter $\la$ and the variable of integration
to the lower part of the contour which is the line parallel to the real axis at $-i\eta/2$
we find
\begin{align}
\log\fb(\la-i\eta/2)&=-\beta e_0(\la)
+i\theta(\la-\la^+)
-i\theta(\la-\la^-)
-\frac{1}{2\pi}\int_{\mathbb{R}}K(\la-\m)\log(1+\fb(\m-i\eta/2))d\m\, ,
\end{align}
where by $\la^\pm$ we understand  $\la^\pm\rightarrow\la^\pm+i\eta/2\, ,$ which now
belong to the upper (lower) half-plane.
The expression (\ref{NLEph}) for the next-largest eigenvalue becomes
\begin{align}\label{int14}
\log\Lambda_{ph}(0)&=\frac{\beta h}{2}+ip_0(\la^+)-ip_0(\la^-)
                  +\frac{1}{2\pi }\int_{\mathbb{R}}p_0'(\m)\log(1+\fb(\m-i\eta/2))d\m\, .
\end{align}
Introducing the  function $u(\la)$ satisfying  $e^{-u(\la)/T}=\fb(\la-i\eta/2),$
the NLIE for the auxiliary function and the integral expression for the next-largest eigenvalues
in the $N/2$ sector at low-temperatures can be written as
\be\label{int15}
u(\la)=e_0(\la)-iT\theta(\la-\la^+)+iT\theta(\la-\la^-)+
\frac{T}{2\pi}\int_{\mathbb{R}}K(\la-\m)\log\left(1+e^{-u(\m)/T}\right)d\m
\, ,
\ee
\be\label{int16}
\log\Lambda_{ph}(0)=\frac{h}{2T}+ip_0(\la^+)-ip_0(\la^-)+\frac{1}{2\pi}\int_{\mathbb{R}}p_0'(\m)\log\left(1+e^{-u(\m)/T}\right)d\m\, ,
\ee
where the  parameters $\la^\pm$ satisfy the constraint  $1+e^{-u(\la^\pm)/T}=0,$
and are
located in the upper (lower) half-plane. While we derived these equations assuming that $\eta\in(0,\pi/2)$
we are going to assume that they are valid also for $\eta\in(\pi/2,\pi)$. The CFT analysis in Sec.\ref{CFT}
shows that this assumption is justified. The obvious generalization of
Eqs.~(\ref{int15}) and (\ref{int16}) in the case of $r$ pairs of Bethe roots and
holes is given by Eqs.~(\ref{NLIEphgf}) and  (\ref{NLEphgf}).

\section{Derivation of the integral equations for the next-largest eigenvalues in the $N/2-1$ sector}\label{Anles}

\subsection{Integral equation for the auxiliary function}

\begin{figure}
\includegraphics[width=1\linewidth]{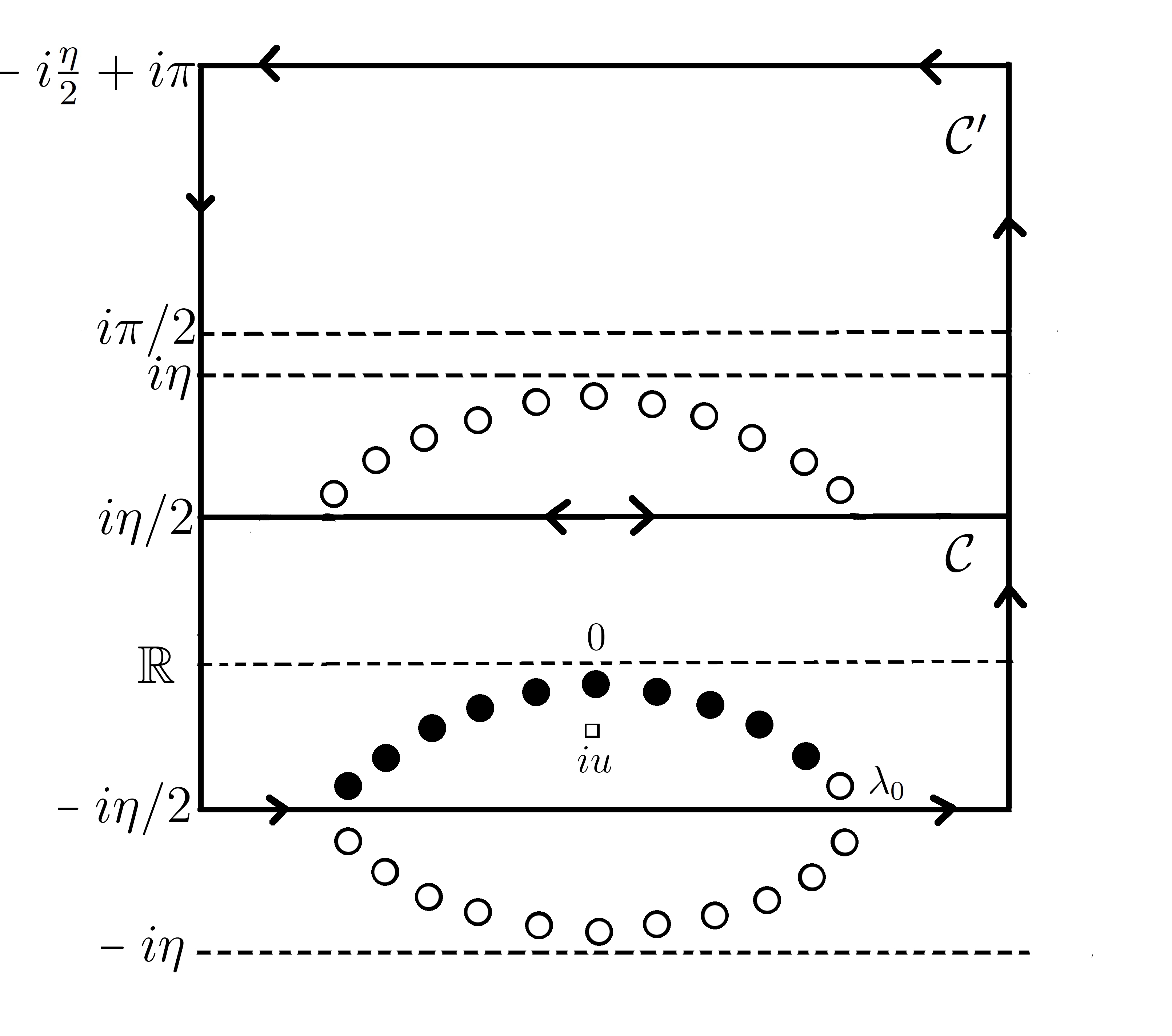}
\caption{
Typical distribution of Bethe roots ($\bullet$) and holes ($\circ$) in the strip $|\Im \lam|<\eta\, ,\eta\in[0,\pi/2)$ characterizing
the next-largest eigenvalues of the QTM in the $N/2-1$ sector without ``particle-hole'' type terms.
All the other roots and holes can be obtained using the $i\pi$ periodicity.
The contour  $\mathcal{C}$ surrounds all the Bethe
roots, the hole located at $\la_0$ and the pole of the auxiliary function $\fc(\la)$ at $iu$.
}
\label{NLEsfig}
\end{figure}

As we have mentioned in Sec.~\ref{Ss} it is sufficient to consider the case
with $N/2-1$ Bethe roots and, possibly, one hole in the relevant strip of the complex plane.
First, we will consider the case with one hole inside the strip.
A typical distribution of the Bethe roots and hole, at low-temperatures and  $\eta\in(0,\pi/2)$,
is presented in Fig.~\ref{NLEsfig},  where we have denoted by $\la_0$ the hole inside the
strip $|\Im \la|<\eta/2.$ The eigenvalue and auxiliary function $\fc(\la)$ corresponding
to the distribution presented in Fig.~\ref{NLEsfig}
are described by formulas similar to
(\ref{Lamle}) and (\ref{defaa}), but in this case $q(\la)$ is defined as
\[
q(\la)=\prod_{j=1}^{\frac{N}{2}-1}\sinh(\la-\la_j)\, ,
\]
where $\{\la_j\}_{j=1}^{N/2-1}$ are the $N/2-1$ Bethe roots. The equation $\fc(\la)+1=0$ has
$3N/2-1$ solutions, of which,  $N/2-1$ are Bethe roots and $N$ are holes
denoted  by $\{\la^{(h)}\}_{j=1}^{N-1}$ and  $\la_0$.

Consider the  contour $\mathcal{C}$ introduced in Sec.~\ref{Sleie}. Inside the
contour, the function $1+\fc(\la)$ has $N/2$ zeros at the Bethe roots $\{\la_j\}_{j=1}^{N/2-1}$
and hole $\la_0$, and a pole of order $N/2$ at $iu$. Therefore, we can
define ($\la$ is
located outside the contour $\mathcal{C}$)
\be\label{fff}
f_{s}(\la)\equiv\frac{1}{2\pi i}\int_{\mathcal{C}}\frac{d}{d\la}\left(\log\sinh(\la-\m)\right)
\log(1+\fc(\m))d\m=\frac{1}{2\pi i}\int_{\mathcal{C}}\log\sinh(\la-\m)
\frac{\fc'(\m)}{1+\fc(\m)}d\m\, ,
\ee
which can be evaluated using Theorem \ref{T} with the result
\be\label{int17}
f_{s}(\la)=\sum_{j=1}^{N/2-1}\log\sinh(\la-\la_j)+\log\sinh(\la-\la_0)-\frac{N}{2}\log\sinh(\la-iu)\, .
\ee
Taking the logarithm in Eq.~(\ref{defaa}), and using (\ref{int17}), which can be rewritten as
$\log q(\la)=f_s(\la)-\log\sinh(\la-\la_0)+\log\phi_-(\la)+N/2\log\sin\eta$ we find
\[
\log\fc(\la)=-\beta h+\log\left(\frac{\phi_+(\la)}{\phi_-(\la)}\frac{\phi_-(\la+i\eta)}
{\phi_+(\la+i\eta)}\right)+ \log\left(\frac{\sinh(\la-\la_0-i\eta)}{\sinh(\la-\la_0+i\eta)}
\right)+f_{s}(\la+i\eta)-f_{s}(\la-i\eta)\, .
\]
Making use of Eq.~(\ref{Trotterl}), we can take the  Trotter limit, $N\rightarrow\infty,$
obtaining the NLIE for the auxiliary function
\begin{align}\label{NLIEas}
\log\fc(\la)&=-\beta e_0(\la+i\eta/2)
\mp i\pi+i\theta(\la-\la_0)
-\frac{1}{2\pi}\int_{\mathcal{C}}K(\la-\m)\log(1+\fc(\m))d\m\, .
\end{align}
In Eq.~(\ref{NLIEas}), the minus (plus) sign in front of the $i\pi$ factor is considered when
$\Re \la_0$ is positive (negative).
The same NLIE is valid also for $\eta\in(\pi/2,\pi)$ if the contour $\mathcal{C} $ is replaced by
a similar rectangular contour with the upper (lower) edges parallel to the
real axis
through $\pm i(\pi-\eta-\epsilon)/2\, , \epsilon\rightarrow 0.$

\subsection{Integral expression for the next-largest eigenvalue in the $N/2-1$ sector}

The starting point of our derivation will be again the representation (\ref{Lh}),
which is also valid  for the eigenvalues  in the $N/2-1$ sector. We need an integral
representation for $
q^{(h)}(\la)=\sinh(\la-\la_0)\prod_{j=1}^{N-1}\sinh(\la-\la_j^{(h)})\,
$.
If we consider the contour $\mathcal{C}'$ introduced in \ref{Sleir}, then the following
identity holds
\be\label{integralrs}
\int_{\mathcal{C}+\mathcal{C}'}d(\la-\m)\frac{\fc'(\m)}{1+\fc(\m)}d\m=0\, .
\ee
For $\la$ close to the real axis, using (\ref{integralrs}), Theorem \ref{T}
and the  fact that inside the contour $\mathcal{C}_{ph}'$ the function
$1+\fc(\la)$ has: $N-1$ zeros at  $\{\la_j^{(h)}\}_{j=1}^{N-1}$ or $\{\la_j^{(h)}\}_{j=1}^{N-1}+i\pi\, ,$
$N/2-1$ simple poles at $\{\la_j\}_{j=1}^{N/2-1}+i\eta$ and a pole of order $N/2$ at
$-iu-i\eta+i\pi$, we find
\begin{align}\label{int18}
\frac{1}{2\pi i}\int_{\mathcal{C}}d(\la-\m)\frac{\fc'(\m)}{1+\fc(\m)}d\m=-
\left(\sum_{j=1}^{N-1}d(\la-\la_j^{(h)})-\sum_{j=1}^{N/2-1}d(\la-\la_j-i\eta)-\frac{N}{2}d(\la+iu+i\eta)\right)\, .
\end{align}
Inside the contour $\mathcal{C}$ the function $1+\fc(\la)$ has
$N/2$ zeros at the Bethe roots $\{\la_j\}_{j=1}^{N/2-1}$ and hole $\la_0$
and a pole of order $N/2$ at $iu$. Using again Theorem \ref{T} we have
\be\label{int19}
\frac{1}{2\pi i}\int_{\mathcal{C}}d(\la-\m-i\eta)\frac{\fc'(\m)}{1+\fc(\m)}d\m=
\sum_{j=1}^{N/2-1}d(\la-\la_j-i\eta)+d(\la-\la_0-i\eta)-\frac{N}{2}d(\la-iu-i\eta)\, .
\ee
Taking the difference of Eqs.~(\ref{int18}) and (\ref{int19}), integrating by parts,
and then integrating w.r.t.~$\la$ we obtain the following representation
\begin{align}\label{int20}
\log q^{(h)}(\la)&=\log\left(\frac{\sinh(\la-\la_0)}{\sinh(\la-\la_0-i\eta)}\right)
                   +\log\left(\phi_+(\la+i\eta)\phi_-(\la-i\eta)\right)\nonumber\\
&\ \ \ \ \ \ \ \ \ \ \ \ \ \ \ \ \ \ \ \ \ \ \ \ \ \ \ \ \
-\frac{1}{2\pi i}\int_{\mathcal{C}}\left[d(\la-\m)-d(\la-\m-i\eta)\right]\log(1+\fc(\m))d\m
+c\, ,
\end{align}
with $c$ a constant of integration. The integral expression for the next-largest eigenvalue
of the QTM in the sector $N/2-1$ is obtained by replacing (\ref{int20}) in (\ref{Lh}) with the result
\be\label{NLEs}
\log\Lambda_s(0)=\frac{\beta h}{2}+\log\left(\frac{\sinh\la_0}{\sinh(\la_0+i\eta)}\right)
+\frac{1}{2\pi }\int_{\mathcal{C}}p_0'(\m+i\eta/2)\log(1+\fc(\m))d\m\, .
\ee
Eq.~(\ref{NLEs}) is also
valid in the domain $\eta\in(\pi/2,\pi)$ if the contour $\mathcal{C}$ is replaced
by a rectangular contour, extending
to infinity, with the edges parallel to the real axis
through $\pm i(\pi-\eta-\epsilon)/2\, ,$ $\epsilon\rightarrow 0$.

\subsection{Final form of the integral equations}

We consider $\eta\in(0,\pi/2)$. Performing the same operations as in  Appendix \ref{Aphlt},
Eq.~(\ref{NLIEas}) is
transformed into
\begin{align}\label{int21}
\log\fc(\la-i\eta/2)&=-\beta e_0(\la) \mp i\pi
+i\theta(\la-\la_0)
-\frac{1}{2\pi}\int_{\mathbb{R}}K(\la-\m)\log(1+\fc(\m-i\eta/2))d\m\, ,
\end{align}
where  $\la_0\rightarrow\la_0+i\eta/2\, $ is in the upper half-plane.
The expression (\ref{NLEs}) for the next-largest eigenvalue becomes
\begin{align}\label{int22}
\log\Lambda_{s}(0)&=\frac{\beta h}{2}-i\pi+ip_0(\la_0)
 +\frac{1}{2\pi }\int_{\mathbb{R}}p_0'(\m)\log(1+\fc(\m-i\eta/2))d\m\, .
\end{align}
Introducing the  function $v(\la)$ satisfying  $e^{-v(\la)/T}=\fc(\la-i\eta/2)$
the NLIE for the auxiliary function and the integral expression for the next-largest eigenvalues
in the $N/2-1$ sector at low-temperatures can be written as
\be\label{int23}
v(\la)=e_0(\la)\pm i\pi T-iT\theta(\la-\la_0)+
\frac{T}{2\pi}\int_{\mathbb{R}}K(\la-\m)\log\left(1+e^{-v(\m)/T}\right)d\m
\, ,
\ee
\be\label{int24}
\log\Lambda_{s}(0)=\frac{h}{2T}-i\pi+ip_0(\la_0)+\frac{1}{2\pi}\int_{\mathbb{R}}p_0'(\m)\log\left(1+e^{-v(\m)/T}\right)d\m\, ,
\ee
where $\la_0$ satisfies the constraint $1+e^{-v(\la_0)/T}=0$. We should mention that
we can discard the $i\pi$ term
on the r.h.s.~of (\ref{int24}) (this has the effect of neglecting an $(-1)^m$ factor
in the asymptotic expansion which is irrelevant in the continuum limit)
On the r.h.s.~of Eq.~(\ref{int23})
we will consider the plus (minus) sign in front of the $i\pi T$ term when $\la_0$ is in the
first (second) quadrant of the complex plane. Again we are going to assume that similar
formulas are valid for $\eta\in(\pi/2,\pi)$. The generalization of  (\ref{int23}) and (\ref{int24})
to the case when $r$ ``particle-hole'' pairs are present is given by Eqs.~(\ref{NLEsgf}) and (\ref{NLIEsgf}).

We still need to derive equations for the case when inside the strip $|\Im \la|<\eta/2$ there
is no hole present. Consider $\la_0$ the hole closest to the line parallel to the real axis
with imaginary part $-i\eta/2$. If we modify  $\mathcal{C}$ adding an indentation
such that $\la_0$ is inside the contour and similarly modifying the upper edge of $\mathcal{C}'$
such that $\la_0+i\pi$ is outside of $\mathcal{C}'$ then all the considerations of the previous
sections still hold. However, when we take the  low-temperature limit of the equations in Eqs.~(\ref{int21})
and (\ref{int22}) the integration contour will  be transformed in the real axis with an indentation
such that $\la_0\rightarrow\la_0+i\eta/2$ (which now belongs to the lower half-plane) is above the contour.
The generalization of this result to the case when $r$ ``particle-hole'' pairs are present is
presented in Sect. \ref{Ss}.


\section{Proof of some identities}\label{Aidentities}

Here we prove some identities used in Sec.~\ref{CFT}. We start with
\be\label{ident1}
\int_{-q}^qp_0'(\la)R(\la,\pm q)\, d\la=p_0'(q)-2\pi\rho(q)\, .
\ee
Using a formal solution of Eq.~(\ref{defres}) on the l.h.s.~of (\ref{ident1}) we have
\begin{align*}
\int_{-q}^qp_0'(\la)R(\la,\pm q)\, d\la&=
\int_{-q}^q\int_{-q}^q\left(1+\frac{1}{2\pi}K\right)^{-1}(\la,\m)\, \frac{1}{2\pi}K(\m\mp q)p_0'(\la)\, d\mu\,  d\la\, ,\\
&=\int_{-q}^qK(q\mp\m)\rho(\m)d\m\, ,
\end{align*}
where in the second line we have used the symmetry of the kernel $K(\lam-\m)=K(\m-\la)$
and the integral equation for the density (\ref{density}). The identity (\ref{ident1})
follows from
\be
\rho(\pm q)+\frac{1}{2\pi}\int_{-q}^qK(q\mp\m)\rho(\m)\, d\m=\frac{1}{2\pi}p_0'(\pm q)\, ,
\ee
and the fact that $\rho(\la)$ and $p_0'(\la)$ are even functions.
Using a similar method we can prove that
\be\label{ident2}
\int_{-q}^q\varepsilon_0(\la)p_0'(\la)\, d\la=2\pi \int_{-q}^q e_0(\la)\rho(\la)\, d\la.
\ee
Making use of the equation for the dressed energy (\ref{dressede}) we can rewrite the l.h.s.
of (\ref{ident2}) as
\begin{align*}
\int_{-q}^q\varepsilon_0(\la)p_0'(\la)\, d\la&=\int_{-q}^q\int_{-q}^q\left(1+\frac{1}{2\pi}K\right)^{-1}
(\la,\m)\,  e_0(\m)\, p_0'(\la)\, d\m\, d\la\, ,\\
=2\pi \int_{-q}^{q}e_0(\m)\rho(\m)d\m\, ,
\end{align*}
where we have used again the symmetry of the kernel and Eq.~(\ref{density}).

\section{Derivation of the asymptotic expansion for the generating functional of density correlators}\label{agenfunc}

In this Appendix we will show how we can derive using our method the results obtained in \cite{KMS1,KMS2}.
The first step in the computation of  the asymptotic expansion for the generating functional
of density correlators in the Bose gas is the derivation of  NLIEs for the eigenvalues
of the twisted QTM $\tr^{QTM}_\varphi=A^{QTM}(0)+e^\varphi D^{QTM}(0)$ in the $N/2$
sector. The eigenvalues of the twisted QTM in the $N/2$ sector
are \cite{GKS1}:
\be
\Lambda_i^{(\varphi)}(\la)=\frac{\phi_-(\la)}{\phi_-(\la-i\eta)}\frac{q(\la-i\eta)}{q(\la)}e^{\frac{\beta h}{2}}+
\frac{\phi_+(\la)}{\phi_+(\la+i\eta)}\frac{q(\la+i\eta)}{q(\la)}e^{-\frac{\beta h}{2}+\varphi}\, ,
\ee
with $q(\la)=\prod_{j=1}^{N/2}\sinh(\la-\la_j)$ and $\{\la_j\}_{j=1}^{N/2}$ satisfying the BAEs
\be
\left(\frac{b(u',\la_j)}{b(\la_j,-u')}\right)^{N/2}=e^{-\beta h+\varphi}
\prod_{j\ne k}^{N/2}
\frac{\sinh(\la_j-\la_k+i\eta)}{\sinh(\la_j-\la_k-i\eta)}\, ,
\ \ \ \ j=1,\cdots,N/2\, .
\ee
The derivation of the NLIEs and integral expressions for the $N/2$ sector eigenvalues
of the twisted QTM (this includes also the largest eigenvalue) is almost identical
with the one presented in Section \ref{SLE} and Appendix \ref{Anleph} (we use a similar
distribution of Bethe roots and holes as in Fig.~\ref{LEfig} and Fig.~\ref{NLEphfig}).
We obtain similar equations as  Eqs.~(\ref{LElta}) and (\ref{LElte}) (for the largest
eigenvalue) and Eqs.~(\ref{NLEphgf}) and (\ref{NLIEphgf}) (for the next-largest
eigenvalues in the $N/2$ sector) with the only difference being the replacement of the
$h/(2T)$ with $h/(2T)+\varphi T$ in the r.h.s.~of Eqs.~(\ref{LElte}) and (\ref{NLIEphgf}).
The reader should note that $\varphi$ does not appear in the integral expressions for the
eigenvalues.

Performing the continuum limit in (\ref{int50}) we find
\begin{align}\label{gfexp}
\langle e^{\varphi\int_{0}^xj(x')\, dx'}\rangle_T
&=\sum_i \tilde C_i e^{-\frac{x}{\xi^{(\varphi)}[\ov u_i^\varphi]}}\, ,\ \ \ \ \ x\rightarrow \infty\, ,
\end{align}
with the correlation lengths defined by
\be\label{aint55}
\frac{1}{\xi^{(\varphi)}[\ov{u}_i^\varphi]}=-\frac{1}{2\pi}\int_{\mathbb{R}}\log
\left(\frac{1+e^{-\overline{u}_i^\varphi(k)/T}}{1+\, e^{-\overline{\varepsilon}(k)/T}}
\right)\, dk-i\sum_{j=1}^r k^+_j+i\sum_{j=1}^r k^-_j\, ,
\ee
and the auxiliary functions $\overline{u}_i^\varphi(k)$ satisfying the NLIEs:
\be\label{aunls}
\overline{u}_i^\varphi(k)=k^2-\mu-\varphi T+iT\sum_{j=1}^r\overline\theta(k-k^+_j)
-iT\sum_{j=1}^r\overline\theta(k-k^-_j)
-\frac{T}{2\pi}\int_{\mathbb{R}}\overline{K}(k-k')
\log\left(1+e^{-\overline{u}_i^\varphi(k')/T}\right)dk'
\, .
\ee
The $2r$ parameters, $\{k^+_j\}_{j=1}^r\,  (\{k^-_j\}_{j=1}^r) $ appearing in Eq.~(\ref{aunls})
belong to the upper (lower) half of the complex plane and satisfy the constraint
$1+e^{\overline u_i^\varphi(k^\pm_j)/T}=0.$ $r$ can take the values $0,1,2,\cdots$
with the $r=0$ term (this means that the sums  in Eq.~(\ref{aint55}), (\ref{aunls}) are zero)
being the dominant contribution in the expansion. Eq.~(\ref{gfexp}) was first derived in \cite{KMS1}.

\section{Low-temperature limit of the asymptotic expansions}\label{ACFTnls}

The low-temperature analysis of the asymptotic expansions (\ref{AEdd}) and (\ref{AEff})
is very similar with the one performed in Sec.~\ref{CFTdd} and \ref{CFTff}.
The only difference is the fact that in the Bose gas case the principal integral operator
is $I-\frac{1}{2\pi}\overline K$ and $p_0(k)=k$ which means that $p_0'(k)=1$. Therefore,
the calculations are almost identical with the ones for the XXZ spin chain
except
for some
sign changes. The integral equations for the zero temperature dressed energy $\overline{\varepsilon}_0(k)$
and the dressed charge $\overline Z(k)$ are given by
\be\label{denls}
\overline{\varepsilon}_0(k)-\frac{1}{2\pi}\int_{-\overline q}^{\overline q}
\overline K(k-k')\overline{\varepsilon}_0(k')\, dk'
=k^2-\mu\equiv \overline e_0(k)\, ,
\ee
and
\be\label{dcnls}
\overline Z(k)-\frac{1}{2\pi}\int_{-\overline q}^{\overline q }\overline K(k-k')\overline Z(k')\, dk'=1
\, ,\ \ \ \ \ \ \overline Z(\pm \overline q)=\overline {\mathcal{Z}}\, .
\ee
The resolvent of the integral operator $I-\frac{1}{2\pi}\overline K$ and the dressed phase equations
are obtained from the XXZ spin chain equivalents, (\ref{defres}) and (\ref{defdp}), by changing
the sign in front of the integral and replacing $K(\la,\mu)\, , \theta(\la)$ and $\pm q$
with $\overline K(k,k')\, , \overline \theta(k)$ and $\pm\overline q.$ The identities (\ref{dpident})and (\ref{ident1})
transform into (note the sign changes)
\be
\overline Z(k)=1+\overline F(k|-\overline q)-\overline F(k|\overline q)\, ,\ \ \ \
\frac{1}{\overline{\mathcal{Z}}}=1-\overline F(\overline q|\overline q)-\overline F(\overline q|-\overline q)\, ,
\ee
\be
\int_{-\overline q}^{\overline q}R(k,\pm\overline q)\, dk=2\pi\overline\rho(\overline q)-1\, ,
\ee
with (\ref{ident2}) still valid in the Bose case.
Additional simplifications occur due to the fact that $\overline Z(k)=2\pi \overline \rho(k)$. The Fermi velocity
can be rewritten as
\be\label{vfnls}
v_F=\frac{\overline\varepsilon_0'(\overline q)}{2\pi\overline \rho(\overline q)}=
\frac{\overline\varepsilon_0'(\overline q)}{\overline{\mathcal{Z}}}\, .
\ee
Using these relations and performing calculations similar with the ones from Sec.~\ref{CFTdd} and \ref{CFTff}
we obtain (\ref{inte1}) and (\ref{inte2}).

\end{appendices}

\clearpage
\newpage

\clearpage
\setcounter{equation}{0}%
\setcounter{page}{1}
\numberwithin{equation}{section}
\setcounter{section}{0}
\renewcommand{\thesection}{\arabic{section}}

\begin{center}
{\Large Supplementary Material for EPAPS \\
Correlation lengths of the repulsive one-dimensional Bose gas }
\end{center}

\section{Algebraic Bethe Ansatz solution of the XXZ spin chain}

The XXZ Hamiltonian (\ref{ham}) was solved by Yang and Yang \cite{YY1,YY2,YY3} with the
help of the coordinate Bethe ansatz.
From the point of view of ABA \cite{KBI}, which provides an alternative method of
solving (\ref{ham}), the XXZ spin-chain is the fundamental spin model associated with
the trigonometric $\R$-matrix defined by
\be\label{rmats}
\R(\la,\m)\equiv\left(\begin{array}{cccc} \R_{11}^{11}(\la,\m) & \R_{12}^{11}(\la,\m) & \R_{21}^{11}(\la,\m) & \R_{22}^{11}(\la,\m)\\
                                     \R_{11}^{12}(\la,\m) & \R_{12}^{12}(\la,\m) & \R_{21}^{12}(\la,\m) & \R_{22}^{12}(\la,\m)\\
                                     \R_{11}^{21}(\la,\m) & \R_{12}^{21}(\la,\m) & \R_{21}^{21}(\la,\m) & \R_{22}^{21}(\la,\m)\\
                                     \R_{11}^{22}(\la,\m) & \R_{12}^{22}(\la,\m) & \R_{21}^{22}(\la,\m) & \R_{22}^{22}(\la,\m)
                  \end{array}\right)
          =   \left(\begin{array}{cccc} 1 & 0         & 0         & 0\\
                                       0 & b(\la,\m) & c(\la,\m) & 0\\
                                       0 & c(\la,\m) & b(\la,\m) & 0\\
                                       0 & 0         & 0         & 1
                  \end{array}\right)\, ,
\ee
where
\be\label{defbcs}
b(\la,\m)=\frac{\sinh(\la-\m)}{\sinh(\la-\m+i\eta)}\, , \ \ \ \ \  c(\la,\m)=\frac{\sinh(i\eta)}{\sinh(\la-\m+i\eta)}\, .
\ee
%
Let us show how we can obtain and solve the XXZ Hamiltonian (\ref{ham}) in the ABA formalism.
The first step is the introduction of the $\Lo$-operators acting on
$\mathbb{C}^2\otimes \mathcal{H}=\mathbb{C}^2\otimes(\mathbb{C}^2)^{\otimes L}$
\be\label{defls}
\Lo_j(\la,0)=\sum_{a,b,a_1,b_1=1}^2\R_{b\, b_1}^{aa_1}(\la, 0)e_{ab}^{(0)}e_{a_1b_1}^{(j)}\, ,
\ \ \ \ \  \Lo_j(\la,0)\in \mbox{End}\left((\mathbb{C}^2)^{\otimes(L+1)}\right)\, ,
\ee
where $e_{ab}^{(i)}$ is the canonical basis in $\mbox{End}\left((\mathbb{C}^2)^{\otimes(L+1)}\right)$, {\it i.e.},
$e_{ab}^{(0)}=e_{ab}\otimes\mathbb{I}_2^{\otimes L}\, $ and $e_{ab}^{(i)}=\mathbb{I}_2\otimes
\mathbb{I}_2^{\otimes (i-1)}\otimes e_{ab}\otimes\mathbb{I}_2^{\otimes(L-i)}$ with $e_{ab}$
the $2$-by-$2$ matrices with all the elements zero except the one at the intersection of the
$a$-th row and $b$-th column which is equal to one. The additional $\mathbb{C}^2$ space on
which the $\Lo$-operators act in addition to the Hilbert space of the spin chain is called the
auxiliary space and, for practical purposes, it  is useful to present them as $2$-by-$2$ matrices
with entries which are operators acting on $\mathcal{H}.$  Using (\ref{defls}) and (\ref{rmats})
we obtain
\be\label{lreps}
\Lo_j(\la,0)=\left(\begin{array}{lr} e_{11}^{(j)}+b(\la,0)e_{22}^{(j)} & c(\la,0)e_{21}^{(j)}\\
                                     c(\la,0) e_{12}^{(j)} & b(\la,0)e_{11}^{(j)}+e_{22}^{(j)}
                   \end{array}\right)\, ,
\ee
where $e_{ab}^{(j)}$ is now the canonical basis in $\mbox{End}(\mathcal{H})$. The monodromy matrix
defined as
\be\label{defms}
\T(\la)=\Lo_L(\la,0)\Lo_{L-1}(\la,0)\cdots\Lo_1(\la,0)\, ,\ \ \ \ \ \
T(\la)=\left(\begin{array}{lr} A(\la) & B(\la) \\
                               C(\la) & D(\la)
              \end{array}\right)\, ,
\ee
provides a representation of the Yang-Baxter algebra
\be\label{YBAs}
\check\R(\la,\m)[\T(\la)\otimes\T(\m)]=[\T(\m)\otimes\T(\la)]\check\R(\la,\m)\, ,
\ee
with  $\check\R_{b_1\, b_2}^{a_1a_2}(\la,\m)=(\Pe\R)_{b_1\, b_2}^{a_1a_2}(\la,\m)=\R_{b_1b_2}^{a_2a_1}(\la,\m)$
and $\Pe$ the permutation matrix $\Pe a\otimes b=b\otimes a$ for $a,b\in\mathbb{C}^2$.
In Eq.~(\ref{YBAs}), which signalizes  the integrability of the model, $\T(\la)\otimes\T(\m)$
should be understood as the usual tensor product between two
square matrices of dimension $2\times 2$
with operator valued entries
as can be seen on the r.h.s.~of (\ref{defms}). Finally, the
transfer matrix is defined as the trace of the monodromy matrix in the auxiliary space
\be
\tr(\la)=\mbox{tr}_{0}\T(\la)=A(\la)+D(\la)\, .
\ee
In Sect. \ref{Ahams} it is shown that the XXZ Hamiltonian (\ref{ham}) can be obtained
as
\be\label{BL1s}
H(J,\Delta,h)=2J\sinh(i\eta)\tr^{-1}(0)\tr'(0)-hS_z\, .
\ee
The eigenvectors and eigenvalues of the transfer matrix $\tr(\la)=A(\la)+D(\la)$ can be
obtained with the help of ABA if we can find a pseudovacuum which is  an eigenvector of
$A(\la)$ and $D(\la)$ and the action of $\T(\la)$ on it is triangular. Using the explicit expression
of the $\Lo$-operators in the auxiliary space (\ref{lreps}) we can see that
\[
|\Omega\rangle=\underbrace{\left(\begin{array}{c} 1\\0 \end{array}\right)\otimes\cdots
\otimes\left(\begin{array}{c} 1\\0 \end{array}\right)}_{\mbox{ L times}}\, ,
\]
satisfies the requirements of a pseudovacuum and
\[
\T(\la)|\Omega\rangle=\left(\begin{array}{cc} A(\la)|\Omega\rangle & B(\la)|\Omega\rangle \\
                                             C(\la)|\Omega\rangle & D(\la)|\Omega\rangle
                       \end{array}\right)
                    =\left(\begin{array}{cr} |\Omega\rangle & B(\la)|\Omega\rangle \\
                                              0 & (b(\la,0))^L|\Omega\rangle
                    \end{array}\right)\, .
\]
The operator $B(\la)$ can be interpreted as a creation operator and we will look for
the eigenvectors of $\tr(\la)$ of the form $|\{\la_j\}_{j=1}^n\rangle=B(\la_1)
\cdots B(\la_n)|\Omega\rangle.$ In Sect. \ref{ABA6s} we show how the eigenvalues
of $\tr(\la)$ can be derived with the result
\be\label{eigenvalues}
\tau(\la|\{\la_j\}_{j=1}^n)=\prod_{j=1}^n\frac{\sinh(\la-\la_j-i\eta)}{\sinh(\la-\la_j)}
+\left(\frac{\sinh(\la-i\eta/2)}{\sinh(\la+i\eta/2)}\right)^L\prod_{j=1}^n\frac{\sinh(\la-\la_j+i\eta)}{\sinh(\la-\la_j)}\, ,
\ee
provided that the $\{\la_j\}_{j=1}^n$ parameters satisfy the Bethe equations
\be\label{BEs}
\left(\frac{\sinh(\la_j-i\eta/2)}{\sinh(\la_j+i\eta/2)}\right)^L=
\prod_{s\ne j}^n\frac{\sinh(\la_j-\la_s-i\eta)}{\sinh(\la_j-\la_s+i\eta)}\, ,
\ \ \ \ \ \ \ j=1,\cdots,n\, .
\ee
In order to make contact with the results in the literature in Eqs.~(\ref{eigenvalues})
and (\ref{BEs}) we have performed the transformation $\la\rightarrow\la-i\eta/2$ and
$\la_j\rightarrow\la_j-i\eta/2\, ,\  j=1,\cdots,n$. This also means that (\ref{BL1s})
should now be understood as $H(J,\Delta,h)=2J\sinh(i\eta)\tr^{-1}(i\eta/2)\tr'(i\eta/2)-hS_z.$
The only thing that remains in order to obtain the spectrum of the Hamiltonian
(\ref{ham}) is to quantify the action of the magnetic operator. $S_z$ commutes with the
transfer matrix therefore they share the same eigenvectors. Using $S_z=1/2\sum_{j=1}^L\sigma_z^{(j)}
=1/2\sum_{j=1}^L(e_{11}^{(j)}-e_{22}^{(j)})$ with $e_{ab}^{(j)}$ the canonical
basis in $(\mathbb{C}^2)^{\otimes L}$ and the explicit expression for the $B(\la)$
operator from (\ref{lreps}) we find $[S_z,B(\la)]=-B(\la)$. The action of the
magnetic operator on the eigenvectors of the transfer matrix is then given
by $S_z|\{\la_j\}_{j=1}^n\rangle=S_zB(\la_1)\cdots B(\la_n)|\Omega\rangle=
(L/2-n)B(\la_1)\cdots B(\la_n)|\Omega\rangle$ where we have used $S_z|\Omega\rangle=L/2|\Omega\rangle$.
Finally, the energy spectrum of the XXZ spin chain in magnetic
field is
\be\label{magnons}
E(\{\la\})=\sum_{j=1}^ne_0(\la_j)-h\frac{L}{2}\, ,\ \ \ \
e_0(\la)=\frac{2J\sinh^2(i\eta)}{\sinh(\la+i\eta/2)\sinh(\la-i\eta/2)}+h\, .
\ee

\section{Derivation of the XXZ Hamiltonian from the transfer matrix}\label{Ahams}

In this Section we are going to show that the logarithmic derivative of the
transfer matrix is identical with $H^{(0)}(J,\Delta)$ (modulo a constant) proving
Eq.~(\ref{BL1s}). We introduce two types of operators, which we will
call $\R$- and $\Pe$-operators, acting on $(\mathbb{C}^2)^{\otimes(L+1)}$ and defined
by
\be\label{defrps}
\R_{j,k}(\la,\m)=\sum_{a_1,b_1,a_2,b_2=1}^2\R_{b_1\, b_2}^{a_1a_2}(\la,\m)e_{a_1b_1}^{(j)}e_{a_2b_2}^{(k)}\, ,\ \ \ \
\Pe_{j,k}=\sum_{a_1,b_1,a_2,b_2=1}^2\Pe_{b_1\, b_2}^{a_1a_2} e_{a_1b_1}^{(j)}e_{a_2b_2}^{(k)}\, ,
\ee
where $\Pe_{b_1\, b_2}^{a_1a_2}=\delta_{a_1b_2}\delta_{a_2b_1}$ and $e_{ab}^{(j)}$ is the canonical
basis in $\mbox{End}\left((\mathbb{C}^2)^{\otimes(L+1)}\right)$. The $\Lo$-operators
are a subset of the  $\R$-operators $\Lo_j(\la,0)=\R_{0,j}(\la,0)$ and $\Pe_{j,k}$ acts
on an arbitrary vector in $(\mathbb{C}^2)^{\otimes(L+1)}$ by permuting the $j$-th and
$k$-th component. Some useful properties of the (permutation) $\Pe$-operators are
\be\label{pps}
\Pe_{j,l}\Pe_{k,j}=\Pe_{k,l}\Pe_{j,l}=\Pe_{k,j}\Pe_{k,l}\, ,\ \ \Pe_{k,l}=\Pe_{l,k}\, ,\ \ \Pe_{k,l}^2=\mathbb{I}\, ,
 \ \ \Pe_{j,k}e_{ab}^{(k)}=e_{ab}^{(j)}\Pe_{j,k}\, .
\ee
First, we are going to calculate $\tr(0)=\mbox{tr}_0\T(0)$. Using the fact that $\R(0,0)=\Pe\, ,$
$\Lo_j(0,0)=\Pe_{0,j}$ and applying successively the first identity of (\ref{pps}) we find
\begin{align*}
\T(0)=&\Pe_{0,L}\Pe_{0,L-1}\cdots\Pe_{0,2}\Pe_{0,1}\, ,\\
        =&\Pe_{L,L-1}\Pe_{L,L-2}\cdots\Pe_{L,2}\Pe_{L,1}\Pe_{0,L}\, ,\\
              &\ \ \ \ \ \ \ \ \ \vdots \\
        =&\Pe_{1,2}\Pe_{2,3}\cdots\Pe_{L-1,L}\Pe_{0,L}\, .
\end{align*}
Taking into account that $\mbox{tr}_0\, \Pe_{0,L}=\mathbb{I}$ we obtain
$\tr(0)=\Pe_{1,2}\Pe_{2,3}\cdots\Pe_{L-1,L}$ with $\tr^{-1}(0)=\Pe_{L-1,L}\cdots\Pe_{2,3}
\Pe_{1,2}$. For the computation of the derivative of the transfer matrix $\tr'(0)=
\mbox{tr}_0 \T'(0),$ we will also need the last identity of (\ref{pps}) and the fact that
$\Pe_{j,k}\6_{\la}\R_{l,m}(\la,0)=\6_{\la}\R_{l,m}(\la,0)\Pe_{j,k}$ if
$j,k\ne l,m$. We find
{\allowdisplaybreaks
\begin{align*}
\T'(0)=        &\sum_{i=1}^L\Pe_{0,L}\cdots\Pe_{0,i+1}\left.\6_\la\R_{0,i}(\la,0)\right|_{\la=0}\Pe_{0,i-1}\cdots \Pe_{0,1}\, ,\\
        =&\sum_{i=1}^L\Pe_{0,L}\cdots\Pe_{0,i+1}\Pe_{0, i-1}\left.\6_\la\R_{i-1, i}(\la,0)\right|_{\la=0}\cdots \Pe_{0, 1}\, ,\\
        =&\sum_{i=1}^L\Pe_{0,L}\cdots\Pe_{0,i+1}\Pe_{0, i-1}\cdots \Pe_{0, 1}\left.\6_\la\R_{i-1, i}(\la,0)\right|_{\la=0}\, ,\\
        =&\sum_{i=1}^L\Pe_{0,1}\Pe_{1,L}\cdots\Pe_{1, i+1}\Pe_{1, i-1}\cdots \Pe_{1, 2}\left.\6_\la\R_{i-1, i}(\la,0)\right|_{\la=0}\, ,\\
        =&\sum_{i=1}^L\Pe_{0,1}\Pe_{1,2}\cdots\Pe_{i-1, i+1}\Pe_{i+1, i+2}\cdots \Pe_{L-1,L}\left.\6_\la\R_{i-1, i}(\la,0)\right|_{\la=0}\, ,
\end{align*}
}
therefore,
$
\tr'(0)=\text{tr}_{0}\T'_{0}(0)=\sum_{i=1}^L\Pe_{1,2}\cdots\Pe_{i-1, i+1}\Pe_{i+1, i+2}\cdots
\Pe_{L-1,L}\left.\6_\la\R_{i-1, i}(\la,0)\right|_{\la=0}\, .
$
Collecting everything we obtain
\[
\tr(0)^{-1}\tr'(0)=\sum_{i=1}^L\Pe_{i-1,i}\left.\6_\la\R_{i-1, i}(\la,0)\right|_{\la=0}
=\sum_{i=1}^L\left.\6_\la\check\R_{i-1, i}(\la,0)\right|_{\la=0}\, ,
\]
where periodic boundary conditions $\R_{0,1}=\R_{L,1}$ are understood. Using the explicit
expression for the $\R$-matrix it is easy to see that
$2J\sinh(i\eta)\left.\6_\la\check\R_{i-1, i}(\la,0)\right|_{\la=0}$ is equal to
$J[\sigma_x^{(i-1)}\sigma_x^{(i)}+\sigma_y^{(i-1)}\sigma_y^{(i)}
+\Delta(\sigma_z^{(i-1)}\sigma_z^{(i)}-1)]$ proving Eq.~(\ref{BL1s}).

\section{Algebraic Bethe ansatz for the generalized XXZ spin chain}\label{ABA6s}

Let us  consider a general monodromy matrix which is intertwined by the
$\R$-matrix (\ref{rmats})
\be\label{YBEAs}
\check\R(\la,\m)[\T(\la)\otimes\T(\m)]=[\T(\m)\otimes\T(\la)]\check\R(\la,\m)\, ,
\ee
and assume the existence of a pseudovacuum such that
\[
T(\la)|\Omega\rangle=\left(\begin{array}{cc} A(\la)|\Omega\rangle & B(\la)|\Omega\rangle \\
                                             C(\la)|\Omega\rangle & D(\la)|\Omega\rangle
                       \end{array}\right)
                    =\left(\begin{array}{cr} a(\la)|\Omega\rangle & B(\la)|\Omega\rangle \\
                                              0 & d(\la)|\Omega\rangle
                    \end{array}\right)\, .
\]
The monodromy matrices of the XXZ spin chain and quantum transfer  matrix  are  particular
cases of this model with $a(\la)=1\, , \ d(\la)=(b(\la,0))^L$ and $a(\la)=b(u',\la)^{N/2}e^{\beta h/2}\, ,\
d(\la)=b(\la,-u')^{N/2}e^{-\beta h/2}$ respectively. We are interested in finding the eigenvalues of the
transfer matrix $\tr(\la)=A(\la)+D(\la)$. Interpreting the $B(\la)$ operator as a creation
operator we are going to look for eigenvectors of the type $|\{\la_j\}_{j=1}^p\rangle=B(\la_1)
\cdots B(\la_p)|\Omega\rangle$ satisfying the eigenvalue equation
\[
\tr(\la)|\{\la_j\}_{j=1}^p\rangle=\tau(\la|\{\la_j\}_{j=1}^p)|\{\la_j\}_{j=1}^p\rangle\, .
\]
This eigenvalue equation will impose a set of equations on the $\{\la_j\}_{j=1}^p$ parameters
which we will call Bethe equations. Before we derive these equations, we need
the commutation relations between the operator valued entries of the monodromy matrix.
Using the  explicit matrix representation of Eq.~(\ref{YBEAs})
\begin{align*}
&\left(\begin{array}{cccc}
         1&0&0&0\\
         0&c(\la,\m)&b(\la,\m)&0\\
         0&b(\la,\m)&c(\la,\m)&0\\
         0&0&0&1
       \end{array}\right)
\left(\begin{array}{cccc}
        A(\la) A(\m) & A(\la) B(\m) &B(\la) A(\m) & B(\la) B(\m)\\
        A(\la) C(\m) & A(\la) D(\m) &B(\la) C(\m) & B(\la) D(\m)\\
        C(\la) A(\m) & C(\la) B(\m) &D(\la) A(\m) & D(\la) B(\m)\\
        C(\la) C(\m) & C(\la) D(\m) &D(\la) C(\m) & D(\la) D(\m)\\
      \end{array}\right)\\
&= \left(\begin{array}{cccc}
        A(\m) A(\la) & A(\m) B(\la) &B(\m) A(\la) & B(\m) B(\la)\\
        A(\m) C(\la) & A(\m) D(\la) &B(\m) C(\la) & B(\m) D(\la)\\
        C(\m) A(\la) & C(\m) B(\la) &D(\m) A(\la) & D(\m) B(\la)\\
        C(\m) C(\la) & C(\m) D(\la) &D(\m) C(\la) & D(\m) D(\la)\\
      \end{array}\right)
\left(\begin{array}{cccc}
         1&0&0&0\\
         0&c(\la,\m)&b(\la,\m)&0\\
         0&b(\la,\m)&c(\la,\m)&0\\
         0&0&0&1
       \end{array}\right)
\end{align*}
the following commutation relations can be obtained from the matrix elements $(1,4), (1,3)$
and $(2,4)$
\begin{align}
B(\la)B(\m)&=B(\m)B(\la)\, ,\label{cr1}\\
A(\la)B(\m)&=f(\m,\la)B(\m)A(\la)-h(\m,\la)B(\la)A(\m)\, ,\label{cr2s}\\
D(\la)B(\m)&=f(\la,\m)B(\m)D(\la)-h(\la,\m)B(\la)D(\m)\, ,  \label{cr3s}
\end{align}
where we have introduced
\be
f(\la,\m)=\frac{1}{b(\la,\m)}\, ,\ \ \ \ \ h(\la,\m)=\frac{c(\la,\m)}{b(\la,\m)}\, .
\ee
Now we can evaluate the action of the transfer matrix $\tr(\la)=A(\la)+D(\la)$ on the
prospective eigenvector $B(\la_1)\cdots B(\la_p)|\Omega\rangle.$ Acting with $A(\la)$ on
$B(\la_1)\cdots B(\la_p)|\Omega\rangle$ and using the commutation relation (\ref{cr2s})
we obtain $2^p$ terms of which only one has the desired form
\[
\prod_{j=1}^pf(\la_j,\la)a(\la)B(\la_1)\cdots B(\la_p)|\Omega\rangle\, .
\]
This term is obtained by using $p$ times  only the first term on the r.h.s.~of (\ref{cr2s}). The other $2^p-1$
terms which should cancel in order to satisfy the eigenvalue equation can be written
as
\[
\sum_{j=1}^p M_j(\la,\{\la\})B(\la_1)\cdots\hat B(\la_j)\cdots B(\la_p)B(\la)|\Omega\rangle\, ,
\]
where the hat means that the corresponding operator is missing. At first impression obtaining
the coefficients $M_j(\la,\{\la\})$ may seem a daunting task, but,
fortunately, the commutativity
of the $B(\la)$ operators simplifies the matter considerably. The $M_1(\la,\{\la\})$ coefficient
is easy to compute: it requires to use first the second term of the r.h.s.~of (\ref{cr2s}) and
then the first term $p-1$ times with the result
\[
-a(\la_1)h(\la_1,\la)\prod_{k=2}^p f(\la_k,\la_1)\, .
\]
Now using the commutativity of the $B(\la)$ operators we argue that the $M_j(\la,\{\la\})$
coefficient can be obtained from $M_1(\la,\{\la\})$ substituting $\la_1$ with $\la_j$ obtaining
\be\label{defMs}
M_j(\la,\{\la\})=-a(\la_j)h(\la_j,\la)\prod_{k\ne j}^p f(\la_k,\la_j)\, .
\ee
Therefore, the action of $A(\la)$ on $B(\la)\cdots B(\la_p)|\Omega\rangle$ can be written as
\be\label{a}
A(\la)B(\la_1)\cdots B(\la_p)|\Omega\rangle=a(\la)\prod_{j=1}^pf(\la_j,\la)B(\la_1)\cdots B(\la_p)|\Omega\rangle+
\sum_{j=1}^p M_j(\la,\{\la\})B(\la_1)\cdots\hat B(\la_j)\cdots B(\la_p)B(\la)|\Omega\rangle\, ,
\ee
with $M_j(\la,\{\la\})$ given by (\ref{defMs}). In an analogous fashion the action of
$D(\la)$ on $B(\la_1)\cdots B(\la_p)|\Omega\rangle$ is computed as
\be\label{d}
D(\la)B(\la_1)\cdots B(\la_p)|\Omega\rangle=d(\la)\prod_{j=1}^pf(\la,\la_j)B(\la_1)\cdots B(\la_p)|\Omega\rangle+
\sum_{j=1}^p N_j(\la,\{\la\})B(\la_1)\cdots\hat B(\la_j)\cdots B(\la_p)B(\la)|\Omega\rangle\, ,
\ee
with
\be\label{defNs}
N_j(\la,\{\la\})=-h(\la,\la_j)\prod_{k\ne j}^p f(\la_j,\la_k)d(\la_j)\, .
\ee
Then Eqs.~(\ref{a}) and (\ref{d}) show that $B(\la_1)\cdots B(\la_m)|\Omega\rangle$ is an
eigenvector of $\tr (\la)=A(\la)+D(\la)$ with eigenvalue
\[
\tau(\la|\{\la\}_{j=1}^p)=a(\la)\prod_{j=1}^pf(\la_j,\la)+d(\la)\prod_{j=1}^pf(\la,\la_j)\, ,
\]
if the parameters $\{\la_j\}_{j=1}^p$ satisfy the equations
$M_j(\la,\{\la\})+N_j(\la,\{\la\})=0\, ,\  j=1,\cdots,p$ or,
equivalently
\[
\frac{a(\la_j)}{d(\la_j)}=\prod_{k\ne j}^p\frac{f(\la_j,\la_k)}{f(\la_k,\la_j)}\, , \ \ \ \ \  j=1,\cdots,p\, .
\]
The following remark is in order:
Strictly speaking our treatment is not
completely rigorous since we have implicitly assumed the linear independence of vectors of the form
$B(\la_1)\cdots B(\la_p)|\Omega\rangle$. The interested reader can find the rigorous solution
of this problem in Chap. XII of \cite{EFGKK}.


\section{The XXZ Spin Chain Quantum Transfer Matrix}\label{aQTMs}

In this Section we  review the principal features of the QTM method. The
considerations below follow closely the presentation in \cite{GKS1}.

We introduce $N$ auxiliary spaces denoted by $\ov 1,\cdots, \ov N$ and two types of monodromy matrices
using $\R$-operators (see Sect. \ref{Ahams}):
\begin{subequations}\label{amms}
\begin{align}
\text{Type 1}& \ \ \ \   \T_{\ov j}(\la)=\R_{\ov j, L}(\la,\m)\cdots\R_{\ov j, 1}(\la,\m)\, ,\\
\text{Type 2}& \ \ \ \ \ov\T_{\ov j}(\la)=\R_{1,\ov j}(\m,\la)\cdots \R_{L,\ov j}(\m,\la)\, .
\end{align}
\end{subequations}
Using the results of Sect. \ref{Ahams} we have
\[
\tr(0)=\text{tr}_{\ov j}\T_{\ov j}(0)=\Pe_{1,2}\Pe_{2,3}\cdots\Pe_{L-1,L}\, ,
\]
and
\[
\tr'(0)=\text{tr}_{\ov j}\T'_{\ov j}(0)=\sum_{i=1}^L\Pe_{1,2}\cdots\Pe_{i-1, i+1}\Pe_{i+1, i+2}\cdots
\Pe_{L-1,L}\left.\6_\la\R_{i-1, i}(\la,0)\right|_{\la=0}\, ,
\]
which implies that
\begin{align*}
\tr(0)^{-1}\tr'(0)&=\sum_{i=1}^L\Pe_{i-1,i}\left.\6_\la\R_{i-1, i}(\la,0)\right|_{\la=0}=\sum_{i=1}^L\left.\6_\la\check\R_{i-1, i}(\la,0)\right|_{\la=0}\, ,\\
 &=\frac{H^{(0)}(J,\Delta)}{2J\sinh(i\eta)}\, ,
\end{align*}
under periodic boundary conditions $\R_{0,1}=\R_{L,1}$. The last relation shows that
\be\label{e1s}
\tr(\la)=\tr(0)\exp\left(\la\frac{H^{(0)}(J,\Delta)}{2J\sinh(i\eta)}+\mathcal{O}(\la^2)\right)\, .
\ee
In a similar fashion we can show that
\[
\ov\tr(0)=\text{tr}_{\ov j}\ov\T_{\ov j}(0)=\Pe_{L,L-1}\cdots\Pe_{3,2}\Pe_{2,1}=\tr^{-1}(0)\, ,
\]
and
\[
\ov\tr'(0)=\text{tr}_{\ov j}\ov\T'_{\ov j}(0)=\sum_{i=1}^L\left.\6_\la\R_{i,i-1}(0,\la)\right|_{\la=0}\Pe_{L,L-1}\cdots\Pe_{i+2, i+1}\Pe_{i+1, i-1}\cdots
\Pe_{2,1}\, ,
\]
implying that
\begin{align*}
\ov\tr'(0)\ov\tr^{-1}(0)&=\sum_{i=1}^L\left.\6_\la\R_{i, i-1}(0,\la)\right|_{\la=0}\Pe_{i,i-1}=-
\sum_{i=1}^L\left.\6_\la\check\R_{i,i-1}(\la,0)\right|_{\la=0}\, ,\\
&=-\frac{H^{(0)}(J,\Delta)}{2J\sinh(i\eta)}\, ,
\end{align*}
where we have used $\R_{k,j}\Pe_{j,k}=\Pe_{j,k}\R_{j,k}=\check\R_{jk}$ and the fact that
$\left.\6_\la\R_{i,i-1}(\la,0)\right|_{\la=0}=-\left.\6_\la\R_{i,i-1}(0,\la)\right|_{\la=0}.$
Again, from the last relation we see that the following expansion is valid
\be\label{ee2s}
\ov\tr(\la)=\exp\left(-\la\frac{H^{(0)}(J,\Delta)}{2J\sinh(i\eta)}+\mathcal{O}(\la^2)\right)\ov\tr(0)\, .
\ee
Employing a special form of the Trotter-Suzuki formula with $u'=-2J\sinh(i\eta)\frac{\beta}{N}$
\be
\lim_{N\rightarrow \infty}\left(\tr^{-1}(0)\tr(u')\right)^N=\lim_{N\rightarrow \infty}\left(1+\frac{1}{N}(-\beta H^{(0)}(J,\Delta)+\mathcal{O}(1/N))\right)^N
=e^{-\beta H^{(0)}(J,\Delta)}\, ,
\ee
we find
\be
\lim_{N\rightarrow \infty}\rho_{N,L}:=\left(\ov\tr(-u')\tr(u')\right)^{N/2}=\left(1+\frac{2}{N}\left(-\beta H^{(0)}(J,\Delta)+\mathcal{O}(1/N)\right)\right)^{N/2}=e^{-\beta H^{(0)}(J,\Delta)}\, .
\ee
This formula is the starting point for the introduction of the QTM. Using Eqs.~(\ref{amms}) we have
\begin{align}\label{defrhos}
\rho_{N,L}=&\text{tr}_{\, \ov 1,\cdots,\ov{N-1}}\left\{ \ov\T_{\ov{N}}(-u')\T_{\ov{N-1}}(u')\cdots\ov\T_{\ov{2}}(-u')\T_{\ov{1}}(u') \right\}\, ,\nonumber\\
          =&\text{tr}_{\, \ov 1,\cdots,\ov{N-1}}\left\{ \ov\T_{\ov{N}}(-u')\T_{\ov{N-1}}^t(u')\cdots\ov\T_{\ov{2}}(-u')\T_{\ov{1}}^t(u') \right\}\, ,\nonumber\\
          =&\text{tr}_{\, \ov 1,\cdots,\ov{N-1}}\left\{\R_{1,\ov{N}}(\m,-\uu) \R_{2,\ov{N}}(\m,-\uu)
                              \cdots \R_{L-1,\ov{N}}(\m,-\uu)\R_{L,\ov{N}}(\m,-\uu)\right.\nonumber\\
           &\ \ \ \ \ \ \ \ \ \ \ \ \ \ \ \  \R_{\ov{N-1}\,, 1}^{t_1}(\uu,\m)\R_{\ov{N-1}\, ,2}^{t_1}(\uu,\m)\cdots
             \check\R_{\ov{N-1}\, ,L-1}^{t_1}(\uu,\m)\R_{\ov{N-1}\, ,L}^{t_1}(\uu,\m)\nonumber\\
           &\ \ \ \ \ \ \ \ \ \ \ \ \ \ \ \ \ \ \ \ \ \ \ \ \ \ \ \ \ \ \ \ \  \vdots   \nonumber \\
           &\ \ \ \ \ \ \ \ \ \ \ \ \ \ \ \  \R_{1,\ov{2}}(\m,-\uu) \R_{2,\ov{2}}(\m,-\uu)
                              \cdots \R_{L-1,\ov{2}}(\m,-\uu)\R_{L,\ov{2}}(\m,-\uu)\nonumber\\
           &\ \ \ \ \ \ \ \ \ \ \ \ \ \ \, \, \left. \left.\R_{\ov{1}\, ,1}^{t_1}(\uu,\m)\R_{\ov{1}\,, 2}^{t_1}(\uu,\m)\cdots
             \check\R_{\ov{1}\, ,L-1}^{t_1}(\uu,\m)\R_{\ov{1}\, ,L}^{t_1}(\uu,\m)\right\}\right|_{\m=0}\nonumber\\
          =&\text{tr}_{\, \ov 1,\cdots,\ov{N-1}}\left\{ \R_{1,\ov{N}}(\m,-\uu)\R_{\ov{N-1}\,, 1}^{t_1}(\uu,\m)\cdots
                                            \R_{1,\ov{2}}(\m,-\uu)\R_{\ov{1}\, ,1}^{t_1}(\uu,\m) \right.\, \nonumber\\
           &\ \ \ \ \ \ \ \ \ \ \ \ \ \ \ \ \ \ \ \ \ \ \ \ \ \ \ \ \ \ \ \ \ \vdots \nonumber\\
           & \ \ \ \ \ \ \ \ \ \ \ \ \ \ \     \left.\left.\R_{L,\ov{N}}(\m,-\uu)\R_{\ov{N-1}\,, L}^{t_1}(\uu,\m)\cdots
                                            \R_{L,\ov{2}}(\m,-\uu)\R_{\ov{1}\,, L}^{t_1}(\uu,\m) \right\}\right|_{\m=0}\, \nonumber\\
          =&\text{tr}_{\, \ov 1,\cdots,\ov{N-1}}\left\{\T_1^{QTM}(0)\cdots\T_L^{QTM}(0)\right\}\, ,
\end{align}
where $({\R}^{t_1})_{bd}^{ac}=\R_{ad}^{bc}$ and we introduced the monodromy matrix of the QTM
\be
\T_j^{QTM}(\la)=\R_{j,\ov{N}}(\la,-\uu)\R_{\ov{N-1}\,, j}^{t_1}(\uu,\la)\cdots
                                            \R_{j,\ov{2}}(\la,-\uu)\R_{\ov{1}\,, j}^{t_1}(\uu,\la)\, .
\ee
The QTM is defined as $\tr^{QTM}(\la)=\text{tr}_j\T_j^{QTM}(\la)$. The usefulness  of Eq.~(\ref{defrhos})
will become obvious if we notice that
\begin{align}\label{azs}
Z_{XXZ}(\beta)&=\text{tr}_{1,\cdots ,L}\,  e^{-\beta H^{(0)}(J,\Delta)}=\lim_{N\rightarrow \infty}\text{tr}_{1,\cdots ,L}\,  \rho_{N,L}\, \nonumber\\
&=\lim_{N\rightarrow\infty}\text{tr}_{1,\cdots ,L}\, \text{tr}_{\ov 1,\cdots ,\ov N}\left(\T_1^{QTM}(0)\cdots\T_L^{QTM}(0)\right)\, \nonumber\\
&=\lim_{N\rightarrow\infty}\text{tr}_{\ov 1,\cdots ,\ov N}\left(\tr^{QTM}(0)\right)^L=\sum_{n=1}^\infty \Lambda_{n}^L(0)\, ,
\end{align}
where the last sum is over all the eigenvalues of the QTM. In the case of the XXZ spin chain QTM
the following assumptions are true:
\begin{itemize}
\item The limits $L\rightarrow \infty$ and $N\rightarrow \infty$ are interchangeable \cite{MS, SI}.
\item The largest eigenvalue of the QTM which is denoted by $\Lambda_0(\la)$ is real,
positive, nondegenerate and separated by a gap from the next-largest eigenvalues in the
Trotter limit $N\rightarrow \infty$.
\end{itemize}
Therefore, using (\ref{azs}) the free energy per lattice site of the XXZ spin chain is
\be
f(\beta)=-\lim_{N,L\rightarrow\infty}\frac{\log Z_{XXZ}(\beta)}{\beta L}=-\frac{\log\Lambda_0(0)}{\beta}\, .
\ee
This result shows that the thermodynamic behavior of the lattice model is
completely determined by the largest eigenvalue of the QTM evaluated at 0
providing an elegant solution to the almost impossible problem of summing
over all the eigenvalues of the Hamiltonian.

Until now we have considered  the case without magnetic field. The presence
of the magnetic field operator $S_z=\frac{1}{2}\sum_{j=1}^L\sigma_z^{(j)}$
in the Hamiltonian (\ref{ham}) can be easily taken into account in the
QTM formalism. The first observation is that
\be
[\R(\la,\m),\Theta\otimes\Theta]=0,\ \ \ \ \Theta=\left(\begin{array}{lr}e^{\beta h/2}& 0\\
                                                                    0& e^{-\beta h/2}
                                                  \end{array}\right)\, ,
\ee
with the obvious consequence that $\Theta\otimes\Theta$ is a  spectral parameter free solution
of the Yang-Baxter algebra with $\R$-matrix (\ref{rmats}),
{\it i.e.,} $\check\R(\la,\m)(\Theta\otimes\Theta)=
(\Theta\otimes\Theta)\check\R(\la,\m)$.
 Therefore
\begin{align}
\lim_{N\rightarrow\infty}\rho_{N,L} e^{\beta hS_z}&=e^{-\beta(H^{(0)}(J,\Delta)-hS_z)}\, ,\nonumber\\
&=\text{tr}_{\ov 1,\cdots, \ov N}\left(\T^{QTM}_1(0)\Theta_1\cdots \T_L^{QTM}(0)\Theta_L\right)\, ,
\end{align}
which shows that the presence of the magnetic field term in the Hamiltonian
is taken into account by the following transformation
\be
\T^{QTM}(\la)\rightarrow\T^{QTM}(\la)
\left(\begin{array}{cc} e^{\frac{\beta h}{2}} & 0\\
 0 & e^{-\frac{\beta h}{2}}
 \end{array}\right)\, .
\ee

\subsection{Correlation functions within the QTM approach}

The QTM method provides considerable simplifications in the treatment of
temperature dependent correlation functions. We are interested in correlation
functions of local operators of the following type
\be
\langle O_1^{(j)}\cdots O_{k-j+1}^{(k)}\rangle_T=\lim_{L\rightarrow\infty}\frac{\text{tr}_{1\cdots L}\, e^{-\beta H(J,\Delta,h)}O_1^{(j)}\cdots O_{k-j+1}^{(k)}}
{\text{tr}_{1\cdots L}\, e^{-\beta H(J,\Delta,h)}}\, ,
\ee
where $j,\cdots,k\in\{1,\cdots,L\}$ with $j<k$ are the local spaces on which
the operators $O^{(j)}=\mathbb{I}_2^{\otimes (j-1)}\otimes O\otimes \mathbb{I}_2^{\otimes (L-j)}$ act.
Using (\ref{defrhos}) we have
\begin{align}\label{c1s}
\langle O_1^{(j)}\cdots O_{k-j+1}^{(k)}\rangle_T
&=
\lim_{L,N\rightarrow\infty}\frac{\text{tr}_{\ov 1\cdots \ov N}\text{tr}_{1\cdots L}
\left(\T_1^{QTM}(0)\cdots T_L^{QTM}(0)O_1^{(j)}\cdots O_{k-j+1}^{(k)}\right)}
{\text{tr}_{1\cdots L}\, e^{-\beta H(J,\Delta,h)}}\, ,\nonumber\\
&=
\lim_{L,N\rightarrow\infty}\frac{\text{tr}_{\ov 1\cdots \ov N}\left\{(\tr^{QTM}(0))^{j-1}\text{tr}(\T^{QTM}(0)O_1)\cdots\text{tr}
(\T^{QTM}(0)O_{k-j+1})(\tr^{QTM}(0))^{L-k}\right\}}
{\text{tr}_{1\cdots L}\, e^{-\beta H(J,\Delta,h)}}\, ,\nonumber\\
&=
\lim_{L,N\rightarrow\infty}\frac{\sum_{n=0}^{2^N-1}\Lambda_n^{L-k+j-1}(0)\langle\Psi_n|\text{tr}(\T^{QTM}(0)O_1)\cdots\text{tr}
(\T^{QTM}(0)O_{k-j+1})|\Psi_n\rangle}
{\sum_{n=0}^{2^N-1}\Lambda_n^{L}(0)}\, ,\nonumber\\
&=
\lim_{N\rightarrow\infty}\frac{\langle\Psi_0|\text{tr}(\T^{QTM}(0)O_1)\cdots\text{tr}
(\T^{QTM}(0)O_{k-j+1})|\Psi_0\rangle}
{\Lambda_0^{k-j+1}(0)}\, ,
\end{align}
Eq.~(\ref{c1s}) was obtained assuming that the QTM is diagonalizable and has ``normalized''
eigenvectors denoted by $|\Psi_n\rangle$. The eigenvector $|\Psi_0\rangle$ corresponds
to the largest eigenvalue $\Lambda_0(0)$. This assumption is valid in the case of the XXZ spin
chain \footnote{F.~G\"ohmann, {\it private communication.}} but in the case of other
lattice models
the rigorous mathematical proof is lacking.
From Sect. \ref{ABA6s} we know that the
eigenvectors of the QTM are of the form $|\Psi\rangle=B^{QTM}(\la_1)\cdots B^{QTM}(\la_p)|\Omega\rangle$.
We will say that an eigenvector $|\Psi\rangle$ is in the $p$ sector if in the previous
expression there are $p$ ``creation'' operators $B^{QTM}(\la)$ ($C^{QTM}(\la)$
can be interpreted as a ``destruction'' operator as a result of $C^{QTM}(\la)|\Omega\rangle=0$).
We will also assume that the eigenvectors in different sectors are orthogonal. Now we
can consider some particular cases of Eq.~(\ref{c1s}).

{\it Transversal correlation function.} First, we will consider the case $O_1=\sigma_-\, , O_{2,\cdots, k-j}=\mathbb{I}_2\, ,O_{k-j+1}=\sigma_+$.
Using $\text{tr}(\T^{QTM}(0)\sigma_-)=B^{QTM}(0)$ and $\text{tr}(\T^{QTM}(0)\sigma_+)=C^{QTM}(0)$, Eq.~(\ref{c1s})
becomes
\be\label{c2s}
\langle\sigma_-^{(j)}\sigma_+^{(k)}\rangle_T=\lim_{N\rightarrow\infty}\frac{\langle\Psi_0|B^{QTM}(0)(A^{QTM}(0)+D^{QTM}(0))^{k-j-1}C^{QTM}(0)|\Psi_0\rangle}
{\Lambda_0^{k-j+1}(0)}\, ,
\ee
Taking into account that $|\Psi_0\rangle$ is in the $N/2$ sector and the interpretation of the
$C^{QTM}(\la)$ as a ``destruction'' operator we are going to assume that the vector $|\Psi'\rangle=C^{QTM}(0)|\Psi_0\rangle$
and its adjoint $\langle\Psi'|=\langle\Psi_0|B^{QTM}(0)$ can be expanded as
\[
 |\Psi'\rangle=\sum_{i\in \frac{N}{2}-1\text{ sector}}c_i|\Psi_i\rangle\, ,\ \ \
  \langle\Psi'|=\sum_{i\in \frac{N}{2}-1\text{ sector}}\ov c_i\langle\Psi_i|\, .
\]
Using these expansions in (\ref{c2s}) we find
\begin{align}
\langle\sigma_-^{(j)}\sigma_+^{(k)}\rangle_T&=\sum_{i\in \frac{N}{2}-1\text{ sector}}B_i
\left(\frac{\Lambda^{(s)}_i(0)}{\Lambda_0(0)}\right)^{k-j}\, ,\nonumber\\
&=\sum_{i\in \frac{N}{2}-1\text{ sector}}B_i\, e^{-\frac{(k-j)}{\xi^{(s)}_i}}\, ,\ \ \ \ k\gg j\, ,
\end{align}
where $B_i$ are unknown constant coefficients, $1/\xi^{(s)}_i=\log(\Lambda_0(0)/\Lambda^{(s)}_i(0))$,
with $\Lambda^{(s)}_i(0)$ the eigenvalues of the QTM in the $N/2-1$ sector.

{\it Longitudinal correlation function.} In this case $O_1=\sigma_z, O_{2,\cdots,k-j}=\mathbb{I}_2, O_{k-j+1}=\sigma_z$
and $ \text{tr}(\T^{QTM}(0)\sigma_z)=A^{QTM}(0)-D^{QTM}(0)$. We obtain
\be\label{c33s}
\langle\sigma_z^{(j)}\sigma_z^{(k)}\rangle_T=\lim_{N\rightarrow\infty}\frac{\langle\Psi_0|(A^{QTM}(0)-D^{QTM}(0))
(A^{QTM}(0)+D^{QTM}(0))^{k-j-1}(A^{QTM}(0)-D^{QTM}(0))|\Psi_0\rangle}
{\Lambda_0^{k-j+1}(0)}\, ,
\ee
The vector $|\Psi'\rangle=(A^{QTM}(0)-D^{QTM}(0))|\Psi_0\rangle$
and its adjoint $\langle\Psi'|=\langle\Psi_0|(A^{QTM}(0)-D^{QTM}(0))$ can be expanded as
\[
 |\Psi'\rangle=\sum_{i\in \frac{N}{2}\text{ sector}}c_i|\Psi_i\rangle\, ,\ \ \
  \langle\Psi'|=\sum_{i\in \frac{N}{2}\text{ sector}}\ov c_i\langle\Psi_i|\, ,
\]
which means that
\begin{align}\label{c3s}
\langle\sigma_z^{(j)}\sigma_z^{(k)}\rangle_T&=\sum_{i\in \frac{N}{2}\text{ sector}}A_i
\left(\frac{\Lambda^{(ph)}_i(0)}{\Lambda_0(0)}\right)^{k-j}\, ,\nonumber\\
&=const+\sum_{i\in \frac{N}{2}\text{ sector}\, , i\ne 0}B_i\, e^{-\frac{(k-j)}{\xi^{(d)}_i}}\, ,\ \ \ \ k\gg j\, .
\end{align}
In the last line of Eq.~(\ref{c3s}) the constant term is the contribution of the
largest eigenvalue (which lies in the $N/2$ sector) and the sum is over all the
eigenvalues in the $N/2$ sector denoted by $\Lambda_i^{(ph)}(0)$
with $1/\xi^{(d)}_i=\log(\Lambda_0(0)/\Lambda^{(ph)}_i(0))$. Comparing with the
conformal result (\ref{asymptd}) we can identify the constant with $\langle \sigma_z^{(j)}\rangle_T^2$.

{\it Generating functional.}  The generating
functional for the $\sigma_z$ correlation functions is obtained for $O_1=O_2\cdots =O_{k-j+1}=\left(\begin{array}{lr}
1&0\\0&e^{\varphi}\end{array}\right)=e^{\varphi e_{22}}$. For $j=1, k=m$ we obtain
\be\label{c4s}
\langle e^{\{\varphi\sum_{n=1}^me^{(n)}_{22}\}}\rangle_T=\lim_{N\rightarrow\infty}\frac{\langle\Psi_0|
(A^{QTM}(0)+e^\varphi D^{QTM}(0))^{m}|\Psi_0\rangle}
{\Lambda_0^{m}(0)}\, .
\ee

In this case compared with (\ref{c2s}) or(\ref{c33s}) in (\ref{c4s}) appears $A^{QTM}(0)+e^\varphi D^{QTM}(0)$
instead of $\tr^{QTM}(0)=A^{QTM}(0)+D^{QTM}(0)$.  This does not pose a serious problem because
$\tr^{QTM}_\varphi=A^{QTM}(0)+e^\varphi D^{QTM}(0)$ is the so-called twisted QTM which can
be easily solved by noting that the considerations of Sect. \ref{ABA6s} applies also in the
case of $\tr^{QTM}_\varphi$ with $a(\la)=b(u',\la)^{N/2}e^{\beta h/2}\, ,\
d(\la)=b(\la,-u')^{N/2}e^{-\beta h/2+\varphi}$. Assuming that
\be\label{c5s}
|\Phi_0\rangle=\sum_{i\in \frac{N}{2}\text{ sector}}c_i|\Psi_i^{(\varphi)}\rangle\, ,\ \ \ \ \
\langle\Phi_0|=\sum_{i\in \frac{N}{2}\text{ sector}}\ov c_i\langle\Psi_i^{(\varphi)}|\, ,
\ee
where the sum is over the $N/2$ sector of the twisted QTM  we find
\begin{align}\label{agfexps}
\langle e^{\{\varphi\sum_{n=1}^me^{(n)}_{22}\}}\rangle_T&=\sum_{i\in \frac{N}{2}\text{ sector}}
C_i\left(\frac{\Lambda_i^{(\varphi)}(0)}{\Lambda_0(0)}\right)^m\, , \nonumber\\
&=\sum_{i\in \frac{N}{2}\text{ sector}}C_i e^{-\frac{m}{\xi^{(\varphi)}_i}}\, ,\ \ \ \ \ m\gg 1\, ,
\end{align}
where we have denoted by $\Lambda_i^{(\varphi)}(0)$ the eigenvalues of the twisted QTM
in the $N/2$ sector and $1/\xi^{(\varphi)}_i=\log(\Lambda_0(0)/\Lambda_i^{(\varphi)}(0))$.

\end{document}